\RequirePackage[2020-02-02]{latexrelease}

\documentclass[twocolumn,preprintnumbers,aps,amsmath,amssymb]{revtex4}

\usepackage{graphicx}
\usepackage{dcolumn}
\usepackage{bm}
\usepackage{color,epsfig,graphics,graphicx}

\begin{document}
	
	\title{
		Production of super-heavy nuclei in hot-fusion reactions
	}%
	
	\author{
		V. Yu. Denisov$^{1,2,3}$
	}%
	
	\affiliation{%
		$^{1}$ INFN Laboratori Nazionali di Legnaro, Legnaro, Italy \\
		$^{2}$ Institute for Nuclear Research, Kiev, Ukraine\\
		$^{3}$ Faculty of Physics, Taras Shevchenko National University of Kiev, Kiev, Ukraine \\
	}%
	
	\date{\today}
	
	\begin{abstract}
		A model for hot-fusion reactions leading to the synthesis of super-heavy nuclei is discussed. The values of the hot-fusion cross-sections obtained in the model agree with the available experimental data. The hot-fusion cross-sections are found for two different models of the fission barrier heights of super-heavy nuclei. The calculations of the cross-sections for the various hot-fusion reactions leading to the 119 and 120 elements are presented. Simple expressions useful for qualitative analysis of the cross-section for forming super-heavy nuclei are obtained. It is shown that the super-heavy nuclei production cross section is proportional to the transmission coefficient of the capture barrier, realization probability of the $xn$-evaporation channel, and exponentially depends on the quasi-elastic barrier, fusion reaction Q-value, compound nucleus formation barrier, neutron separation energies, and fission barrier heights.
	\end{abstract}
	
	\maketitle
	
	\section{Introduction}
	
	The synthesis of super-heavy nuclei (SHN) is a very interesting, exciting, and puzzling task, as much for experimentalists as for theoreticians. The elements beyond Nh with proton numbers $Z= 114-118$ have been synthesized in the hot heavy ion fusion reactions [1-26]. The Og, with $Z=118$, is the heaviest element that has been synthesized to date, discovered by Oganessian et al. \cite{prc74,prc98}. Recently, experiments aimed at the synthesis of isotopes of elements Z = 119 and 120, or the study of properties of related reactions have been performed \cite{119,120,120a,120b,120c,120d,120e,120f,119a}. Still, no decay chains consistent with the fusion-evaporation reaction products have been observed. 
	
	The synthesis of SHN with proton numbers $Z= 108-118$ in collisions of a $^{48}$Ca projectile nucleus with heavy target nuclei from Ra to Cf [1-26] is called a hot fusion reaction because the excitation energies of the compound nucleus formed in these reactions are relatively high. The excited compound nucleus synthesized in the hot fusion reaction is cooled by the emission of 2-5 neutrons [1-26]. The neutron emission competes with the fission of the excited compound nucleus.
	
	Recently, many various models dedicated to the description of the SHN production cross sections in hot fusion reactions have been presented, see, for example, Refs. [36-55] and papers cited therein. The calculation of the production cross-section is the product of the capture cross-section, the probability of the compound nucleus formation, and the survival probability of the evaporation residue. 
	
	The survival probability is proportional to the ratio of the neutron emission width $\Gamma_n$ to the fission width $\Gamma_f$. In the model of constant temperature level density $\rho(E) \propto \exp{\left(\varepsilon /{\cal T}\right)}$, where $\varepsilon$ is the excitation energy of the nucleus, ${\cal T}$ is the temperature, the ratio of widths \cite{vh} is
	\begin{eqnarray}
		\Gamma_n/\Gamma_f \approx c_{\cal T} \exp{[(B_f-S)/{\cal T}]}.
	\end{eqnarray} 
	Here $B_f$ is the fission barrier height, $S$ is the neutron separation energy from the nucleus, $c_{\cal T}=0.2 {\cal T} A^{2/3}$, and $A$ is the number of nucleons in the SHN. As a result, the SHN production cross-section strongly depends on $S_n$ and $B_f$.
	
	The results of the first systematic calculations of the neutron separation energies and the fission barrier heights for a wide range of the SHN in the framework of one model are presented in Refs. \cite{frdm,frdm_fb}. The results of similar systematic calculations performed in another model have been presented recently in Ref. \cite{jks}. Note that the experimental neutron separation energy can be only extracted for some SHN by using the recent atomic mass table \cite{be}. The theoretical values of the neutron separation energy can be found using the atomic mass table obtained in various nuclear mass models too, for example, see Refs. \cite{ms,ws4}. 
	
	The difference between the neutron separation energies obtained in different models \cite{frdm,jks,be,ms,ws4} for most SHN with $110 \leq Z \leq 122$ are close or smaller than 0.8 MeV, see Fig. 1. In contrast to this, the difference between the barrier values calculated in the models \cite{frdm_fb} and \cite{jks} can reach $\sim$4 MeV for SHN formed in hot-fusion reactions, see Fig. 2. According to Eq. (1), the survival probability depends exponentially on the difference $B_f-S_n$, therefore, the model with higher values of the fission barrier heights leads to strongly higher values of the ratio $\Gamma_n/\Gamma_f$, the survival probability, and the SHN production cross sections in hot fusion reactions. The difference between the values of the SHN production cross sections obtained for SHN with $Z = 120$ using the different fission barriers reach 2 orders \cite{kaas}. This result agrees with the exponential dependence of the survival probability on the difference $B_f-S_n$, see Eq. (1). So, the use of different fission barrier values leads to a strong change in the reaction mechanisms in the description of the existing experimental data for the SHN production cross-section. 
	
	The theory of the capture stage of the SHN production is well-known because there are experimental data for the capture cross section and capture barrier heights \cite{fus_exp,qelastbar}. Note that various models and a huge number of experimental data are available for heavy-ion fusion reactions leading to the light, medium, and heavy nuclei \cite{nrv,dhrs,montstef,d22subfus}. The description of the capture process applied for SHN production models in Refs. [36-55] are similar to the ones used for lighter heavy ion systems. Therefore, various approaches to the capture process presented in Refs. [35-54] give close results. 
	
	The calculation of the survival probability of the SHN is based on the nuclear evaporation model, which is well backgrounded. The description of the survival probability considered in the models [36-55] is similar and leads to similar results.
	
	In contrast to these, the stage of the compound nucleus formation is less known and the experimental data is insufficient to fix well the values of the compound nucleus formation probability. As a result, there are many various approaches to the description of the SHN formation probability in heavy-ion reactions, see, for example, Refs. [35-54] and papers cited therein. 
	
	\begin{figure}
		\includegraphics[width=8.5cm]{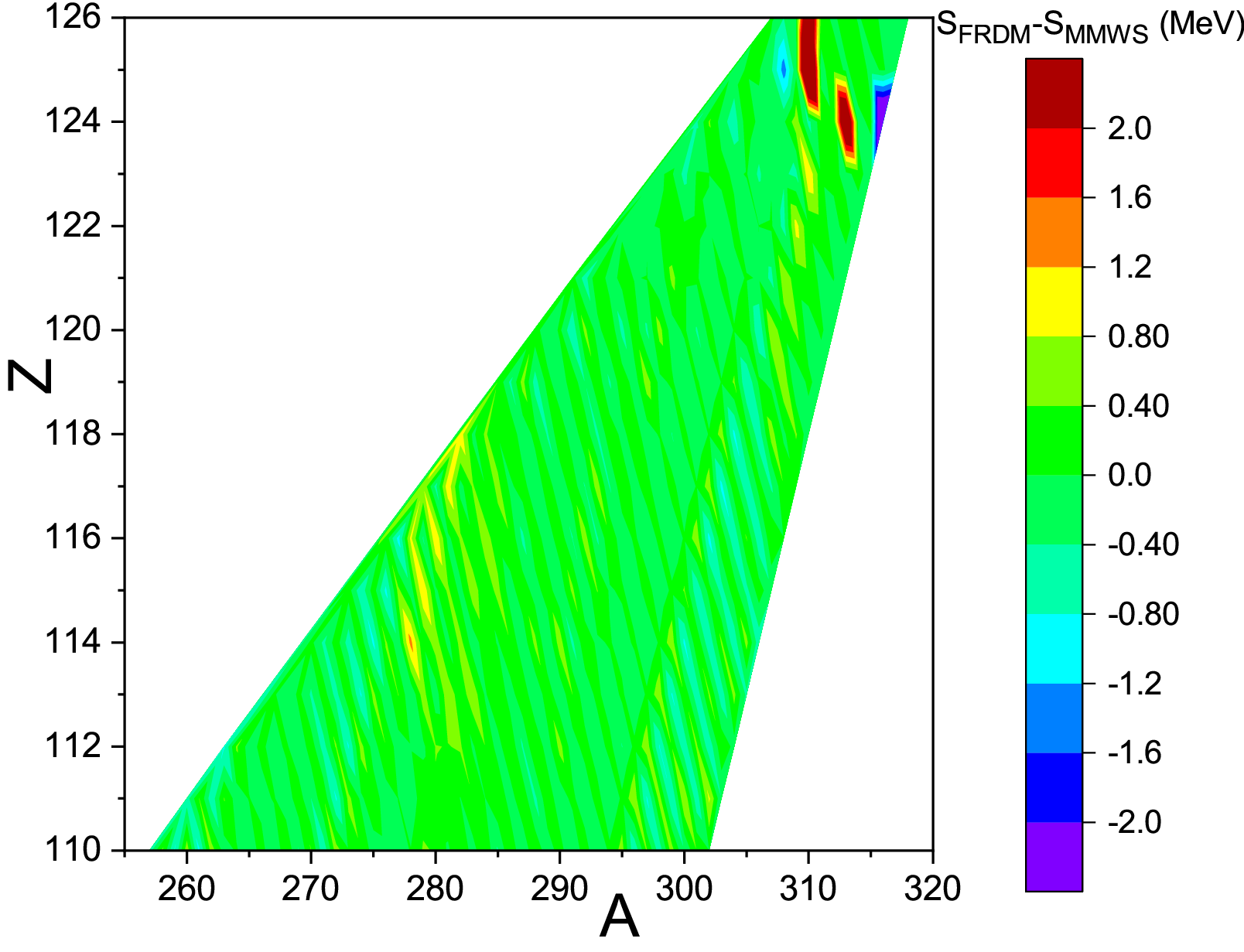}
		\caption{\label{fig1} Mass $A$ and charge $Z$ dependencies of the difference of the neutron separation energies $S_{\rm FRDM} -S_{\rm MMWS}$, where the values of neutron separation energies $S_{\rm FRDM}$ and $S_{\rm MMWS}$ are taken from Refs. \cite{frdm_fb} and \cite{jks} correspondingly. }
	\end{figure}
	
	\begin{figure}
		\includegraphics[width=8.5cm]{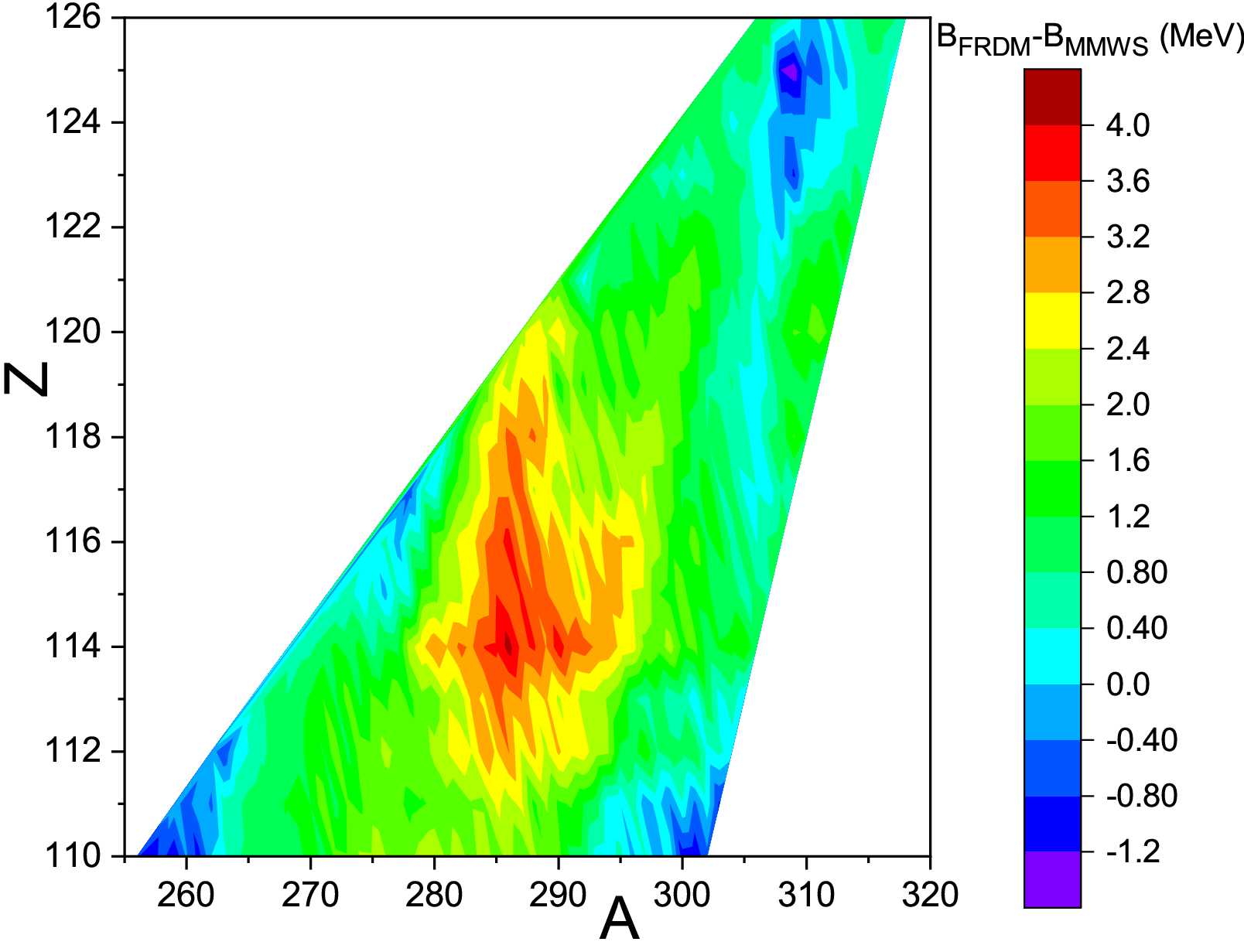}
		\caption{\label{fig2} Mass $A$ and charge $Z$ dependencies of the difference of the fission barriers $B_{\rm FRDM} -B_{\rm MMWS}$ for $\ell=0$, where the values of barrier heights $B_{\rm FRDM}$ and $B_{\rm MMWS}$ are taken from Refs. \cite{frdm_fb} and \cite{jks} correspondingly. }
	\end{figure}
	
	Therefore, it is interesting to consider the SHN production cross-sections calculated for various collisional systems in the framework of the same model of the synthesis by using the different values of the fission barriers of SHN from Refs. \cite{frdm,frdm_fb} and \cite{jks}. Note that the fission barrier heights from model \cite{frdm,frdm_fb} have been used in most parts of the models for the SHN production cross-sections. The changes in the model parameters induced by the different values of the fission barrier height in the case of experimentally known values of the SHN production cross-sections should be discussed too. These changes can be only coupled to the description of the probability of the compound nucleus formation because the other stages of SHN production are well-defined. The predictions of the production cross-sections for SHN with $Z > 118$ in different fission barrier models are important for planning future experiments. The study of these items is the goal of this paper. 
	
	The simple expression for the SHN production cross sections obtained using the model of constant temperature for the level density is also discussed in the paper. This expression is applied for the qualitative analysis of the SHN cross-section dependence on the model parameters.
	
	The models introduced for the SHN production in the cold fusion reactions \cite{ds21} and compound nucleus formation in heavy-ion reactions \cite{d23d} are extended for the case of the hot fusion reactions now. Remind that the projectile and target are spherical in the case of the cold fusion reactions, while the spherical projectile and deformed target collide in the case of the hot fusion reactions. The excitation energies of the SHN formed in the cold fusion reactions are smaller than the ones in the hot fusion reactions.
	
	The model is described in Sec. 2. The results and conclusions are given in sections 3 and 4, respectively. 
	
	\section{Model}
	
	The cross-section of the SHN synthesis in collisions of nuclei, with the subsequent emission of $x$ neutrons from the formed compound nucleus in competition with fission, is given as
	\begin{eqnarray}
		\sigma_{xn}(E) &=& \frac{\pi \hbar^2}{2\mu E} \sum_\ell \sigma_{xn\ell}(E) 
		\\ &=& \frac{\pi \hbar^2}{2\mu E} \sum_\ell (2 \ell +1) T_\ell(E) \times P_\ell(E) \times W^{xn}_\ell(E). \nonumber
	\end{eqnarray}
	Here $\sigma_{xn\ell}(E)$ is the partial cross-section, $\mu$ is the reduced mass, and $E$ is the collision energy of the incident nuclei in the center of the mass system. $T_\ell(E)$ is the transmission coefficient through the fusion (capture) barrier formed by the Coulomb, centrifugal, and nuclear parts of the nucleus-nucleus interaction, $P_\ell(E)$ is the probability of compound nucleus formation, and $W^{xn}_\ell(E)$ is the survival probability of the compound nucleus related to the evaporation of $x$ neutrons in competition with fission. In the cases of experimentally measured hot fusion reactions [1-25], $x$ equals $3$ or $4$ and rarely $2$ or $5$ because the excitation energy of the compound nucleus is relatively high. In comparison to this, the number of nucleons evaporated from the compound nucleus in the cold fusion reactions is 1 or 2 as a rule \cite{ds21,hm}. 
	
	The collision of the spherical projectile nucleus with the deformed axial-symmetric actinide target nucleus occurs in the hot fusion reactions [1-25]. Therefore, the total transmission coefficient is averaged on all orientations of the deformed nucleus
	\begin{eqnarray}
		T_\ell(E) = \frac{1}{2}\int_{0}^\pi d\theta_2 \sin(\theta_2) T_\ell(E,\theta_2),
	\end{eqnarray}
	where $\theta_2$ is the orientation angle of the deformed nuclei \cite{d23a}. The orientation is linked to the space angle between the line connecting the mass centers of the colliding nuclei and the axial-symmetry axis of the deformed nucleus. The transmission coefficient $T_\ell(E,\theta_2)$ through the capture barrier $V(r,\theta_2)$ can be calculated using the Ahmed formula \cite{ahmed} 
	\begin{eqnarray}
		T_\ell(E,\theta_2) = \frac{1-\exp{\left(-4 \pi \alpha_\ell(\theta_2)\right)}}{1+\exp{\left[ 2\pi \left( \beta_\ell(\theta_2)-\alpha_\ell(\theta_2) \right) \right]}},
	\end{eqnarray}
	where $\alpha_\ell(\theta_2)=\frac{2 (B_\ell(\theta_2) E)^{1/2}}{\hbar \omega_\ell(\theta_2)}$ and $\beta_\ell (\theta_2) = \frac{2B_\ell(\theta_2)}{\hbar \omega_\ell(\theta_2)}$. Here $B_\ell(\theta_2) = V^{\rm t}_\ell(r_b(\theta_2),\theta_2)$ is the capture barrier height, $r_b(\theta_2)$ is the radius of the barrier, and $\left. \hbar \omega_\ell(\theta_2)=\left(-\frac{\hbar^2}{\mu} \frac{d^2V^{\rm t}_\ell(r,\theta_2)}{dr^2} \right)^{1/2} \right|_{r=r_b(\theta_2)}$ is the barrier curvature. Ahmed obtained the exact expression for the transmission coefficient through the Morse potential barrier \cite{morse}. The total nucleus-nucleus potential between the spherical and deformed nuclei $V^{\rm t}_\ell(r,\theta_2)$ is determined in Ref. \cite{d23a,d23d}. It consists of the Coulomb, nuclear, and centrifugal interactions of the incident nuclei. The shape of the realistic total nucleus-nucleus potential is closer to the shape of the Morse potential than the parabolic one, see for details Ref. \cite{d23b} and papers cited therein. Therefore, Ahmed's expression for the transmission coefficient is more suitable than the corresponding expression for the parabolic barrier \cite{kemble,hw}. The difference between the Ahmed and parabolic transmission coefficients is important for sub-barrier energies \cite{d23b}. 
	
	After penetration of the capture barrier, the stuck-together nuclei are formed in the capture well \cite{dn}. The stuck-together nuclei are populated states at the capture well of the total potential as a rule \cite{dn}. As a rule, the kinetic energy of relative motion is completely dissipated into inner degrees of freedom due to the strong dissipation caused by overlapping some parts of approaching nuclei during the collision \cite{gk,frobrich,fl}. The uniform temperature of nuclear matter is quickly set in the system of the stuck-together nuclei. The most statistically important shape of the stuck-together nuclei is axial-symmetric because such a shape has the lowest interaction potential energy and, therefore, the highest value of the inner excitation energy. Due to this, all subsequent evolution stages of the stuck-together nuclei can be considered statistically using the Bohr-Wheeler transition state approximation \cite{bw} and they are independent of the orientation angle $\theta_2$.
	
	\subsection{The probability of the compound nucleus formation}
	
	The probability of the compound nucleus formation in the partial wave $\ell$ in the statistical approach is \cite{ds21,d23d}
	\begin{eqnarray}
		P_\ell(E) = \frac{\Gamma^{\rm cnf}_\ell(E) }{\Gamma^{\rm s}_\ell(E) } = \frac{1}{1+G_\ell(E)}.
	\end{eqnarray} 
	Here $\Gamma^{\rm cnf}_\ell(E) $ is the decay width of the stuck-together nuclei to the compound nucleus states, $\Gamma^{\rm s}_\ell(E) =\Gamma^{\rm cnf}_\ell(E) +\Gamma^{\rm d}_\ell(E)$ is the total decay width of the stuck-together nuclei, and $G_\ell(E)= \frac{\Gamma^{\rm d}_\ell(E)}{\Gamma^{\rm cnf}_\ell(E)}$. $\Gamma^{\rm d}_\ell(E) $ is the decay width of the stuck-together nuclei into all channels leading to the two separated nuclei.
	
	The width $\Gamma^{\rm cnf}_\ell(E)$ can be linked to the compound nucleus formation barrier $B^{\rm cnf}_\ell$, which takes place at the smooth shape evolution from the stuck-together nuclei to the spherical or near-spherical compound nucleus. This barrier $B^{\rm cnf}_\ell$ is related to the one-body shape evolution as the ordinary fission barrier \cite{vh,strut4}. The minimal value of $B^{\rm cnf}$ may be estimated similar to the height of the fission barrier because the probability of both the fission and the compound-nucleus formation is related to the trajectory of minimal action \cite{strut4}, which connects the compound nucleus and two separated nuclei. However, during heavy-ion fusion and fission, the collective coordinates describing these processes are changed in opposite directions. Therefore, the width $\Gamma^{\rm cnf}_\ell(E)$ can be found similar to the fission width applying the Bohr-Wheeler approximation of the transition state \cite{bw}. The detailed description of the expressions for the width $\Gamma^{\rm cnf}_\ell(E)$ is presented in Ref. \cite{ds21,d23d}. Due to this, the expressions for the calculation of the width with short comments are given below only. 
	
	The width for passing the compound nucleus formation barrier $B^{\rm cnf}_\ell(\varepsilon)$, which takes place between the stuck-together incident nuclei and compound nucleus in the equilibrium shape, can be written as
	\begin{eqnarray}
		\Gamma^{\rm cnf}_\ell(E)=\frac{2}{2\pi \rho_{\rm sn}(E)} \int_0^{\varepsilon_{\rm max}^{\rm cn}} d\varepsilon \frac{\rho_A(\varepsilon)}{N_{\rm tot}^{\rm cn}} N^{\rm cnf}(\varepsilon).
	\end{eqnarray}
	Here the barrier height $B^{\rm cnf}_\ell(\varepsilon)$ depends on the excitation energy, $\rho_{\rm sn}(E)$ is the level density of the stuck-together nuclei (the level density of the initial state) and $\rho_A(\varepsilon)$ is the level density of the compound nucleus with $A$ nucleons formed in the heavy-ion collision. The ratio $\rho_A(\varepsilon)/N_{\rm tot}^{\rm cn}$ is the probability of finding the nuclear system passing through the barrier with the intrinsic (thermal) excitation energy $\varepsilon$ in the over-barrier transition states.
	\begin{eqnarray}
		N_{\rm tot}^{\rm cn}= \int_0^{\varepsilon_{\rm max}^{\rm cn}} d\varepsilon \rho_A(\varepsilon)
	\end{eqnarray}
	is the total number of states available for barrier passing in the case of the energy-dependent barrier of compound nucleus formation $B^{\rm cnf}(\varepsilon)$,
	\begin{eqnarray}
		N^{\rm cnf}(\varepsilon)
		= \int^{E+Q-B^{\rm cnf}_\ell(\varepsilon)}_{\varepsilon} de \rho_A(e) \;\;
	\end{eqnarray}
	is the number of states available for the nuclear system passing through the barrier at the thermal excitation energy $\varepsilon$, and $Q$ is the fusion reaction Q-value. Note that $B^{\rm cnf}_\ell(\varepsilon)$ and $E + Q$ are, respectively, the barrier height and the excitation energy of the compound nucleus evaluated relatively the ground-state of the compound nucleus formed in the fusion reaction. $\varepsilon_{\rm max}^{\rm cn}$ is the maximum value of the thermal excitation energy of the compound nucleus at the saddle point, which is determined as the solution of the equation
	\begin{eqnarray}
		\varepsilon_{\rm max}^{\rm cn} + B^{\rm cnf}_\ell(\varepsilon_{\rm max}^{\rm cn})= E + Q.
	\end{eqnarray}
	
	The back-shifted Fermi gas model \cite{bsfgm,ripl3} is used for a description of the level density $\rho_A(\varepsilon)$ of the nucleus with $A$ nucleons. The level density in this model is given by
	\begin{eqnarray}
		\rho_A(\varepsilon)=\frac{\pi^{1/2}\exp{\left[2 \sqrt{a_A(\varepsilon-\Delta) \times (\varepsilon-\Delta)}\right]}}{12 [a_A(\varepsilon-\Delta)]^{1/4} (\varepsilon-\Delta)^{5/4}},
	\end{eqnarray}
	where
	\begin{eqnarray}
		a_A(\varepsilon) = a^0_A \left\{1+\frac{E_{\rm shell}^{\rm emp}}{\varepsilon}[1-\exp{(-\gamma \varepsilon)}] \right\}
	\end{eqnarray}
	is the level density parameter \cite{ist,ripl3}. Here
	\begin{eqnarray}
		a^0_A=0.0722396 A+0.195267 A^{2/3} \; {\rm MeV}^{-1}
	\end{eqnarray}
	is the asymptotic level density parameter obtained at high excitation energies, when all shell effects are damped \cite{ist,ripl3}, $E_{\rm shell}^{\rm emp}$ is the empirical shell correction value \cite{ripl3,mn}, $\gamma=0.410289/A^{1/3}$ MeV$^{-1}$ is the damping parameter \cite{ist,ripl3}, and $A$ is the number of nucleons in the nucleus. According to the prescription of Ref. \cite{ripl3}, the value of empirical shell correction $E_{\rm shell}^{\rm emp}$ is calculated as the difference between the experimental value of nuclear mass and the liquid drop component of the mass formula \cite{ripl3,mn}. The back shift energy is described by the following expression $\Delta=12n/A^{1/2}+0.173015$ MeV \cite{ripl3}, where $n = -1, 0$ and 1 for odd-odd, odd-$A$, and even-even nuclei, respectively. 
	
	The width $\Gamma^{\rm d}_\ell(E)$ includes the contributions of the elastic $\Gamma^{\rm e}_\ell(E)$, quasi-elastic $\Gamma^{\rm qe}_\ell(E)$, single- and many-particle transfers $\Gamma^{\rm t}_\ell(E)$, deep-inelastic $\Gamma^{\rm di}_\ell(E)$, and quasi-fission $\Gamma^{\rm qf}_\ell(E)$ decays of the stuck-together nuclei \cite{ds21}. As a result, $\Gamma^{\rm d}_\ell(E) = \Gamma^{\rm e}_\ell(E) + \Gamma^{\rm qe}_\ell(E) + \Gamma^{\rm t}_\ell(E) + \Gamma^{\rm di}_\ell(E) + \Gamma^{\rm qf}_\ell(E).$ The quasi-elastic barrier $B^{\rm qe}_\ell$, which separates the contacting and well-separated deformed incident nuclei, has the lowest barrier height among all barriers related to processes accounted to $\Gamma^{\rm d}_\ell(E)$ \cite{ds21}. Therefore, the width of the quasi-elastic decay of the stuck-together nuclei is the leading contribution to $\Gamma^{\rm d}_\ell(E)$, i.e. $\Gamma^{\rm d}_\ell(E)\approx \Gamma^{\rm qe}_\ell(E)$ and $G_\ell(E) \approx \frac{\Gamma^{\rm qe}_\ell(E)}{\Gamma^{\rm cnf}_\ell(E)}$. 
	
	The expressions for the width $\Gamma^{\rm d}_\ell(E)$ is
	\begin{eqnarray}
		\Gamma^{\rm qe}_\ell(E) = \frac{1}{2\pi \rho_{\rm sn}(E)} \int_0^{E-B^{\rm qe}_\ell} d\varepsilon \int_0^{\varepsilon} d\epsilon \; \rho_{A_1}(\epsilon) \times \nonumber \\ \rho_{A_2}(\varepsilon-\epsilon) .
	\end{eqnarray}
	Here $B^{\rm qe}_\ell$ the value of the quasi-elastic barrier calculated relatively the interaction potential energy of two nuclei on the infinite distance between them. $A_i$ is the number of nucleons in incident nucleus $i$, $i=1,2$, $A=A_1+A_2$ is the number of nucleons in the compound nucleus. The height of the quasi-elastic barrier $B^{\rm qe}$ of the total interaction potential $V^{\rm t}_\ell(r,\{\beta_1\},\{\beta_2\})$ separates the sticking and well-separated nuclei. The total interaction potential energy of two nuclei consists of the Coulomb $V_{\rm C}(r,\{\beta_1\},\{\beta_2\}$, nuclear $V_{\rm N}(r,\{\beta_1\},\{\beta_2\})$, centrifugal $V_\ell(r,\{\beta_1\},\{\beta_2\})$, and deformation energies of the both nuclei $E_{\rm def}^i(\{\beta_i\})$ \cite{ds21,d23d}, i.e. 
	\begin{eqnarray}
		V^{\rm t}_\ell(r,\{\beta_1\},\{\beta_2\})=V_{\rm C}(r,\{\beta_1\},\{\beta_2\}) + \nonumber \\ V_{\rm N}(r,\{\beta_1\},\{\beta_2\}) + V_\ell(r,\{\beta_1\},\{\beta_2\}) + \nonumber \\ E_{\rm def}^1(\{\beta_1\})+ E_{\rm def}^2(\{\beta_2\}).
	\end{eqnarray}
	Here $\{\beta_i\}=\beta_{i2},\beta_{i3}$ are the surface deformation parameters of nucleus $i$ with the surface radius $R_i(\theta)=R_{0i}\left[ 1+\sum_{L=2,3} \beta_{iL} Y_{L0}(\theta)\right]$, $i=1,2$, $R_{0i}$ is the radius of the spherical nucleus, and $Y_{L0}(\theta)$ is the spherical harmonic function \cite{vmk}. The detail description of the $V^{\rm t}_\ell(r,\{\beta_1\},\{\beta_2\})$ is given in Ref. \cite{ds21,d2022f,d23d}. The deformation energy of the nucleus induced by the surface multipole deformations is
	\begin{eqnarray}
		E_{\rm def}^{i}(\{ \beta_{iL} \}) = \sum_{L=2}^3 \left[ C^{\rm ld}_{L A_i Z_i} + C^{\rm sh}_{L A_i Z_i}\right]\frac{\beta_{iL}^2}{2}. \;\;\;
	\end{eqnarray}
	Here $C^{\rm ld}_{LA_iZ_i}= \frac{(L-1)(L+2) b_{\rm surf} A^{2/3}_i}{4 \pi} - \frac{3(L-1) e^2 Z^2_i}{2\pi(2L+1)R_{0i}}$ is the surface stiffness coefficient obtained in the liquid-drop approximation \cite{bm,wong68}, and $b_{\rm surf}$ is the surface coefficient of the mass formula \cite{frdm}. $C_{\rm sc}$ is the shell-cor\-rection contribution to the stiffness coefficient. It is possible to approximate $C^{\rm sc} \approx - 0.05 \; \delta E \; C^{\rm ld}$ \cite{ds21,d2022f}, where $\delta E$ is the phenomenological shell-correction value in MeV. Note that experimental values of the surface stiffness coefficient for different nuclei are distributed around the value $C^{\rm ld}$ \cite{bm,wong68}. The approximation for the surface stiffness coefficient used in Eq. (15) is crude, but it is taken into account by the shell effect. This approximation corresponds to the experimental tendency of the values of the surface stiffness coefficient and simplifies further calculations strongly.
	
	So, the widths $\Gamma^{\rm cnf}_\ell(E)$ and $\Gamma^{\rm qe}_\ell(E)$ depend on the heights of the quasi-elastic $B^{\rm cnf}$ and the compound nucleus formation $B^{\rm qe}_\ell$ barriers correspondingly. The barrier heights $B^{\rm cnf}_\ell$ and $B^{\rm qe}_\ell$ determine the probability of the compound nucleus formation $P_\ell(E)$. $P_\ell(E)$ does not depend on $\rho_{\rm sn}(E)$ because it depends on the ratio of the width, see Eqs. (5), (6) and (13). The height of the compound nucleus formation barrier $B^{\rm cnf}_\ell-Q$ is much higher than the height of the quasi-elastic barrier $B^{\rm qe}_\ell$ for reactions leading to super-heavy systems \cite{ds21}. Therefore, the formation of the compound nucleus for super-heavy systems is strongly suppressed. 
	
	In the collision of identical or almost identical falling nuclei, the compound nucleus formation barrier $B^{\rm cnf}$ is close to the fission barrier because fusion and fission are to some extent mutually inverse processes \cite{d23d}. For very asymmetric collision systems, the height of this barrier is close to the barrier height of the corresponding cluster emission \cite{d23d} because the cluster emission barrier is related to strongly asymmetric fission \cite{spg,rgd,mirea,wzr,matheson}. The height of the cluster barrier is approximately four times higher than the ordinary fission barrier in actinides as a rule \cite{mirea,rgd,wzr}. The height of the cluster emission barrier for SHN decreases with the rise of the number of protons $Z$ in SHN \cite{wzr}. The fission and cluster emission barriers are equal for SHN with the number of protons $Z\gtrsim 112$ \cite{wzr,matheson}. Therefore, the compound nucleus formation barrier height can be defined as $B^{\rm cnf}_\ell=B^{\rm f}_\ell + b^{\rm cnf}$, where $B^{\rm f}_\ell$ is the fission barrier height and $b^{\rm cnf}$ is the difference between the compound nucleus formation and the fission barrier heights. The value of parameter $b^{\rm cnf}$ smoothly decreases with an increase of $Z$ and should be close to zero for $Z\gtrsim 112$ according to Refs. \cite{wzr,matheson}. 
	
	The approach for calculating the quasi-elastic barrier height $B^{\rm qe}$ is described in detail in Refs. \cite{ds21,d23d}. Due to this, the description of this approach is omitted here. 
	
	The ratio of widths $\Gamma^{\rm cnf}_\ell(E) / \Gamma^{\rm qe}_\ell(E)$ can be simplified using the exponential dependence of the level density on the excitation energy \cite{d23d}. In the approach of constant temperature ${\cal T}$ for the level density and taking into account that $B^{\rm cnf}_\ell-Q$ is sufficiently larger $B^{\rm qe}_\ell$ in the case of the hot fusion reactions, the ratio of the widths $\Gamma^{\rm cnf}_\ell(E) / \Gamma^{\rm qe}_\ell(E) \ll 1$ and the probability of compound nucleus formation can be presented in the simple form 
	\begin{eqnarray}
		P_\ell(E) \approx \frac{\Gamma^{\rm cnf}_\ell(E)}{\Gamma^{\rm qe}_\ell(E)} \approx \exp{\{[B^{\rm qe}_\ell-(B^{\rm cnf}_\ell-Q)]/{\cal T} \} } = \nonumber \\
		\exp{\left\{ \left[B^{\rm qe}_\ell-\left(B^{\rm cnf}_0 +\frac{\hbar^2 \ell(\ell+1)}{2J^{\rm cnf}}-Q\right) \right]/{\cal T} \right\} }. \;
	\end{eqnarray}
	Here $J^{\rm cnf}=\frac{2}{5}M R_{0}^2 A \left( 1+\sqrt{\frac{5}{16 \pi}} \beta_{\rm cnf} + \frac{135 }{84 \pi} \beta_{\rm cnf}^2 \right)$ is the moment of inertia of the nucleus at the compound nucleus formation barrier, where $R_0=r_0 A^{1/3}$ is the radius of spherical compound nucleus, $\beta_{\rm cnf}$ is the quadrupole deformation of the nucleus at the barrier, and $M$ is the nucleon mass. The value of $\beta_{\rm cnf}$ coincides with the value of the quadrupole deformation parameter in the fission barrier saddle point. The solid-state expression for the moment of inertia is used because the excitation energy of the nucleus formed in the hot fusion reaction is sufficiently high. 
	
	Note that the moment of inertia of the system in the compound nucleus barrier saddle point $J^{\rm cnf}$ is smaller than the one in the quasi-elastic barrier $J^{\rm qe} \approx \mu r_{\rm qe}^2$, where $r_{\rm qe}$ is the radius of the quasi-elastic barrier. Therefore, the values of $B^{\rm qe}_\ell + Q-B^{\rm cnf}_\ell=B^{\rm qe}_0 + Q-B^{\rm cnf}_0+\frac{\hbar^2 \ell(\ell+1)}{2} \left( \frac{1}{J^{\rm qe}} -\frac{1}{J^{\rm cnf}} \right)$ and $P_\ell(E)$ decrease with increasing of $\ell$, see also Ref. \cite{d23d}. Since $B ^{\rm cf}_\ell$ is sufficiently larger than $B^{\rm qe}_\ell -Q$ in the case of hot fusion reactions for any $\ell$, then $P_\ell(E)\ll 1$. 
	
	The simplified expression for the compound nucleus formation (16) directly shows the dependence of the probability of compound nucleus formation to the barrier heights $B^{\rm qe}_\ell$ and $(B^{\rm cnf}_\ell-Q)$. So, the compound nucleus formation probability depends on the fusion reaction $Q$-value. In contrast to this, the dependence on the probability of compound nucleus formation on $Q$ does not appear in the dinuclear models of SHN productions \cite{kaas,aal,b,lwz,dz} because the formation of compound nucleus in this model links to the sequential nucleon transfer from the heavy nuclei to light the ones. The various phenomenological functions, which are different from Eq. (16), are used for parametrization of the probability of the compound nucleus formation in Refs. \cite{zg,zwr,ss,msmdss,lv,rkb,zs}.
	
	In the case of fermi gas level density, the simplified equation for the probability of compound nucleus formation, which is similar to Eq. (16), is obtained in Ref. \cite{d23d}.
	
	\subsection{The survival probability of the compound nucleus}
	
	The survival probability of the compound nucleus formed in the hot-fusion reaction is related to the competition between the evaporation of $x$ neutrons and fission. The survival probability can be approximated as
	\begin{eqnarray}
		W^{xn}_\ell(E) = R_{xn}(E_{{\rm CN}\ell}^\star) \frac{\Gamma_{1n}(E^\star_1,\ell)}{\Gamma_{1n}(E^\star_{1n},\ell)+\Gamma_{\rm f}(E^\star_1,\ell)} \nonumber \\ \times \frac{\Gamma_{2n}^{A-1}(E^\star_2,\ell)}{\Gamma_{2n}^{A-1}(E^\star_2,\ell)+\Gamma_{\rm f}^{A-1}(E^\star_2,\ell)} \times .. \nonumber \\ \times \frac{\Gamma_{xn}^{A-x+1}(E^\star_x,\ell)}{\Gamma_{xn}^{A-x+1}(E^\star_x,\ell)+\Gamma_{\rm f}^{A-x+1}(E^\star_x,\ell)}.
	\end{eqnarray}
	Here $R_{xn}(E^\star)$ is the realization probability of the $xn$-evaporation channel \cite{jack}, $E^\star_{{\rm CN}\ell}= E-Q-\hbar^2 \ell(\ell+1)/(2J_{\rm gs})$, and $J_{\rm gs}$ is the ground-state moment of inertia. $E^\star_1=E-Q$ is the excitation energy of the compound nucleus formed in the heavy-ion fusion reaction.
	$\Gamma_{yn}^{A-y+1}(E^\star_y,\ell)$ and $\Gamma_{\rm f}^{A-y+1}(E^\star_y,\ell))$ are, respectively, the width of neutron emission and the fission width of the compound nucleus formed after emission of $(y-1)$ neutrons. $E^\star_y=E^\star_{y-1}-S_{y-1} - 2{\cal T}_{y-1}$ is the excitation energy before evaporation of the $y$-th neutron, where $S_{y-1}$ is the separation energy of the $(y-1)$-th neutron. ${\cal T}_{y-1}$ is the temperature of the compound nucleus after evaporation of $(y-1)$ neutrons and is obtained from $E^\star_{y-1}= a_A {\cal T}_{y-1}^2$. Here $a_A$ is the level density $\rho_A(\varepsilon)$ parameter, see Eq. (11).
	
	The width of neutron emission from a nucleus with $A$ nucleons is given as \cite{vh}
	\begin{eqnarray}
		\Gamma_n(E^\star,\ell) = \frac{2 g_n M R_n^2}{\pi\hbar^2 \rho_{A}(E^\star)} \int_0^{E^\star-S} d\varepsilon \; \varepsilon \nonumber \\ \times \rho_{A-1}(E^\star-S-\varepsilon), \;\;
	\end{eqnarray}
	where $S$ is the neutron separation energy from the nucleus, $\rho_{A}(E^\star)$ and $\rho_{A-1}(E^\star)$ are, correspondingly, the energy level densities of the compound nuclei before and after neutron emission, $g_n$ is the neutron intrinsic spin degeneracy, and $R_n$ is the radius of the neutron-nucleus interaction. 
	
	The width for passing the fission barrier was introduced by Bohr and Wheeler in 1939 \cite{bw}. Emphasize that the Bohr-Wheeler fission width is obtained for the fission barrier height independent of the thermal energy of the fissioning system. 
	
	As was shown by Strutinsky in 1966 the fission barrier height consists of the liquid-drop and shell-correction contributions \cite{strut1,strut2,strut3,strut4}. It was found in 1973 that the shell correction energy decreases strongly with an increase of the inner energy $\varepsilon$ of the system \cite{ach}. Due to this, the height of the fission barrier depends drastically on the inner energy $\varepsilon$ of the system \cite{ach,bq,dah,lpc,snp,pnsk,dh,ds21,ds18gg,ds18,dds22}. The dependence of the barrier height on the inner energy $\varepsilon$ for $\ell=0$ \cite{dh,ds21} is
	\begin{eqnarray}
		B^{\rm f}_0(\varepsilon) = B^{\rm ld}_0 + B^{\rm shell}_0 \exp{(-\gamma_D \varepsilon)},
	\end{eqnarray}
	where $B^{\rm ld}_0$ and $B^{\rm shell}_0$ are the liquid-drop and shell-correction contributions to the barrier height, respectively, and $\gamma_D \approx 1.15/(0.0722396 A+0.195267 A^{2/3})=1.15/a_A^0$ MeV$^{-1}$ is the damping coefficient \cite{d23d}. 
	
	For the sake of simplicity, the values of $B^{\rm f}_\ell(\varepsilon)$ for $\ell>0$ is
	\begin{eqnarray}
		B^{\rm f}_\ell(\varepsilon)=B^{\rm f}_0(\varepsilon)+\hbar^2 \ell (\ell+1)[1/(2J^{\rm cnf})-1/(2J^{\rm gs})].
	\end{eqnarray}
	Here $J^{\rm gs}=\frac{2}{5}M R_{0}^2 A \left( 1+\sqrt{\frac{5}{16 \pi}} \beta_{\rm gs} + \frac{135 }{84 \pi} \beta_{\rm gs}^2 \right)$ is the solid-state moment of inertia of the nucleus at the ground state and $\beta_{\rm gs}$ is the quadrupole deformation of the nucleus at the ground state. The excitation energy of the compound nucleus formed in the hot fusion reaction is sufficiently high. The nucleon pairing is broken down at a nuclear temperature close to 0.5 MeV \cite{ds18}. Due to this the influence of the pairing force on both the moment of inertia and the fission barrier height is neglected here.
	
	The values $B^{\rm ld}_0$ and $B^{\rm shell}_0$ can be obtained directly using the values in tables in Ref. \cite{jks}. In contrast to this, the values $B^{\rm ld}_0$ and $B^{\rm shell}_0$ are not given in the tables in the model \cite{frdm,frdm_fb}. However, the values of total barrier $B^{\rm f}_0(\varepsilon)$ and the ground state shell correction energy $E^{\rm shell}_0$ are given in Refs. \cite{frdm_fb} and \cite{frdm}, correspondingly. In this case, the value of $B^{\rm shell}_0= E^{\rm shell}_{\rm b} -E^{\rm shell}_0$ may be approximated as $B^{\rm shell}_0 \approx -E^{\rm shell}_0$ because the value of the shell correction in the barrier point $E^{\rm shell}_{\rm b}$ is small as a rule, see, for example, \cite{jks}. Then, the liquid-drop contribution to the barrier height in this model is $B^{\rm ld}_0 \approx B^{\rm f}_0(\varepsilon)+E^{\rm shell}_0$. 
	
	Note that the values of $B^{\rm ld}<0$ for some SHN, see, for example, Ref. \cite{jks}. In such cases due to the reduction of $B^{\rm shell}$ at high excitation energies $\varepsilon$ the value of $B^{\rm f}(\varepsilon)$ may be below zero. However, the lowest value of the barrier height of the SHN is zero because this relates to the instability of the nucleus with the equilibrium shape concerning fission degrees of freedom. Therefore, if the values of $B^{\rm f}(\varepsilon) < 0$ for some $\varepsilon$ according to Eq. (19) then the value $B^{\rm f}(\varepsilon) = 0$ is used in the calculations.
	
	The fission width of an excited nucleus with the fission barrier dependent on the excitation energy is derived in Ref. \cite{ds18}. It is given by
	\begin{eqnarray}
		\Gamma_{\rm f}(E)=\frac{2}{2\pi \rho(E)} \int_0^{\varepsilon_{\rm max}} d\varepsilon \frac{\rho(\varepsilon)}{N_{\rm tot}} N_{\rm saddle}(\varepsilon).
	\end{eqnarray}
	Here the ratio $\rho(\varepsilon)/N_{\rm tot}$ is the probability of finding the fissioning nucleus with the intrinsic thermal excitation energy $\varepsilon$ in the fission transition state,
	\begin{eqnarray}
		N_{\rm tot}= \int_0^{\varepsilon_{\rm max}} d\varepsilon \rho(\varepsilon)
	\end{eqnarray}
	is the total number of states available for fission in the case of the energy-dependent fission barrier,
	\begin{eqnarray}
		N_{\rm saddle}(\varepsilon)
		= \int^{E-B^{\rm f}(\varepsilon)}_{\varepsilon} de \rho(e)
	\end{eqnarray}
	is the number of states available for the fission at the thermal excitation energy $\varepsilon$. $\varepsilon_{\rm max}$ is the maximum value of the intrinsic thermal excitation energy of the compound nucleus at the saddle point, which is determined as the solution of the equation
	\begin{eqnarray}
		\varepsilon_{\rm max} + B^{\rm f}(\varepsilon_{\rm max})= E.
	\end{eqnarray}
	This equation is related to the energy conservation law, i.e. the sum of thermal $\varepsilon_{\rm max} $ and potential $B^{\rm f}(\varepsilon_{\rm max})$ energies at the saddle point equals to the total excitation energy $E$. At the zero-excitation energy, the fission width (21) coincides with the Bohr-Wheeler width \cite{bw}. The expression obtained in Ref. \cite{ds18} leads to a good description of the experimental values of the ratio $\Gamma_{\rm f}(E)/\Gamma_{\rm n}(E)$ and the fission barrier heights in various nuclei \cite{ds18gg,dds22}. Note that Eqs. (6)-(9) and (21)-(24) are similar.
	
	\section{Discussion}
	
	\subsection{Comparison with available experimental data}
	
	The comparison of the experimental cross-sections of the SHN synthesis $\sigma^{xn}(E)$ [1-26] obtained in the collisions of $^{48}$Ca with the $^{226}$Ra, $^{232}$Th, $^{238}$U, $^{237}$Np, $^{238}$U, $^{239,240,242,244}$Pa, $^{243}$Am, $^{245,248}$Cm, $^{249}$Bk, $^{249}$Cf targets with the calculated ones is presented in Figs. 3 -- 9. The calculations of the SHN production cross-section performed using the fission barriers taken from the macroscopic-microscopic finite-range liquid-drop model \cite{frdm,frdm_fb} are marked as FRDM. The same calculations done applying the fission barriers picked up from the microscopic–macroscopic method with the deformed Woods–Saxon single-particle potential and the Yukawa-plus-exponential macroscopic energy \cite{jks} are pointed as MMWS. So, both these models for the fission barrier are based on the Strutinsky shell-correction method \cite{strut1,strut2,strut3,strut4}, however different prescriptions of both the macroscopic and microscopic parts are used in them. Due to this, the values of the fission barriers, neutron separation energies, nuclear binding energies, ground state quadrupole deformations, and other quantities calculated in the frameworks of the FRDM and MMWS models are different. The differences between the values of the fission barriers and nuclear binding energies are most important for SHN production and, therefore, will be discussed below in detail. 
	
	The intrinsic excitation energies $E^\star$ in Figs. 3 -- 9 are obtained using the experimental collisional energies and the experimental binding energy \cite{be}. If the value of the experimental binding energy of SHN is unknown then the model binding energy of SHN from Ref. \cite{jks} is used. 
	
	The values of ground state deformation of the compound nucleus $\beta_{\rm gs}$ are taken from Refs. \cite{frdm} or \cite{jks} in dependence of the model used for fission barrier. The values of $\beta_{\rm cnf}$ are not given in the FRDM, therefore, for the sake of unity of the calculations in the present model, the value of $\beta_{\rm cnf} = \beta_{\rm gs}+0.25$. The values of $\beta_{\rm cnf}$ obtained in such an approach roughly agree with the values of $\beta_{2}$ in the fission saddle point for various SHN given in the model MMWS \cite{jks}. 
	
	The equilibrium deformation of the deformed target nucleus should be taken into account according to the presented approach for the capture stage of the reaction \cite{d23a}. The equilibrium deformation of the even-even target nuclei is taken from Ref. \cite{be2}. The equilibrium deformation of the odd target nuclei is averaged using the equilibrium deformations of the nearest even-even nuclei.
	
	The radius of the nuclei used in the calculation of the nucleus-nucleus interaction potential \cite{ds21,d23d} is enlarged on 7\%. The other fitting parameter values are given in Table 1. The values of $b^{\rm cnf}$ have two values for the same reaction because they correspond to the different models of fission barrier heights. Note that the values of $b^{\rm cnf}$ obtained in the MMWC model are smaller than the corresponding ones in the FRDM, see Table 1. 
	
	\begin{figure}
		\includegraphics[width=6.6cm]{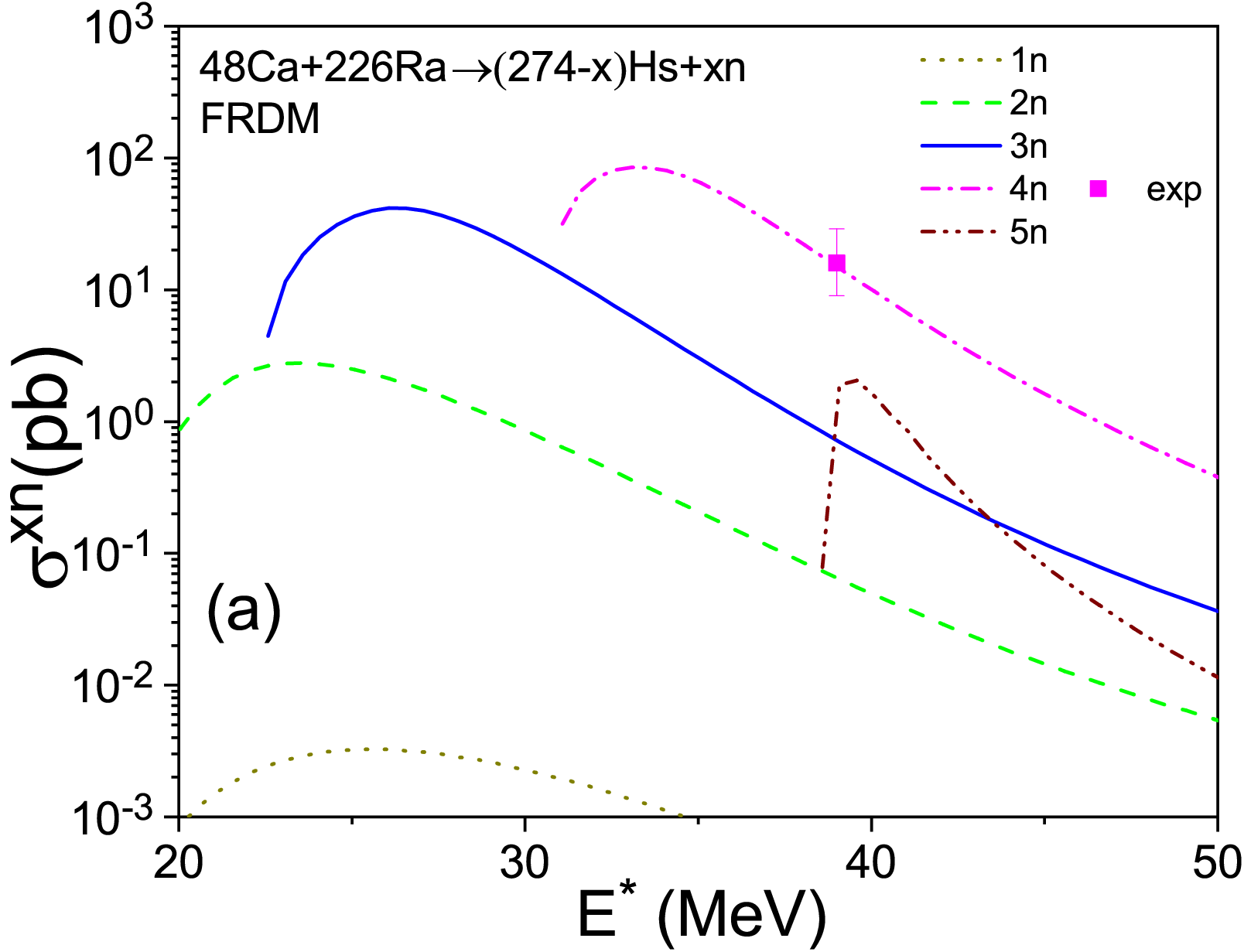}
		\includegraphics[width=6.6cm]{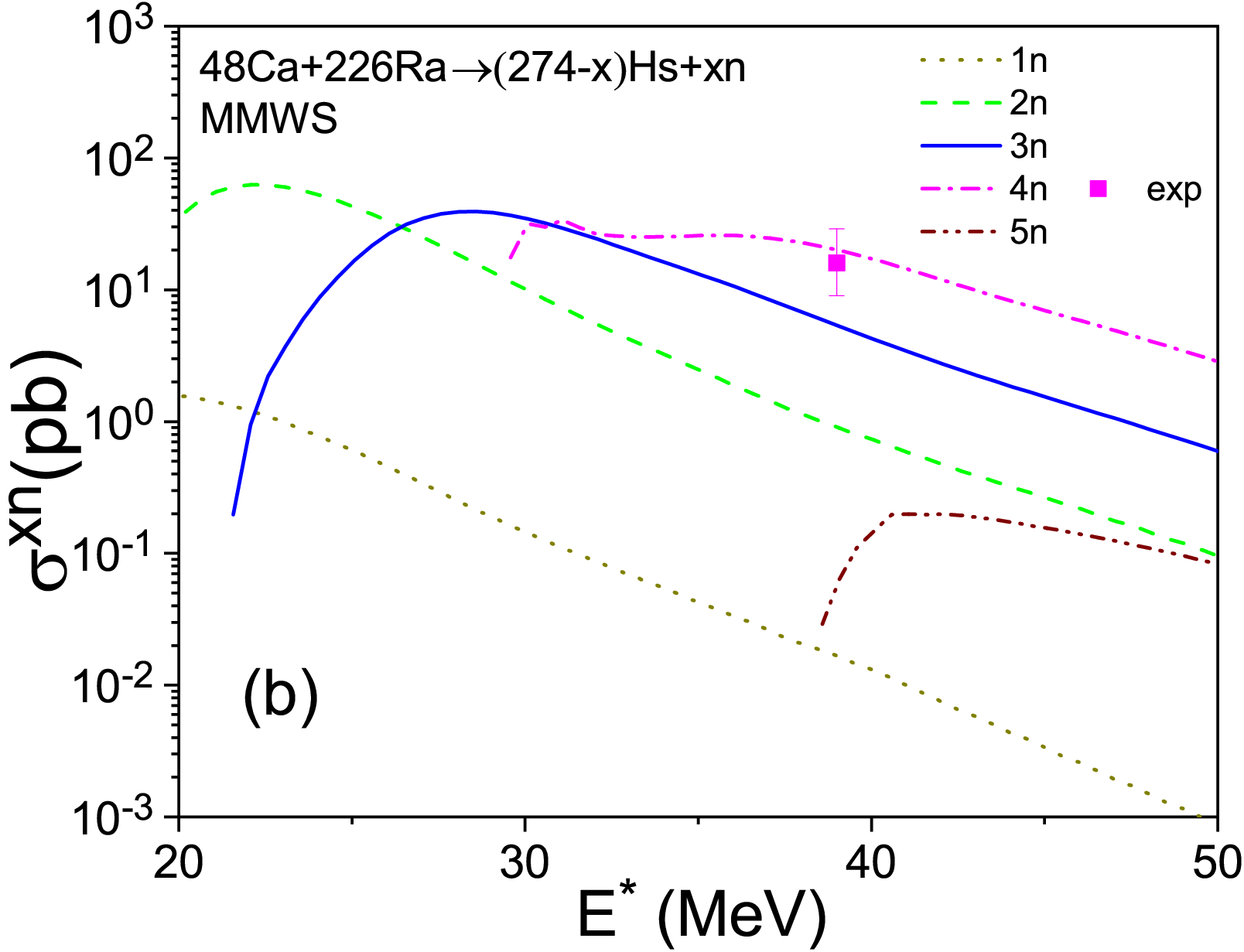}
		\includegraphics[width=6.6cm]{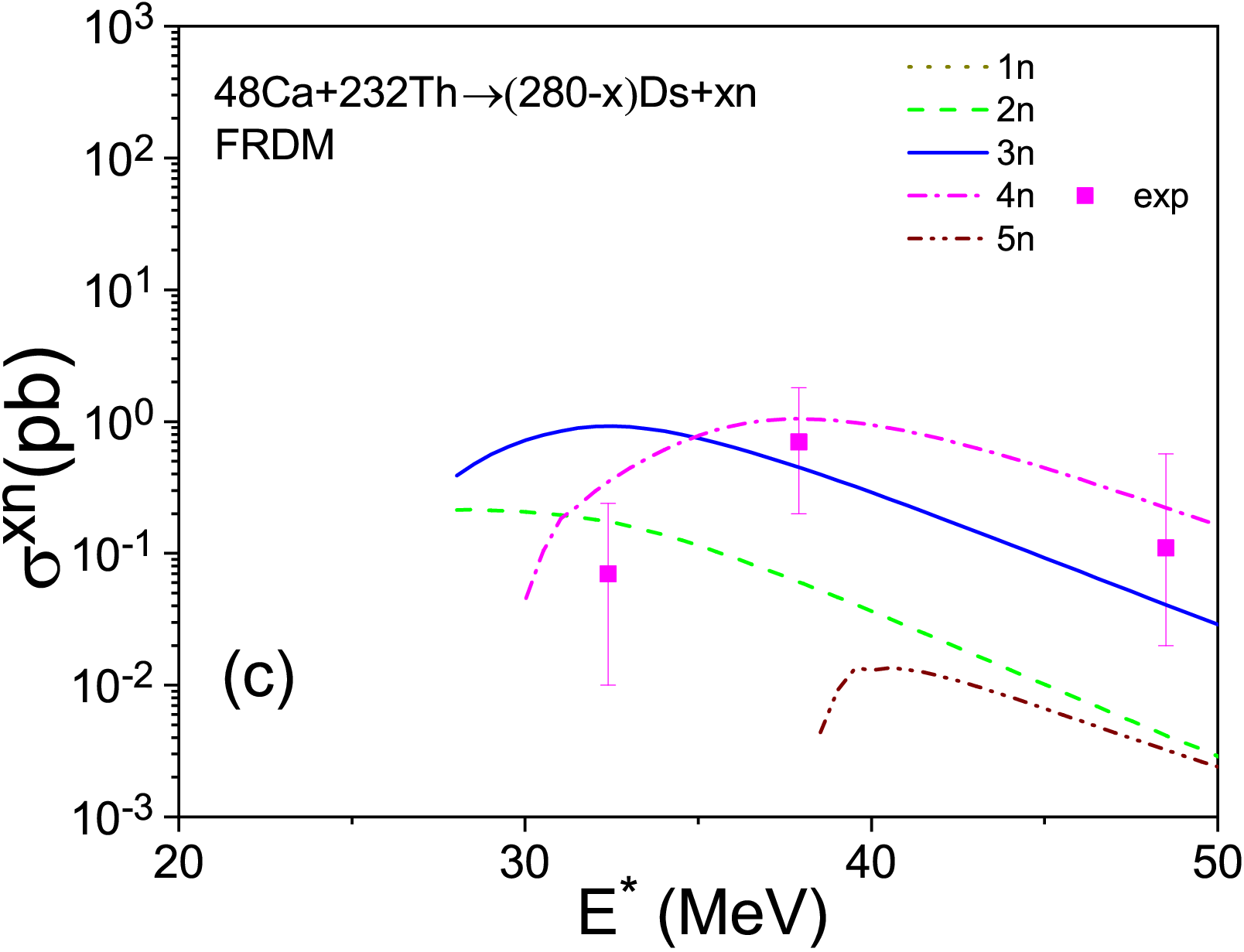}
		\includegraphics[width=6.6cm]{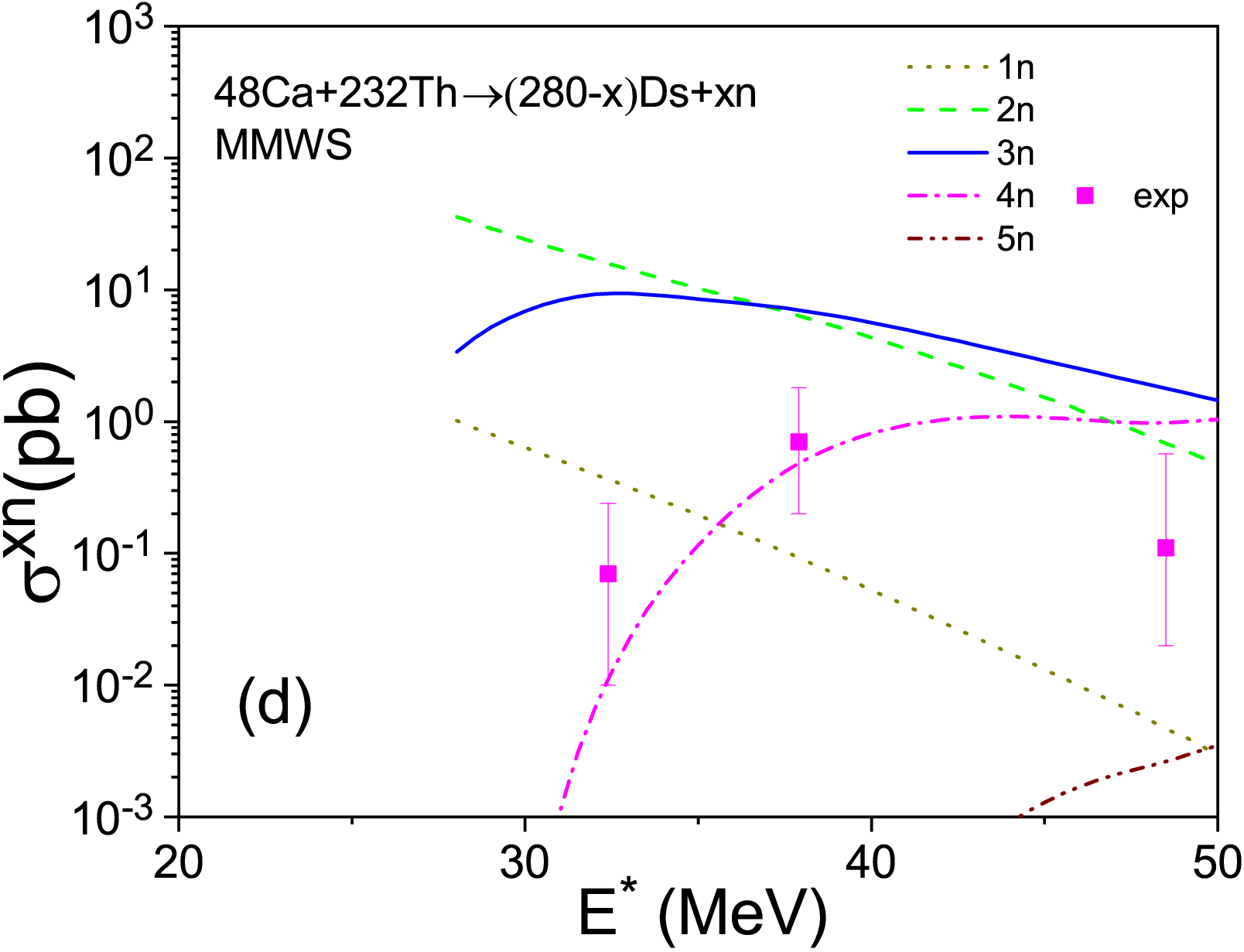}
		\caption{\label{fig3} The comparison of the experimental and theoretical cross-sections for the reactions $^{48}$Ca$+^{226}$Ra$\rightarrow ^{274-x}$Hs+$xn$ (a,b) and $^{48}$Ca$+^{232}$Th$\rightarrow ^{280-x}$Ds+$xn$ (c,d). The experimental data (dots) for the reaction $^{48}$Ca$+^{226}$Ra$\rightarrow ^{274-x}$Hs+$xn$ are taken from Ref. \cite{prc87_Ra} and for the reaction $^{48}$Ca$+^{232}$Th$\rightarrow ^{280-x}$Ds+$xn$ from Ref. \cite{prc108}. The theoretical cross-sections (lines) are calculated for the FRDM (a,c) \cite{frdm,frdm_fb} and MMWS (b,d) \cite{jks} fission barrier models. }
	\end{figure}

	\begin{figure}
		\includegraphics[width=6.9cm]{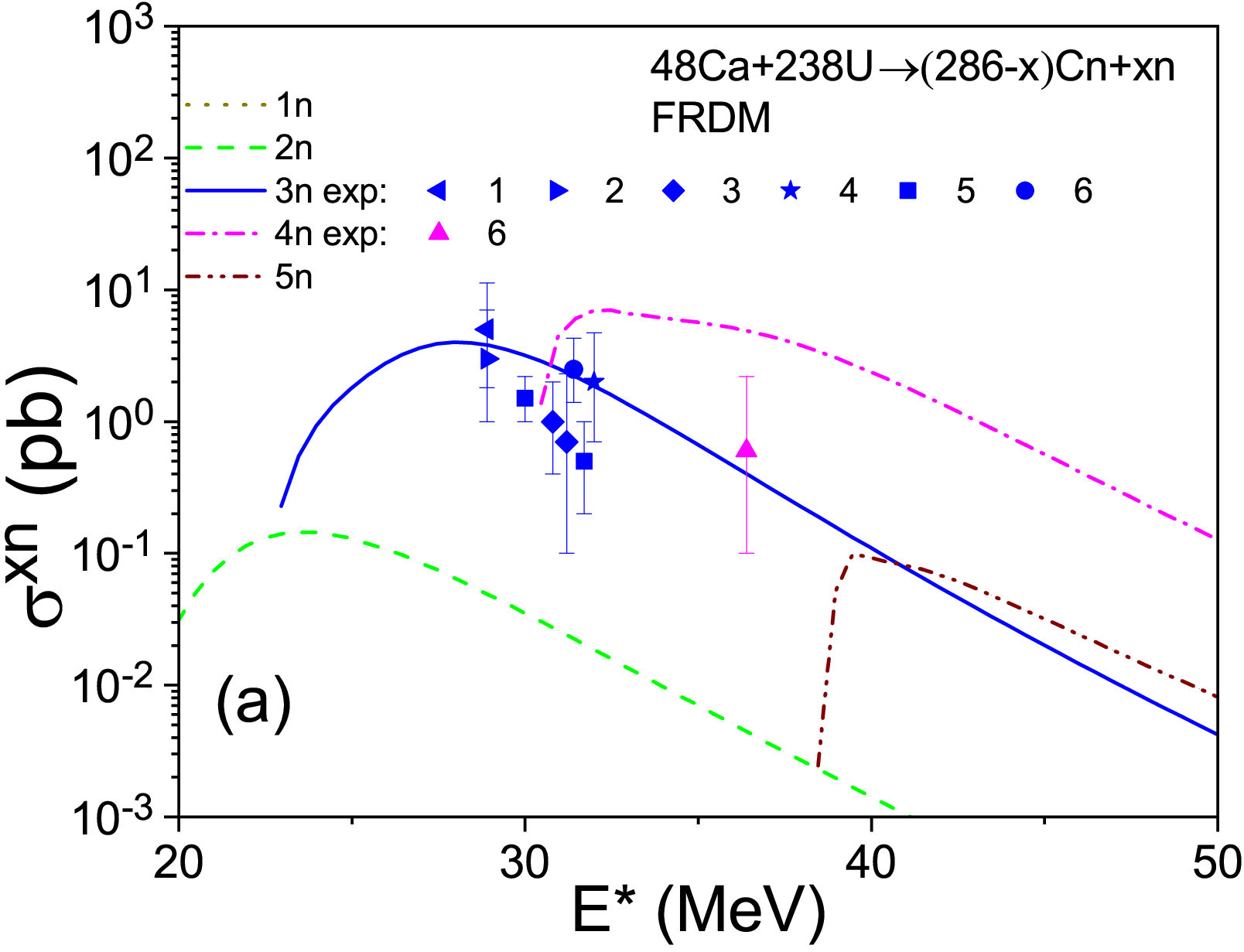}
		\includegraphics[width=6.9cm]{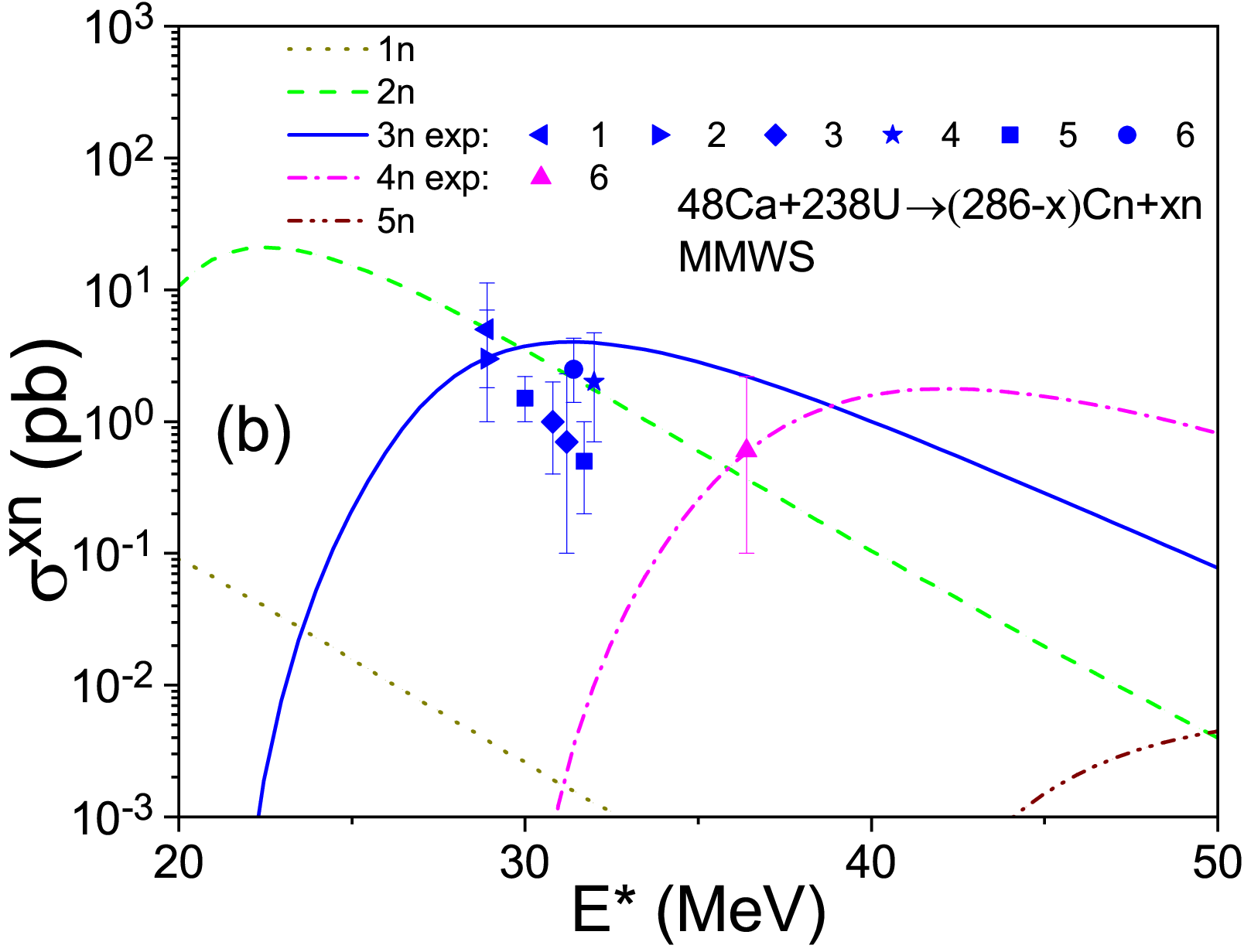}
		\includegraphics[width=6.9cm]{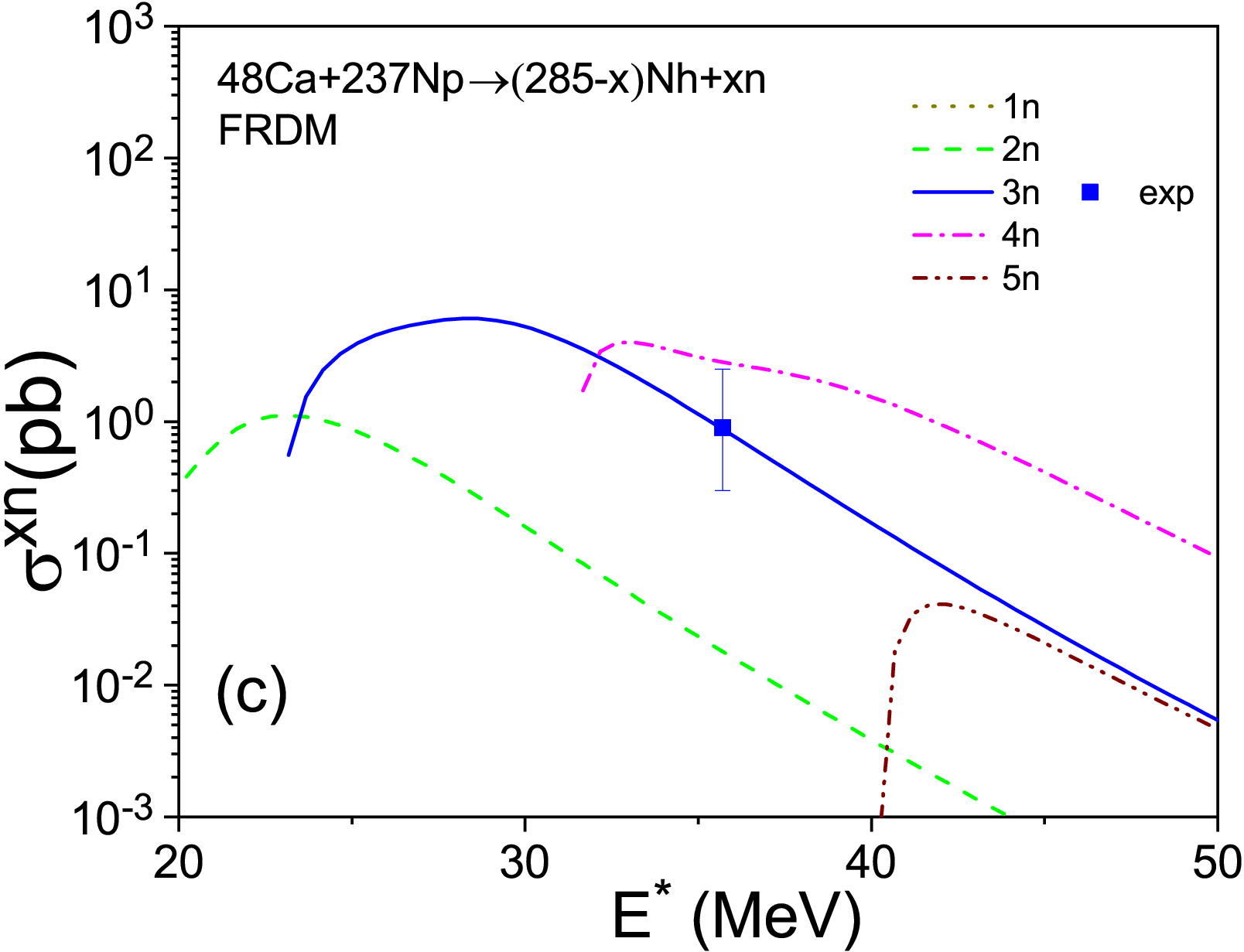}
		\includegraphics[width=6.9cm]{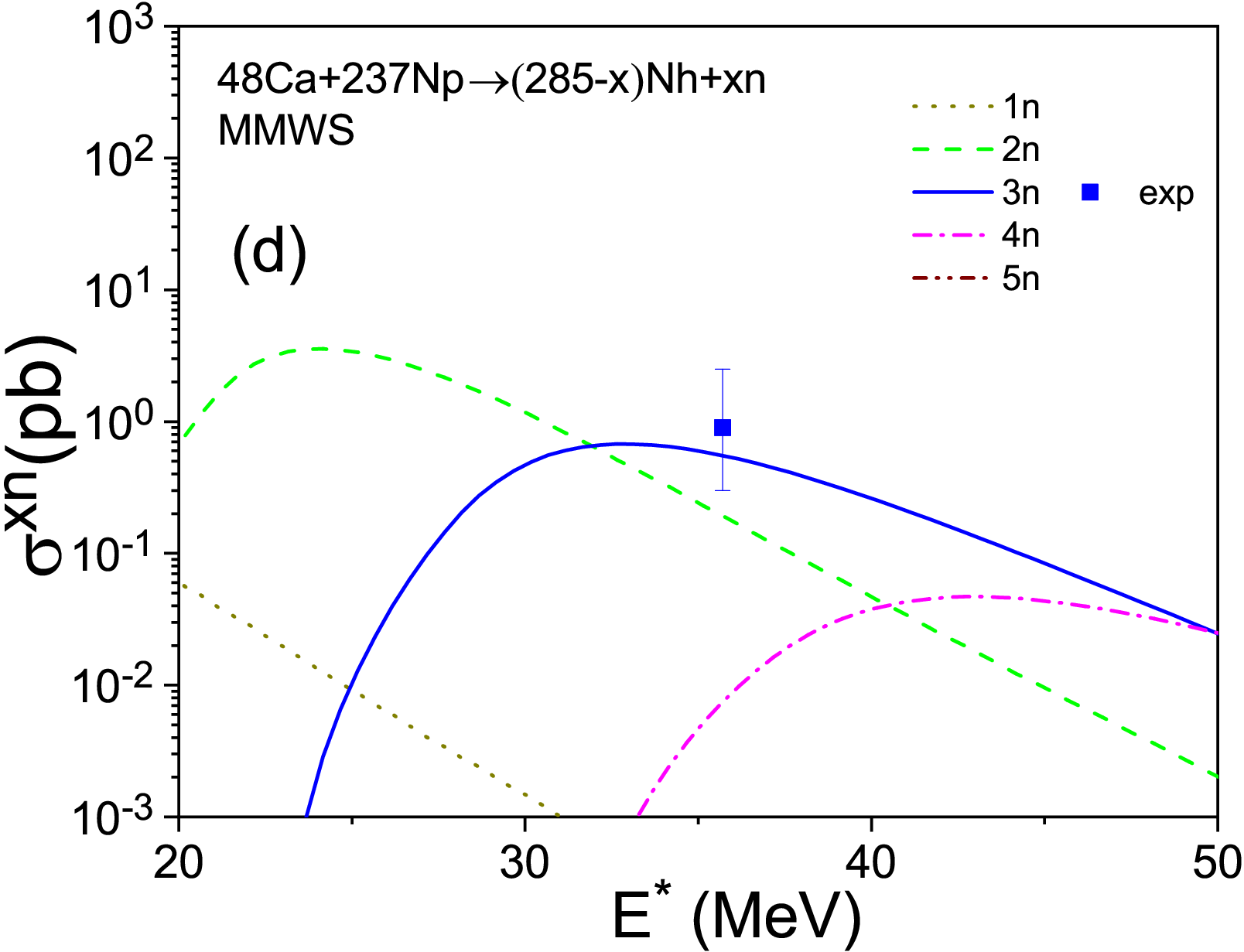}
		\caption{\label{fig4} The same as in Fig. 3 for the reactions $^{48}$Ca$+^{238}$U$\rightarrow ^{286-x}$Cn+$xn$ (a,b) and $^{48}$Ca$+^{237}$Np$\rightarrow ^{285-x}$Nh+$xn$ (c,d). The experimental data (dots) for the reaction $^{48}$Ca$+^{238}$U$\rightarrow ^{286-x}$Cn+$xn$ are taken from Refs. \cite{epja5} - 1, \cite{epja19} - 2, \cite{epja32} - 3, \cite{jpsj86_U} - 4, \cite{prc106_Pu_U} - 5, \cite{prc70} - 6, and for the reaction $^{48}$Ca$+^{237}$Np$\rightarrow ^{285-x}$Nh+$xn$ from Ref. \cite{prc76}.}
	\end{figure}

	\begin{figure}
		\includegraphics[width=6.9cm]{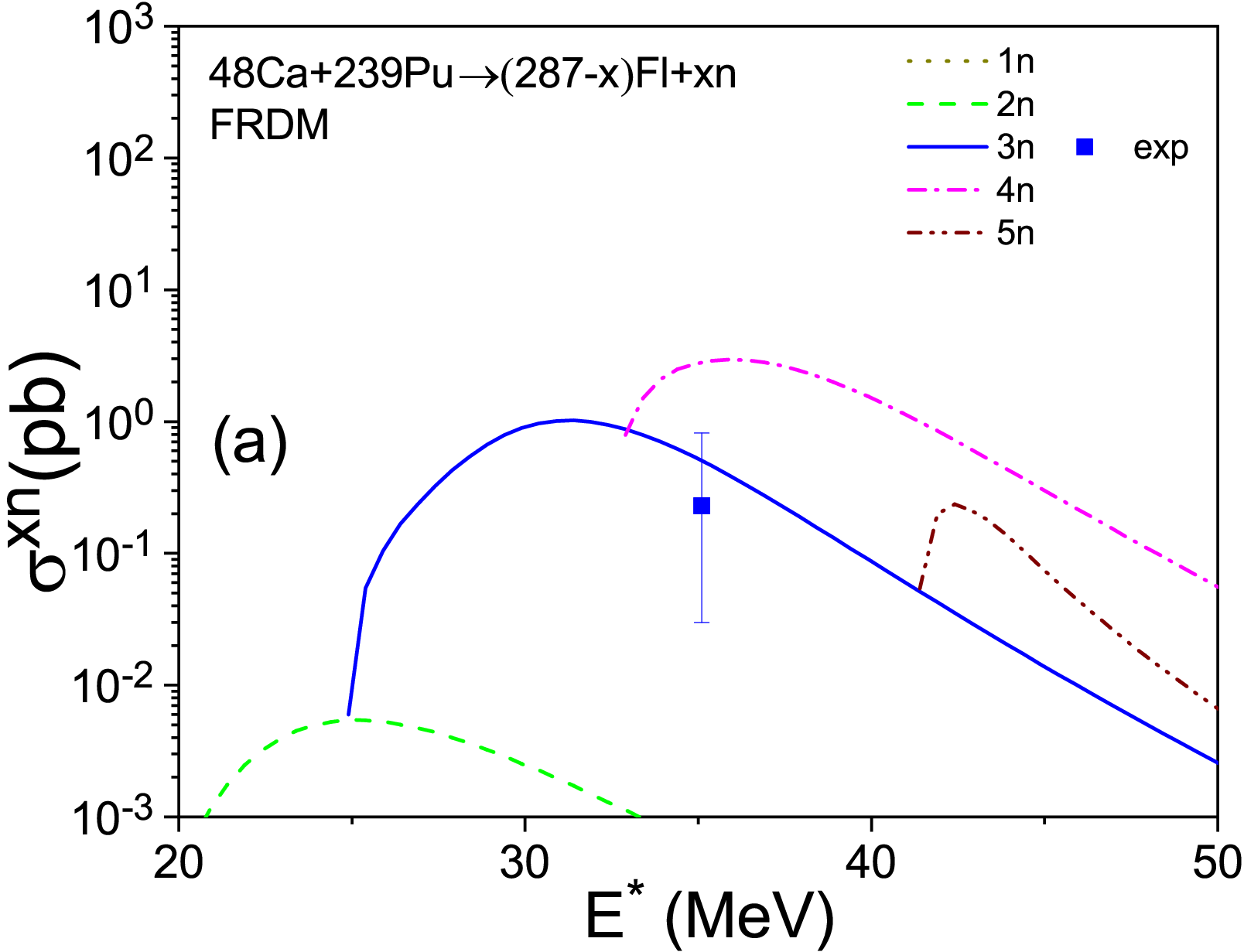}
		\includegraphics[width=6.9cm]{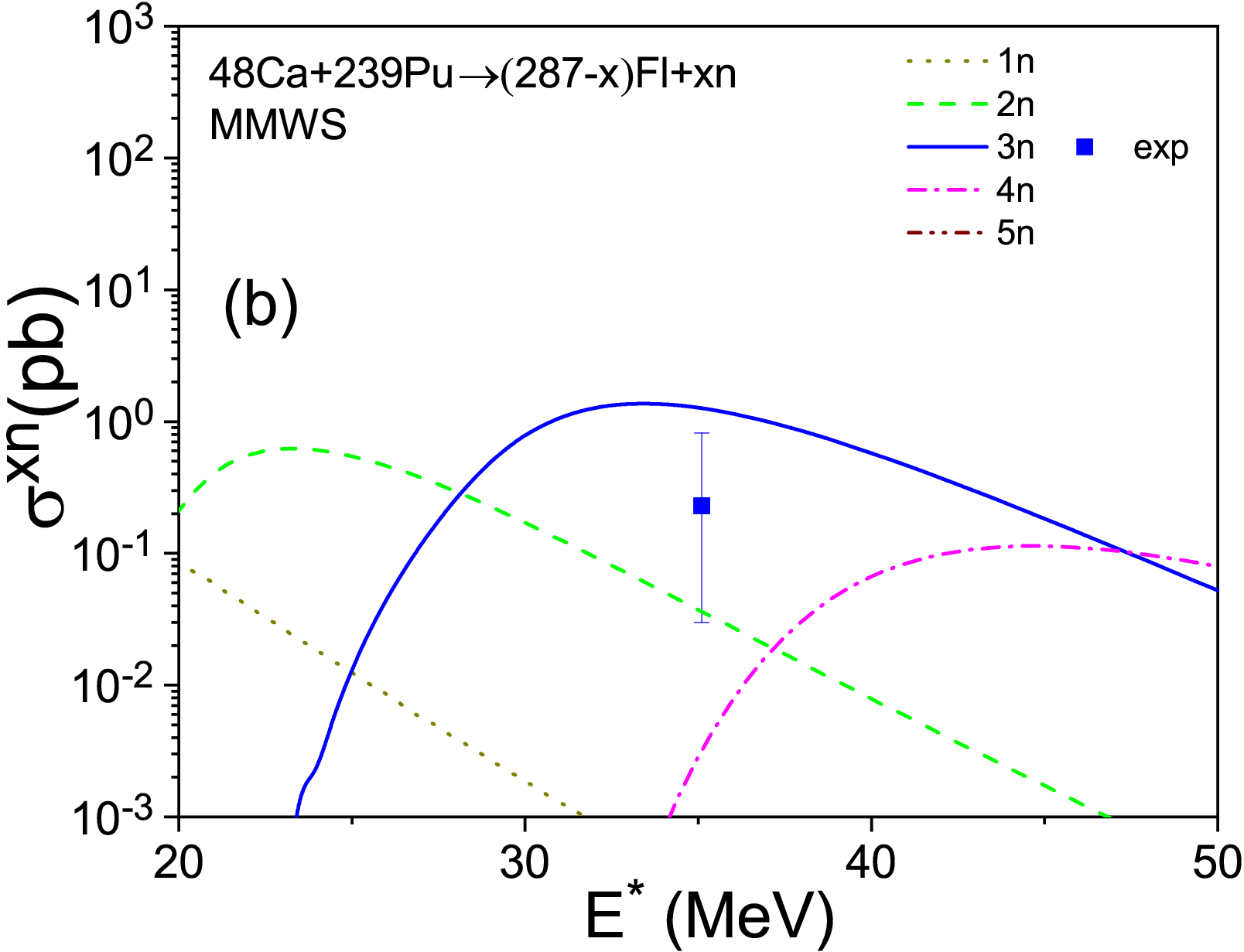}
		\includegraphics[width=6.9cm]{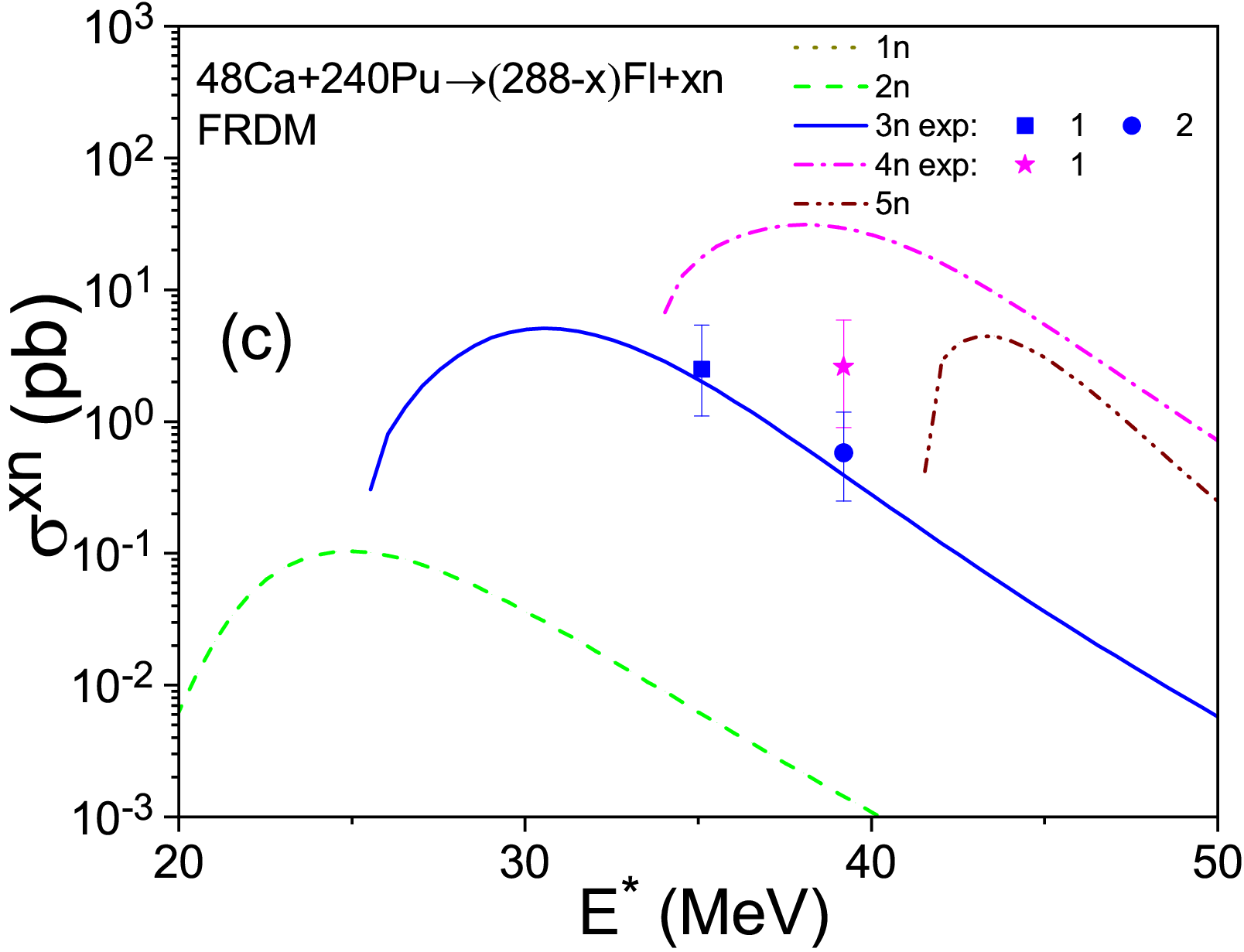}
		\includegraphics[width=6.9cm]{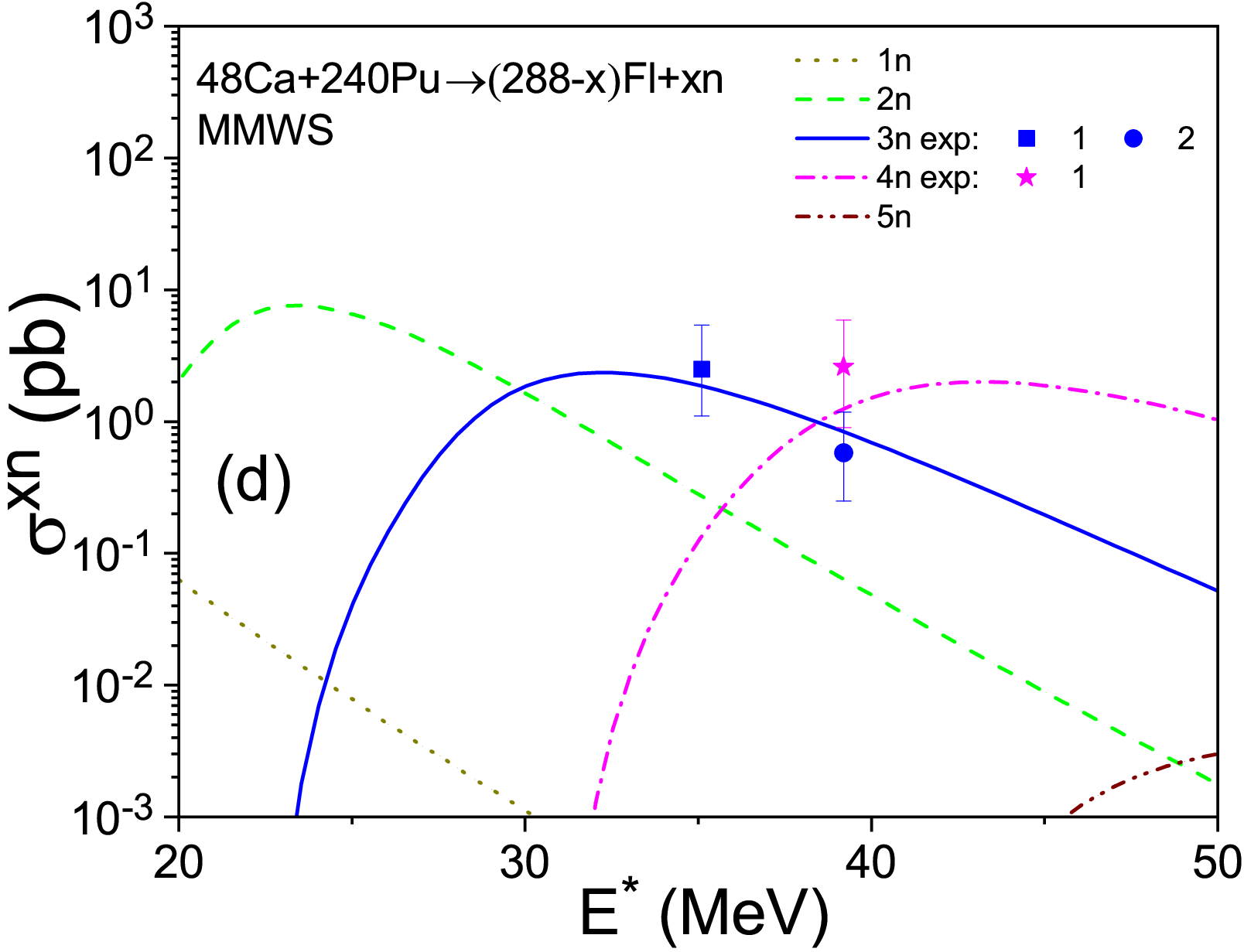}
		\caption{\label{fig5} The same as in Fig. 3 for the reactions $^{48}$Ca$+^{239}$Pu$\rightarrow ^{287-x}$Fl+$xn$ (a,b) and $^{48}$Ca$+^{240}$Pu$\rightarrow ^{288-x}$Fl+$xn$ (c,d). The experimental data (dots) for the reaction $^{48}$Ca$+^{239}$Pu$\rightarrow ^{287-x}$Fl+$xn$ are taken from Ref. \cite{prc92}, and for the reaction $^{48}$Ca$+^{240}$Pu$\rightarrow ^{288-x}$Fl+$xn$ from Refs. \cite{prc92} - 1, \cite{prc97} - 2. }
	\end{figure}

	\begin{figure}
		\includegraphics[width=6.9cm]{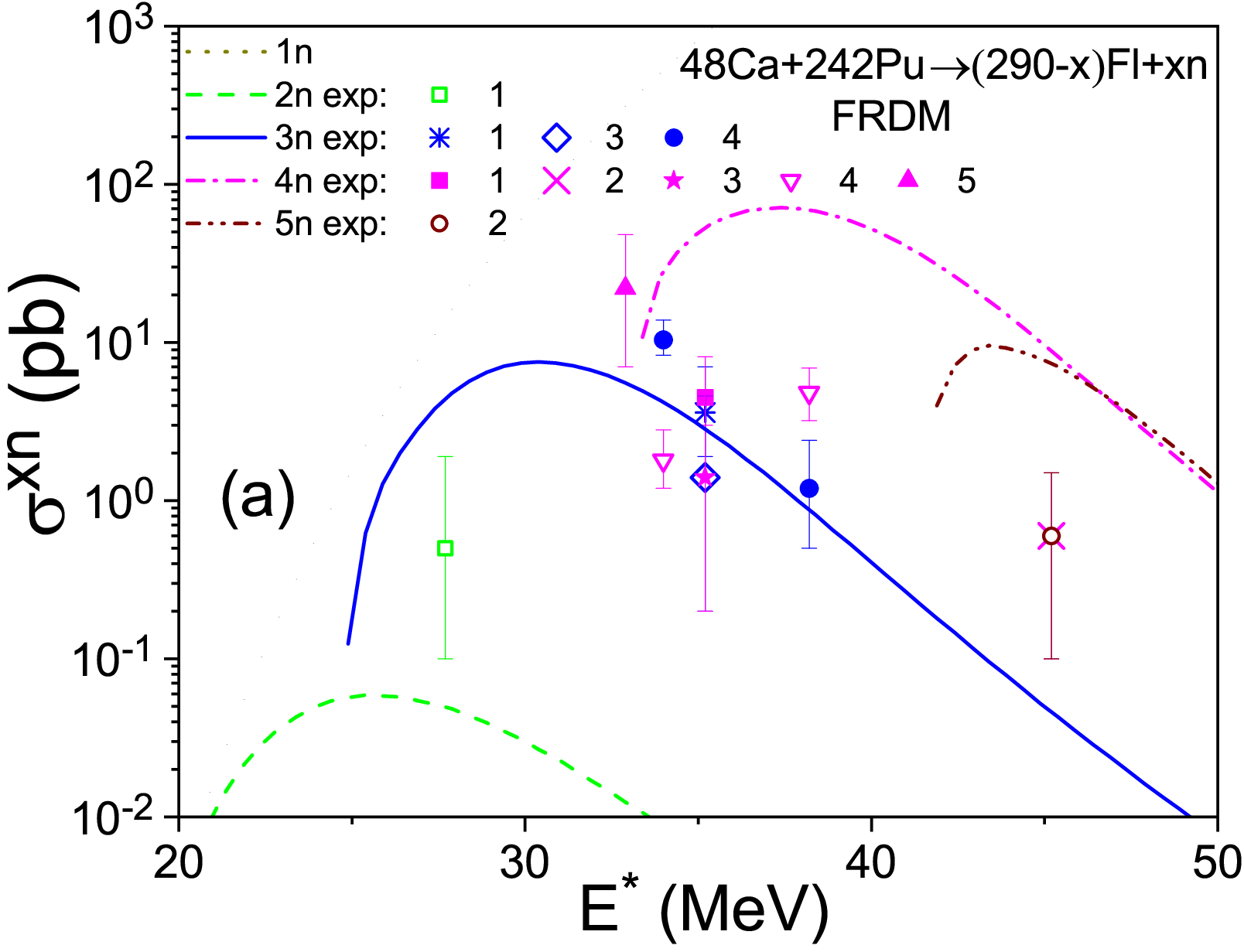}
		\includegraphics[width=6.9cm]{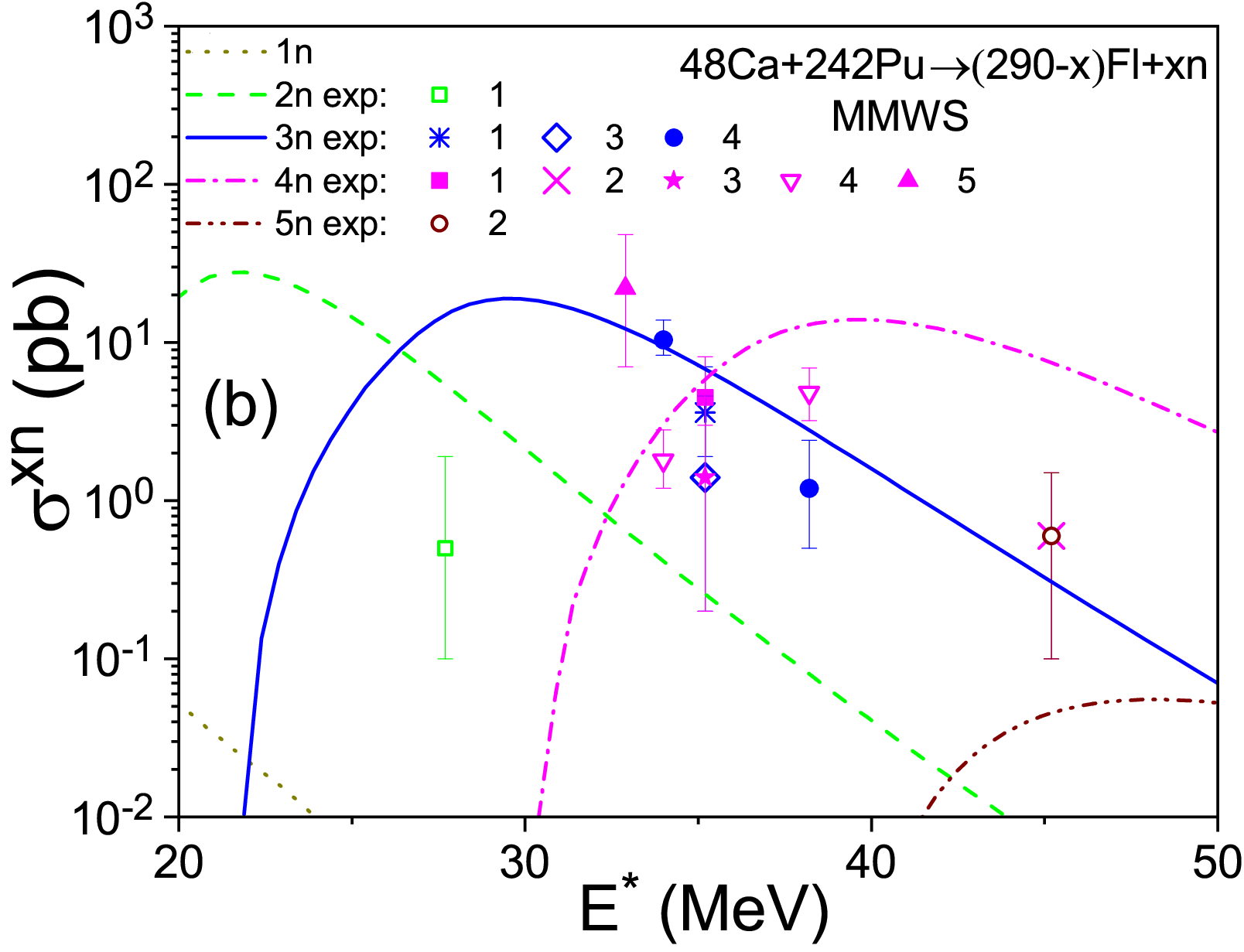}
		\includegraphics[width=6.9cm]{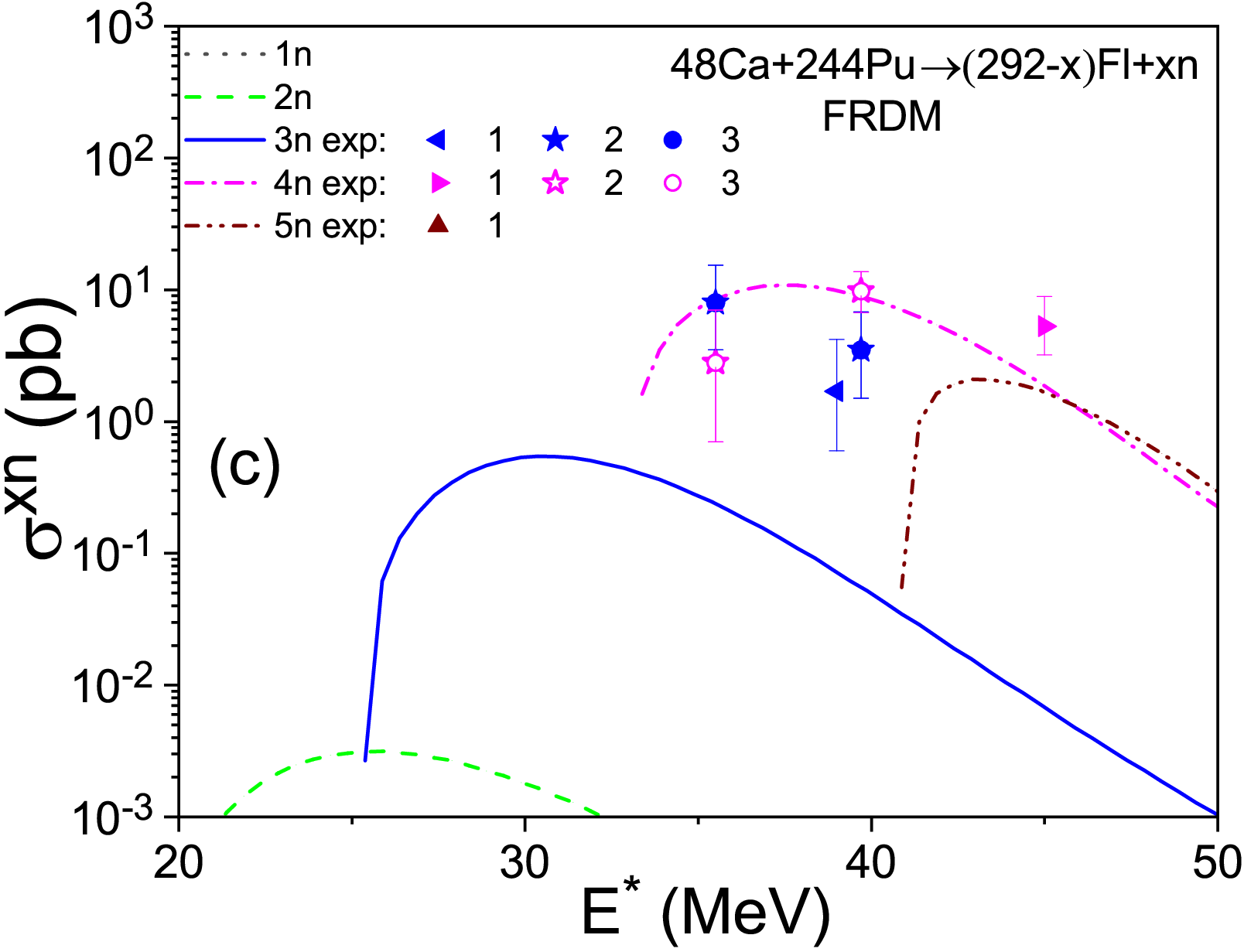}
		\includegraphics[width=6.9cm]{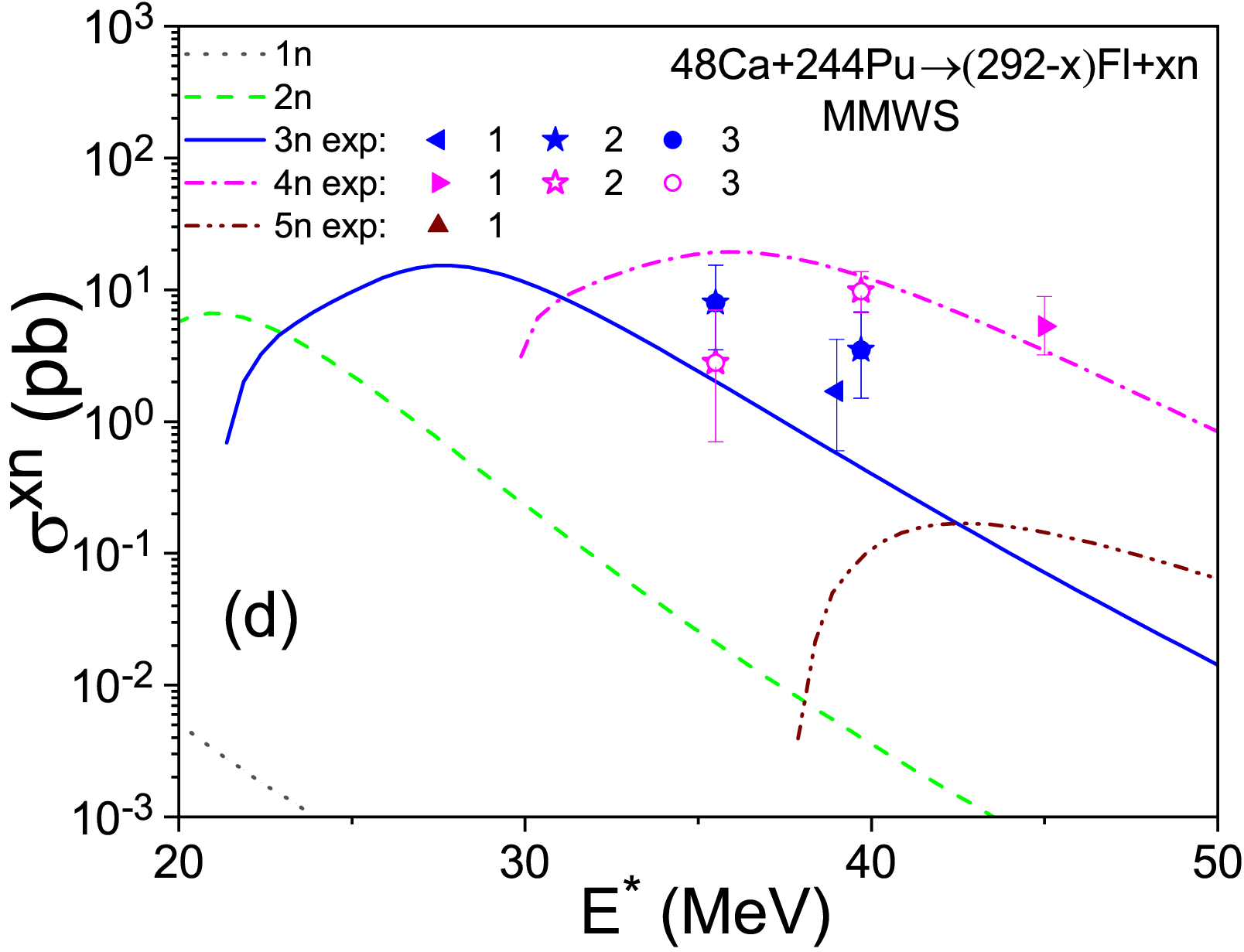}
		\caption{\label{fig6} The same as in Fig. 3 for the reactions $^{48}$Ca$+^{242}$Pu$\rightarrow ^{290-x}$Fl+$xn$ (a,b) and $^{48}$Ca$+^{244}$Pu$\rightarrow ^{292-x}$Fl+$xn$ (c,d). The experimental data (dots) for the reaction $^{48}$Ca$+^{242}$Pu$\rightarrow ^{290-x}$Fl+$xn$ are taken from Refs. \cite{prc70} - 1, \cite{prl105} - 2, \cite{prl103} - 3, \cite{prc106_Pu_U} - 4, \cite{prc107} - 5, and for the reaction $^{48}$Ca$+^{244}$Pu$\rightarrow ^{292-x}$Fl+$xn$ from Refs. \cite{prc69} - 1, \cite{prc83} - 2, \cite{prl104} - 3.}
	\end{figure}

	\begin{figure}
		\includegraphics[width=6.9cm]{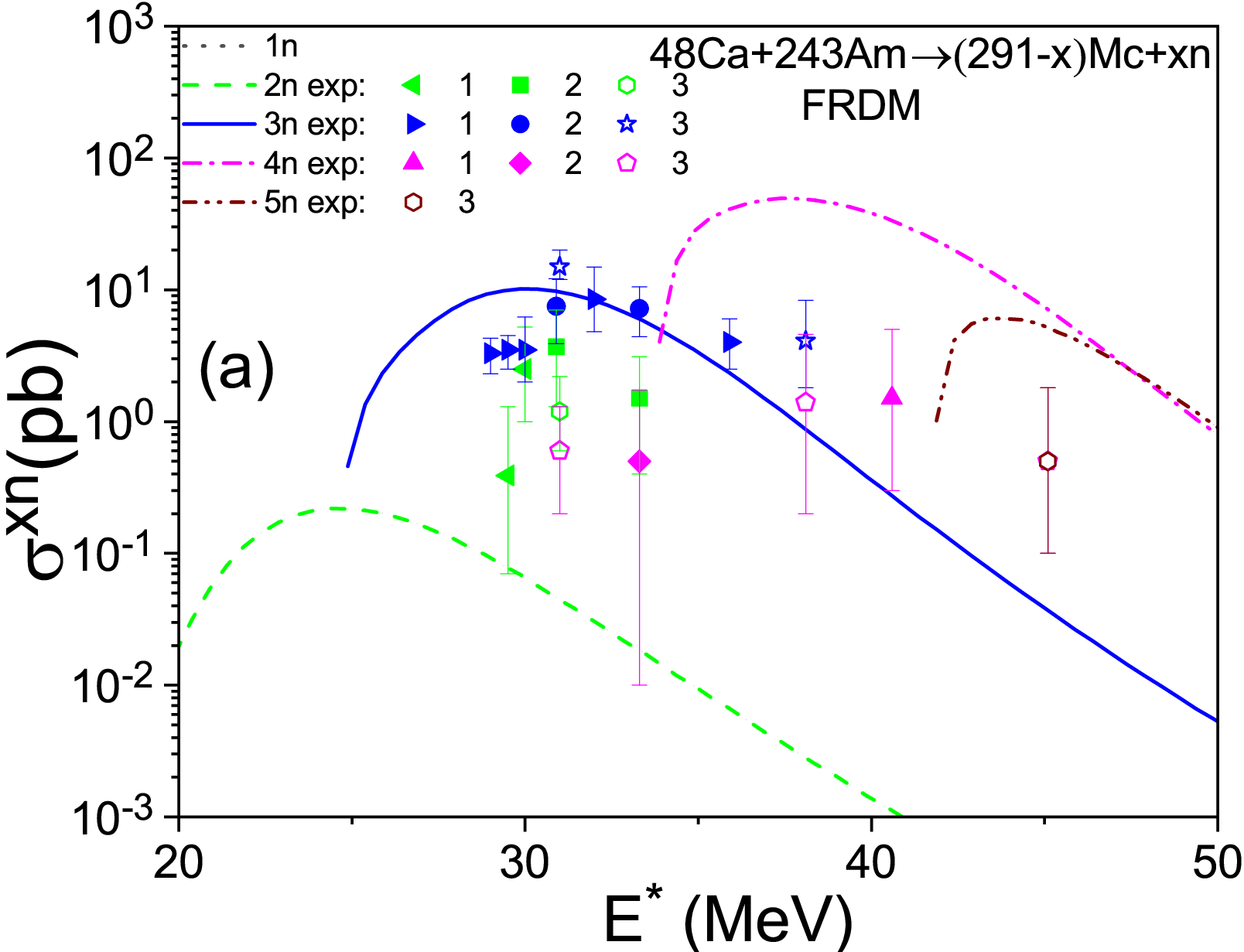}
		\includegraphics[width=6.9cm]{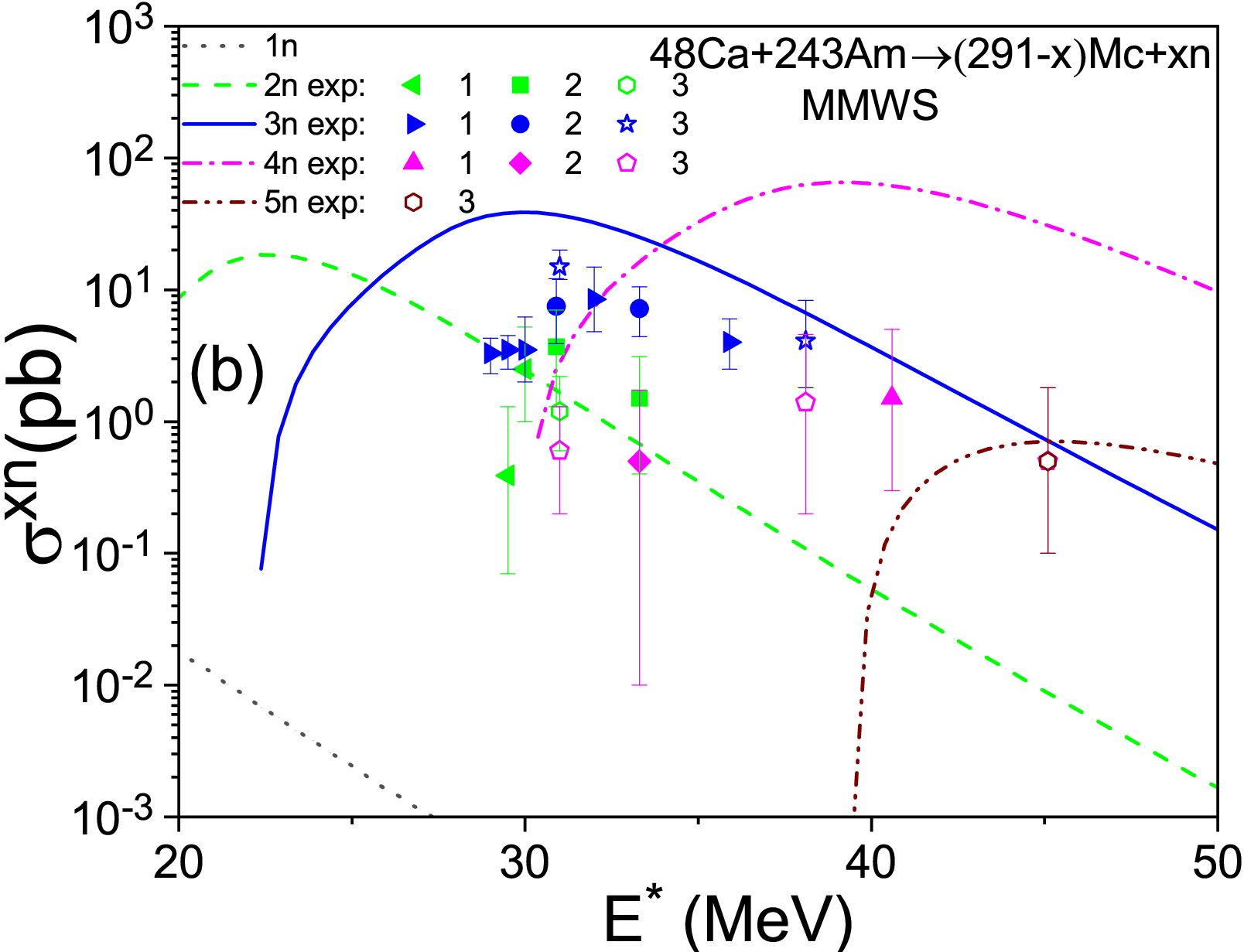}
		\includegraphics[width=6.9cm]{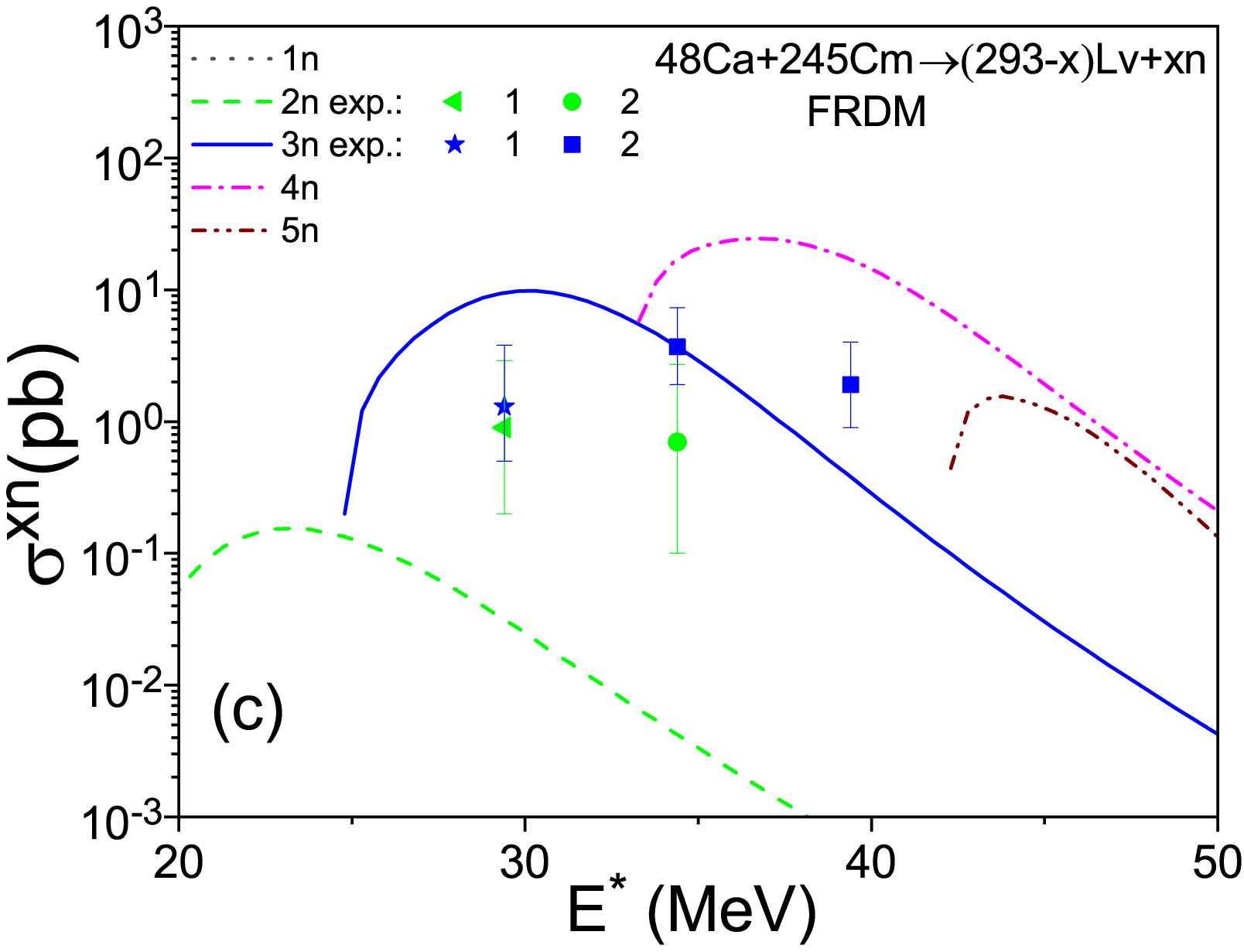}
		\includegraphics[width=6.9cm]{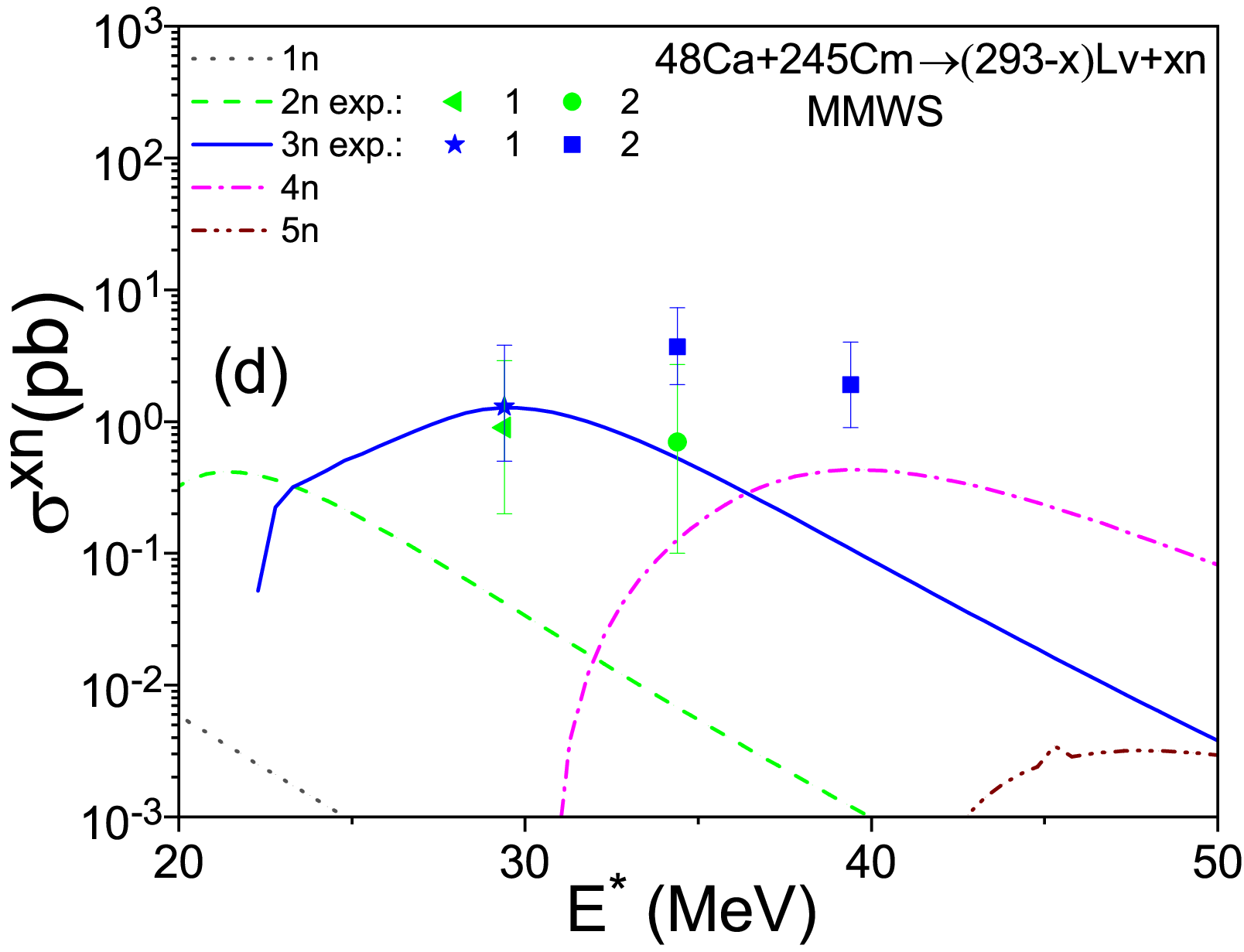}
		\caption{\label{fig7} The same as in Fig. 3 for the reactions $^{48}$Ca$+^{243}$Am$\rightarrow ^{291-x}$Mc+$xn$ (a,b) and $^{48}$Ca$+^{245}$Cm$\rightarrow ^{293-x}$Lv+$xn$ (c,d). The experimental data (dots) for the reaction $^{48}$Ca$+^{243}$Am$\rightarrow ^{291-x}$Mc+$xn$ are taken from Refs. \cite{prc87_Am} - 1, \cite{npa953} - 2, \cite{prc106_Am} - 3, and for the reaction $^{48}$Ca$+^{245}$Cm$\rightarrow ^{293-x}$Lv+$xn$ from Refs. \cite{prc69} - 1, \cite{prc74} - 2.}
	\end{figure}

	\begin{figure}
		\includegraphics[width=6.9cm]{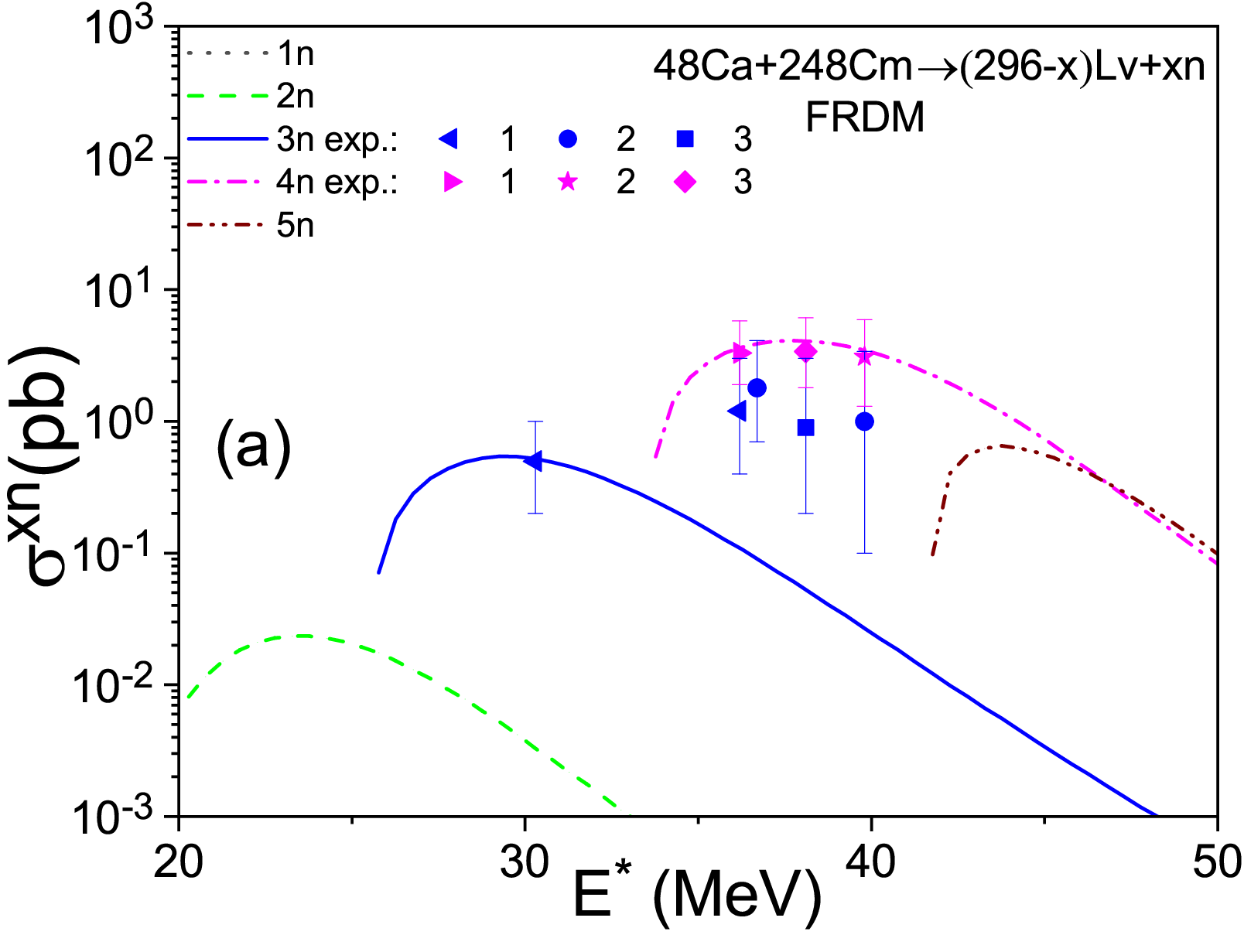}
		\includegraphics[width=6.9cm]{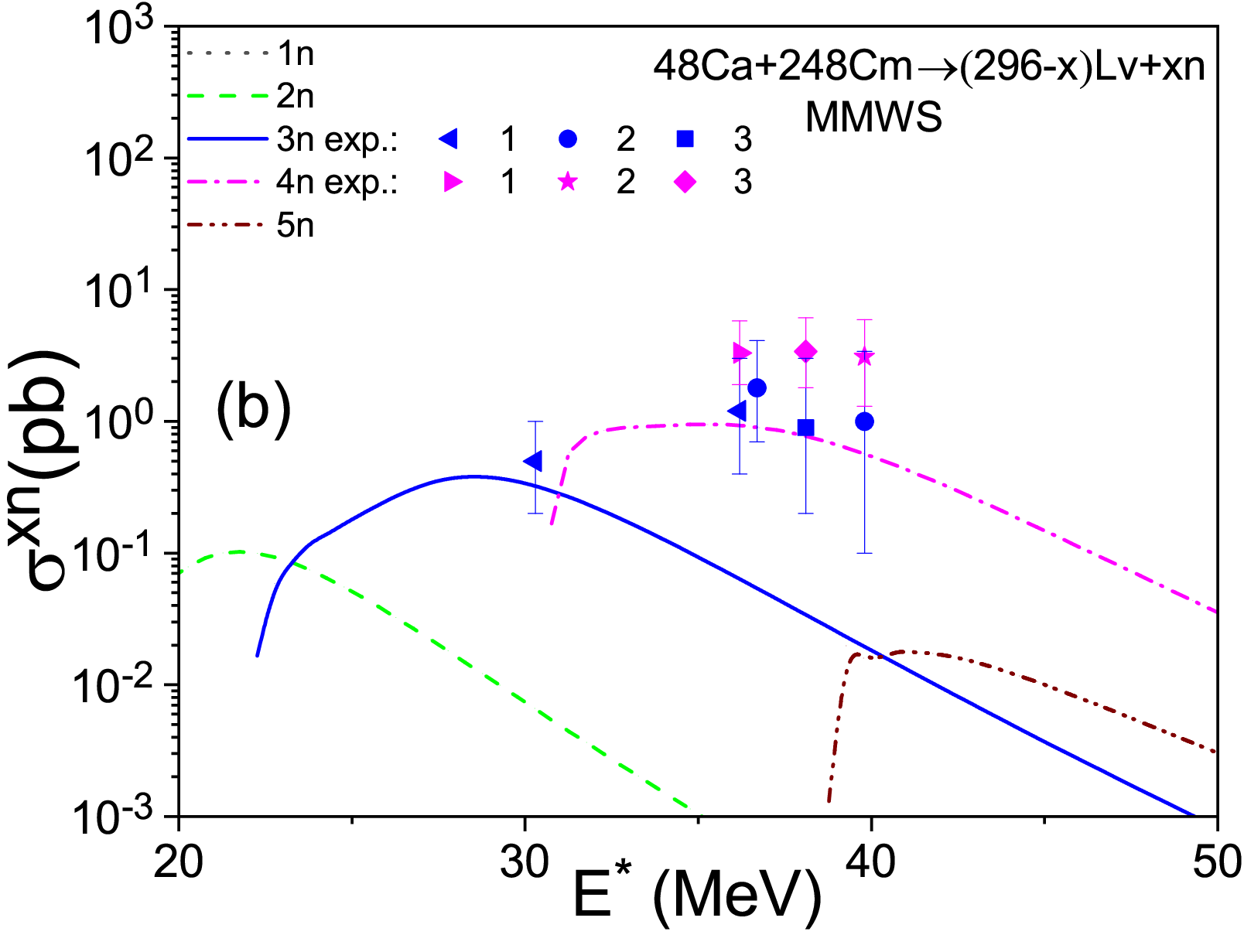}
		\includegraphics[width=6.9cm]{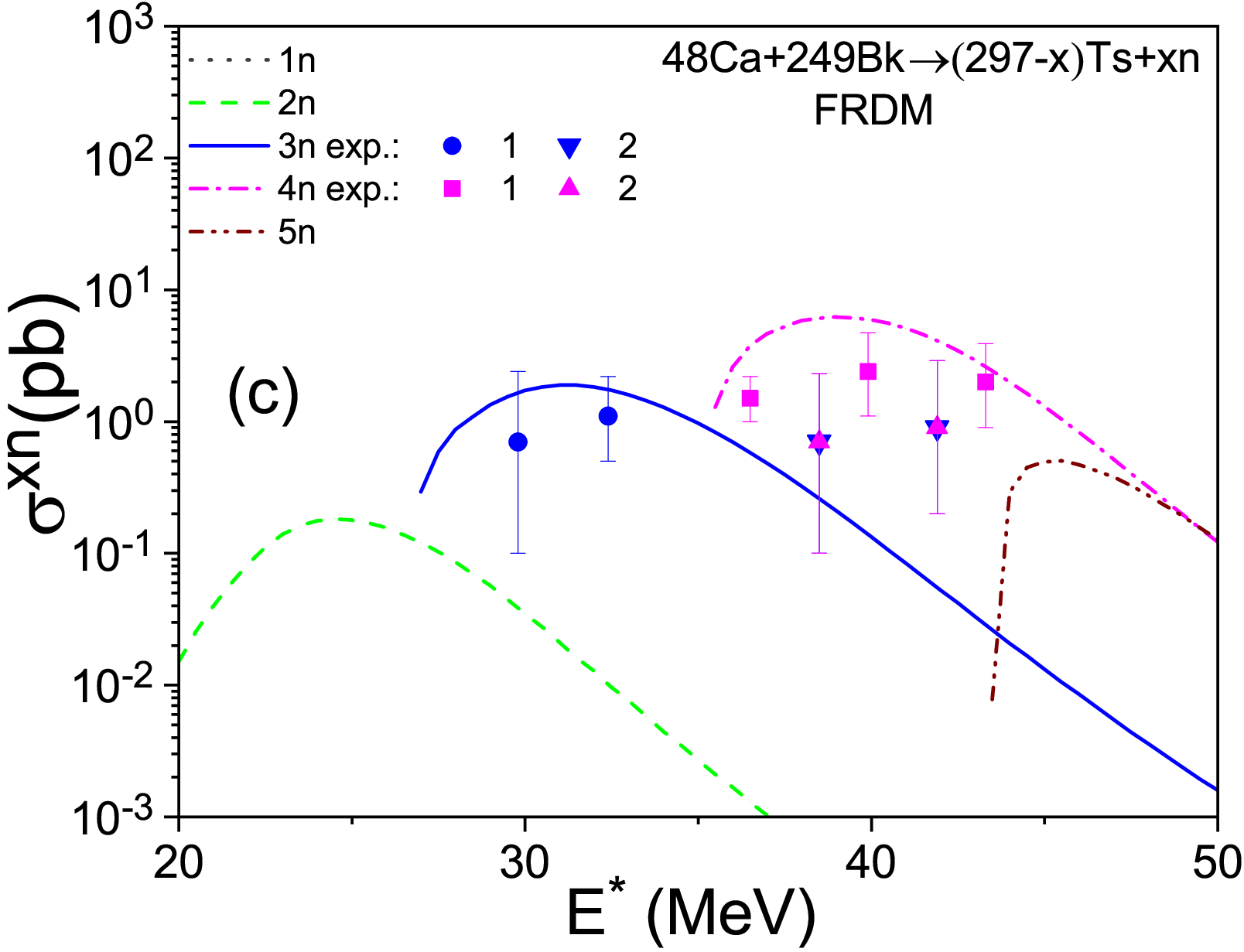}
		\includegraphics[width=6.9cm]{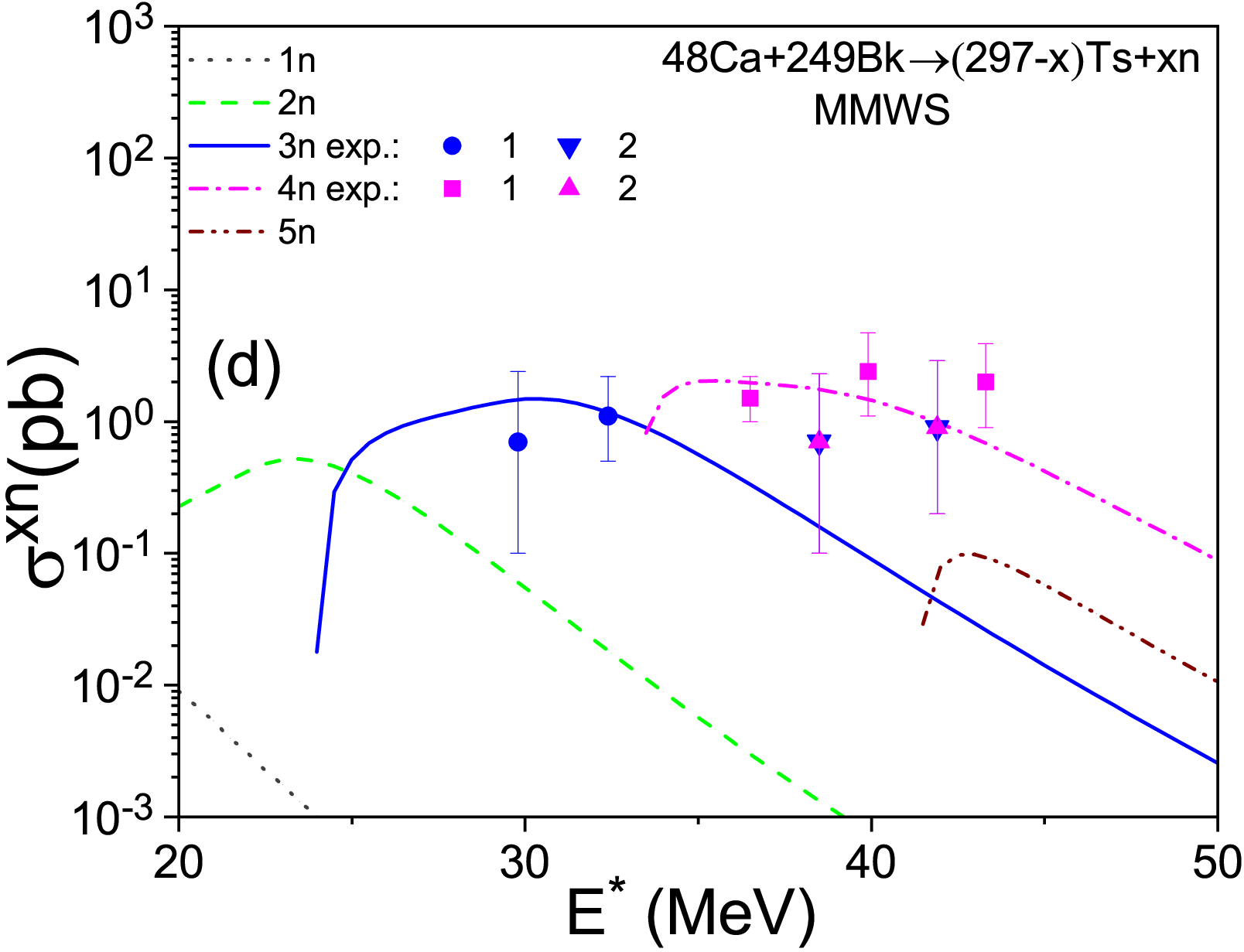}
		\caption{\label{fig8} The same as in Fig. 3 for the reactions $^{48}$Ca$+^{248}$Cm$\rightarrow ^{296-x}$Lv+$xn$ (a,b) and $^{48}$Ca$+^{249}$Bk$\rightarrow ^{297-x}$Ts+$xn$ (c,d). The experimental data (dots) for the reaction$^{48}$Ca$+^{248}$Cm$\rightarrow ^{296-x}$Lv+$xn$ are taken from Refs. \cite{prc70} - 1, \cite{jpsj86_Cm} - 2, \cite{epja48} - 3, and for the reaction $^{48}$Ca$+^{249}$Bk$\rightarrow ^{297-x}$Ts+$xn$ from Refs. \cite{prc87_Bk} - 1, \cite{prc99} - 2.}
	\end{figure}
	
	\begin{figure}
		\includegraphics[width=6.9cm]{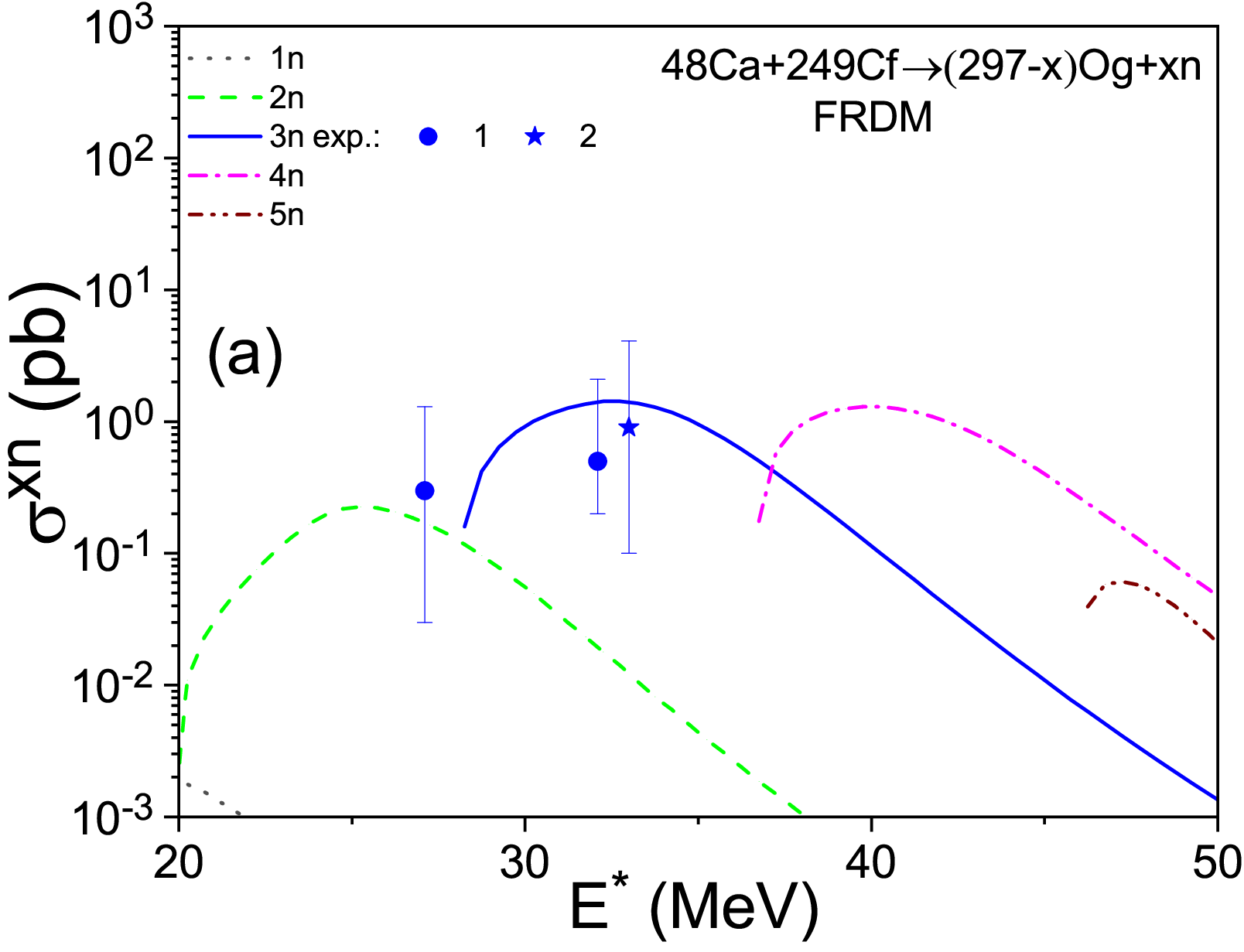}
		\includegraphics[width=6.9cm]{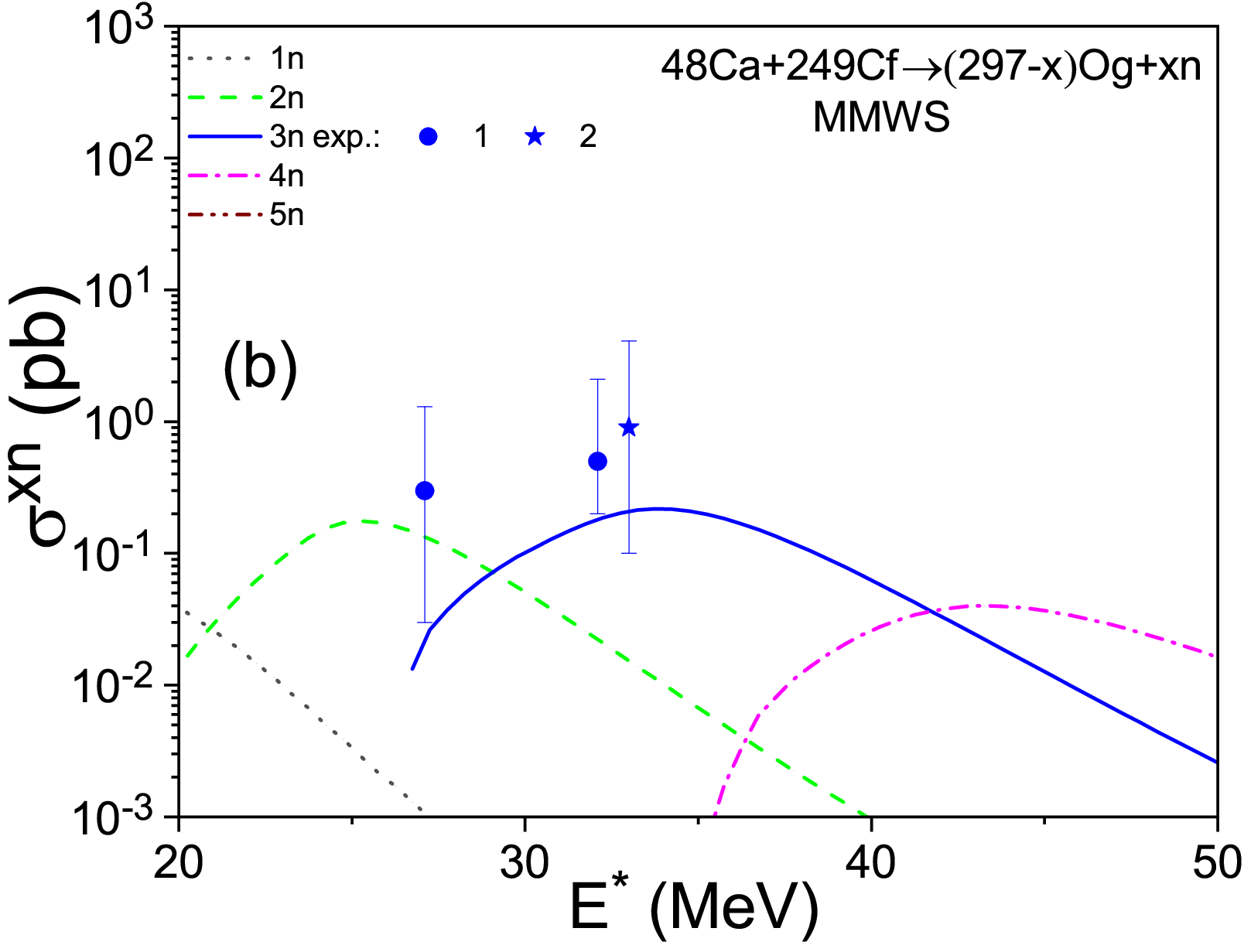}
		\caption{\label{fig9} The comparison of the experimental and theoretical cross-sections for the reaction $^{48}$Ca$+^{249}$Cf$\rightarrow ^{297-x}$Og+$xn$ (a,b). The experimental data (dots) are taken from Refs. \cite{prc74} - 1, \cite{prc98} - 2. The theoretical cross-sections (lines) are calculated for the FRDM (a) \cite{frdm,frdm_fb} and MMWS (b) \cite{jks} fission barrier models.}
	\end{figure}
	
	\begin{table}
		\caption{The values of parameters $b^{\rm cnf}$ and $\delta Q$ for corrections of the compound nucleus formation barrier and the fusion reaction $Q$-value, respectively, for calculation with the FRDM \cite{frdm,frdm_fb} and MMWS \cite{jks} fission barrier models.}
		\begin{tabular}{|c|c|c|c|c|}
			\hline
			Compound & \multicolumn{2}{c|}{FRDM} &\multicolumn{2}{c|}{MMWS} \\
			nucleus & $b^{\rm cnf}$ (MeV) & $\delta Q$ (MeV) & $b^{\rm cnf}$ (MeV) & $\delta Q$ (MeV) \\
			\hline
			$^{274}$Hs & 8.5 & 0 & 6.0 & 0 \\
			$^{280}$Ds & 10. & 0 & 5.0 & 0 \\
			$^{286}$Cn & 5.0 & 0 & 1.0 & 0 \\
			$^{285}$Nh & 3.0 & 0 & 1.0 & 0 \\
			$^{287}$Fl & 5.5 & 0 & 0.0 & 0 \\
			$^{288}$Fl & 4.0 & 0 & 0.0 & 0 \\
			$^{290}$Fl & 4.0 & 0 & 0.0 & 0 \\
			$^{292}$Fl & 4.0 & 0 & 0.0 & 0 \\
			$^{291}$Mc & 3.0 & 0 & 0.0 & 0 \\
			$^{293}$Lv & 2.0 & 0 & 0.0 & 0 \\
			$^{296}$Lv & 2.0 & 0 & 0.0 & 0 \\
			$^{297}$Ts & 2.0 & 2.0 & 0.0 & 2.0 \\
			$^{297}$Og & 2.0 & 3.5 & 0.0 & 3.5 \\
			\hline
		\end{tabular}
	\end{table}
	
	The mutual ratio of the cross sections $\sigma_{xn}(E)$ related to different channels $xn$ are different for different fission barrier models. For example, in the framework of the FRDM the maximal values of the channel cross sections satisfy the non-equality $\sigma_{2n}^{\rm max}(E) \ll \sigma_{3n}^{\rm max}(E) \ll \sigma_{4n}^{\rm max}(E)$ for the reaction $^{48}$Ca+$^{242}$Pu, see Fig. 6. In contrast to this, the result of the MMWS model for this reaction leads to opposite tendency, i.e. $\sigma_{2 n}^{\rm max}(E) > \sigma_{3 n}^{\rm max}(E) > \sigma_{4 n}^{\rm max}(E)$. So, the strong dependence of the cross-section $\sigma_{xn}(E)$ on the fission barrier models is seen.
	
	Due to the strong sensitivities of the channel cross-section $\sigma_{xn}(E)$ on the fission barrier values the careful fit of the cross-section values makes no sense. The fit of the experimental cross-sections is made by eye.
	
	The irregular behavior of the channel cross-section $\sigma_{xn}(E)$ calculated in the MMWS model occurs at small values of excitation energy sometimes. This happens due to the proximity of the fission barrier $B_f$ and neutron separation energy $S_n$ values in this model and the dependence of $B_f$ on $\varepsilon$. The values of the fission barrier $B_f$ obtained in the FRDM are much larger than the ones in the MMWS model, see Fig. 2, therefore, the energy dependence of the cross sections $\sigma_{xn}(E)$ is more regular.
	
	Let us consider the present model of SHN production qualitatively. Using Eqs. (1), (2), (16), and (17), the value of the $1n$ channel partial cross-section can be approximated as 
	\begin{eqnarray}
		\sigma_{1n \ell}(E) &\approx& (2\ell+1) T_\ell(E) \times R_{1n}(E) \times c_{{\cal T}_0} \nonumber \\ &\times&\frac{\exp{ \left( \frac{ B^{\rm qe}_\ell + Q - B^{\rm cnf}_\ell }{{\cal T}_0} + \frac{B^{\rm f}_{1\ell} - S_{1}}{{\cal T}_0} \right) }}{1 + \exp{\left( \frac{B^{\rm f}_{1\ell} - S_{1}}{{\cal T}_0}\right) }} .
	\end{eqnarray}
	Here, for the sake of shortening the length of the expression, the dependence of barriers on the excitation energy of the compound nucleus is omitted. Then, the value of the $xn$ channel partial cross-section in the same approximation is 
	\begin{eqnarray}
		\sigma_{xn\ell}(E) &\approx& (2\ell+1) T_\ell(E) \times R_{xn}(E) \times c_{{\cal T}_0} \times .. \times c_{{\cal T}_{x-1}} \nonumber
		\\ &\times& \frac{ \exp{\left( \frac{ B^{\rm qe}_\ell + Q - B^{\rm cnf}_\ell}{{\cal T}_{0}} +\sum_{y=1}^x \frac{B^{\rm f}_{y\ell} - S_{y}}{{\cal T}_{y-1}} \right) }} {\prod_{y=1}^x \left[ 1 + \exp{\left( \frac{B^{\rm f}_{y\ell} - S_{y}}{{\cal T}_{y-1}}\right) } \right] }, 
	\end{eqnarray}
	where $B^{\rm f}_{y\ell}$, $S_{y}$, and ${\cal T}_{y-1}$ are, respectively, the fission barrier, neutron separation energy, and temperature of the initial compound nucleus after evaporation of the $y-1$ neutrons. The temperature of the compound nucleus ${\cal T}_{y}$ decreases during sequential evaporation of neutrons. Note that at high collision energies $T_\ell(E) \approx 1$ at small values of $\ell$, therefore, the SHN production cross section is determined by the realization probability of the $xn$-evaporation channel $R_{xn}(E)$, quasi-elastic barrier, reactions Q-value, compound nucleus formation barrier, neutron separation energies, and fission barrier heights. 
	
	The expression (26) in the case of fermi-gas level density is 
	\begin{eqnarray}
		\sigma_{xn\ell}(E) \propto (2\ell+1) T_\ell(E) \times R_{xn}(E) \nonumber
		\\ \times \exp{\left[\sqrt{2a_A \left(E+Q-B^{\rm cnf}_\ell\right)} -\sqrt{2a_A \left(E-B^{\rm qe}_\ell\right)}\right] } \nonumber \\ \times 
		\frac{ \exp{\left( \sum_{y=1}^x D_y \right) }} {\prod_{y=1}^x \left[ 1 + \exp{\left( \sum_{y=1}^x D_y \right) } \right] }. 
	\end{eqnarray}
	Here $D_y= \sqrt{2 a_{A-y+1} (E^\star_{y-1}- S_{y} )} - \sqrt{2 a_{A-y+1} (E^\star_{y-1}- B^{\rm f}_{y\ell})}$ and $E^\star_{y-1}$ is the compound nucleus excitation energy after evaporation of the $y-1$ neutrons. 
	
	The values of fission barrier $B^{\rm f}_i$ are different in the FRDM and MMWS models. Therefore, to obtain the values of $\sigma_{xn}(E)$, which are close to the experimental data, using different fission barrier models, it is necessary to change the value of $b^{\rm cnf}$ to compensate the corresponding difference of $\sum_{i=1}^x B^{\rm f}_i$ in the FRDM and MMWS models. This compensation is done by assignment of the different values of $b^{\rm cnf}$ (see Table 1) in the present model when the FRDM and MMWS fission barrier models are used. Without variation of $b^{\rm cnf}$ the cross-section values $\sigma_{xn}(E)$ calculated in the FRDM fission model with the value of $b^{\rm cnf}$ used in the MMWS model are several orders higher than the experimental data. 
	
	As pointed out in Sec. 2.A, the cluster emission barrier height is close to the fission barrier height for SHN. Therefore, the compound nucleus formation barrier height should be close to the fission barrier height in SHN. The proximity of the compound nucleus formation barrier and fission barrier heights takes place in the present model in the case of using the MMWS fission barrier model because $b^{\rm cnf}=0$ for most cases, see Table 1. In contrast to this, the values $b^{\rm cnf}>0$ in the case of using the FRDM fission model. It takes place even for very heavy SHN. 
	
	Note that the values of the cross-section $\sigma_{xn}(E)$ calculated in the present model for reactions leading to the Ts and Og are several orders smaller than the available experimental values. To describe the cross section for these reactions the Q-values of the reactions are modified in the present model. Note that the experimental binding energies are absent for these SHN \cite{be}. Due to this, the model values of the binding energy of the compound nucleus should be used for the calculation of the fusion Q-value. However, the fusion reaction Q-value is not well defined for such super-heavy systems. For example, the fusion reaction Q-value for reaction $^{48}$Ca$+^{249}$Cf$\rightarrow ^{297}$Og calculated using the experimental binding energies for projectile and target nuclei, and the theoretical values of the binding energy of $^{297}$Og from Refs. \cite{frdm}, \cite{jks}, \cite{ms}, and \cite{ws4} are, respectively, -174.3, -178.3, -177.4, -177.1 MeV. The difference between the maximal $Q_{\rm max}=-174.3$ and minimal $Q_{\rm min} = -178.3$ values is 4 MeV. The SHN production cross section depends on $Q$ exponentially, see Eqs. (25)-(26). To describe this reaction in the framework of the present model the Q-value $Q_{\rm calc}=Q_{\rm MMWS}+\delta Q=-178.3+3.5=-174.8$ MeV is used in the calculation. Due to this, the correction value of the fusion reaction Q-value $\delta Q=3.5$ MeV is pointed out for this reaction in Table 1. Note, that $Q_{\rm calc} \in [Q_{\rm min},Q_{\rm max}]$. The correction value of the fusion reaction Q-value is also used for the reaction $^{48}$Ca$+^{249}$Bk$\rightarrow ^{297}$Ts. Then, the important role of the reaction Q-value in SHN synthesis is demonstrated. 
	
	\subsection{Cross-section prediction for reactions leading to SHN with Z=119 and 120}
	
	As has been pointed out in the Introduction the experiments aimed at the synthesis of isotopes of elements Z = 119 and 120, or the study of properties of related reactions have been performed \cite{119,120,120a,120b,120c,120d,120e,120f,119a}. However, these experiments have not led to the synthesis of the elements with Z = 119 and 120. Therefore, it is very interesting to calculate the SHN production cross-sections for different targets and projectiles, which leads to the synthesis of the elements with Z = 119 and 120, in the framework of the present model.
	
	The way to use $^{48}$Ca projectile and Es or Fm targets, which was used for synthesizing Fl, Mc, Lv, Ts, and Og, is questionable due to difficulties in collecting material for such targets now. Therefore, the practical way to use the reactions between Sc or Ti projectiles and $^{249}$Bk or $^{249}$Cf targets. Scandium has only one naturally available stable isotope $^{45}$Sc while titanium has five naturally available stable isotopes $^{46,47,48,49,50}$Ti. Therefore, the cross sections for the reactions $^{45}$Sc$+^{249}$Cf$\rightarrow ^{294}$119, $^{46,48,50}$Ti$+^{249}$Bk$\rightarrow ^{295,297,299}$119, and $^{46,48,50}$Ti$+^{249}$Cf$\rightarrow ^{295,297,299}$120 are presented in Figs. 10 -- 12. The calculations of the cross-sections of SHN for the reactions leading to the compound nucleus with $Z>118$ with $^{249}$Bk and $^{249}$Cf targets are done using the parameter values fixed for reactions $^{48}$Ca$+^{249}$Bk$\rightarrow ^{297}$Ts and $^{48}$Ca$+^{249}$Cf$\rightarrow ^{297}$Og, respectively.
	
	\begin{figure}
		\includegraphics[width=6.9cm]{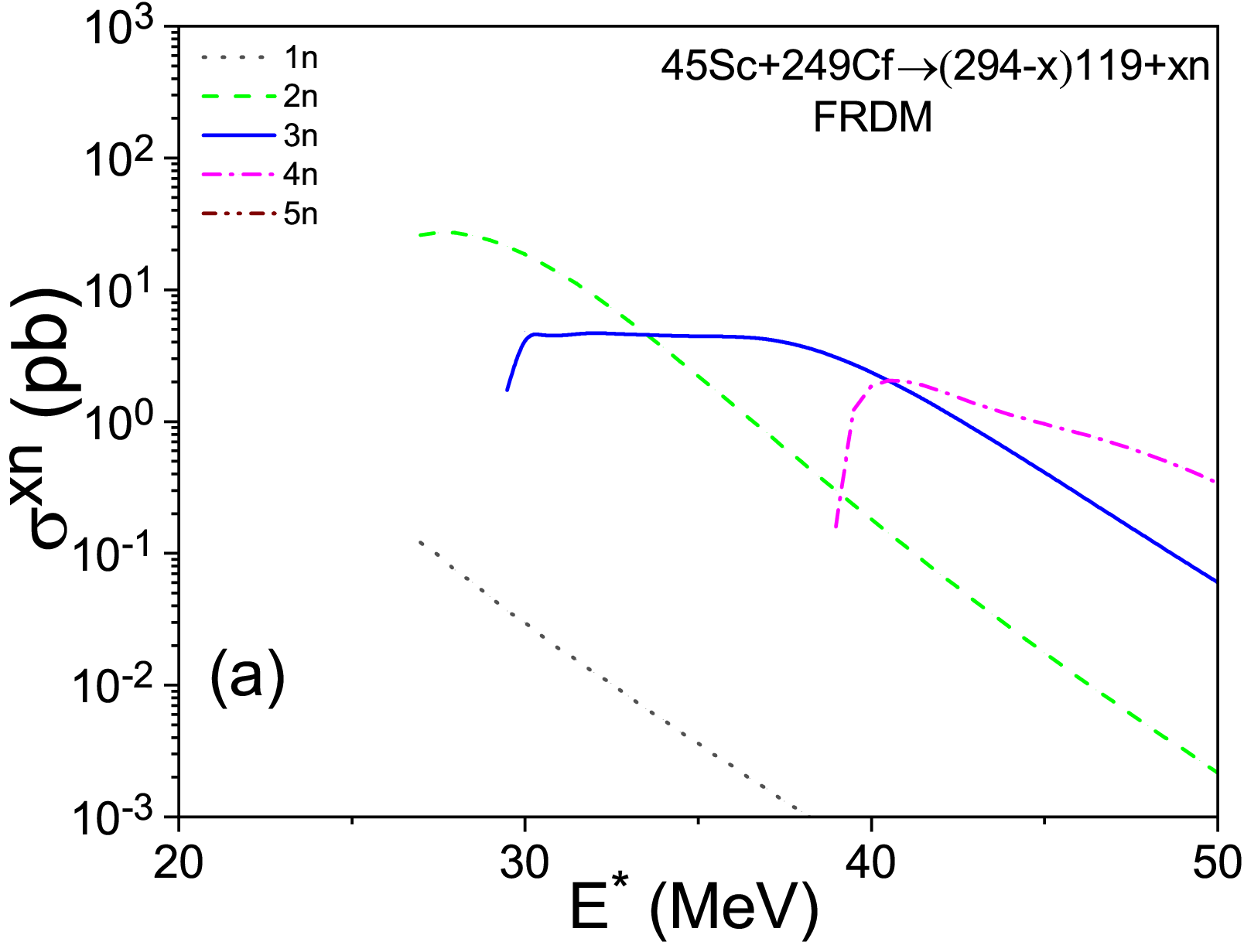}
		\includegraphics[width=6.9cm]{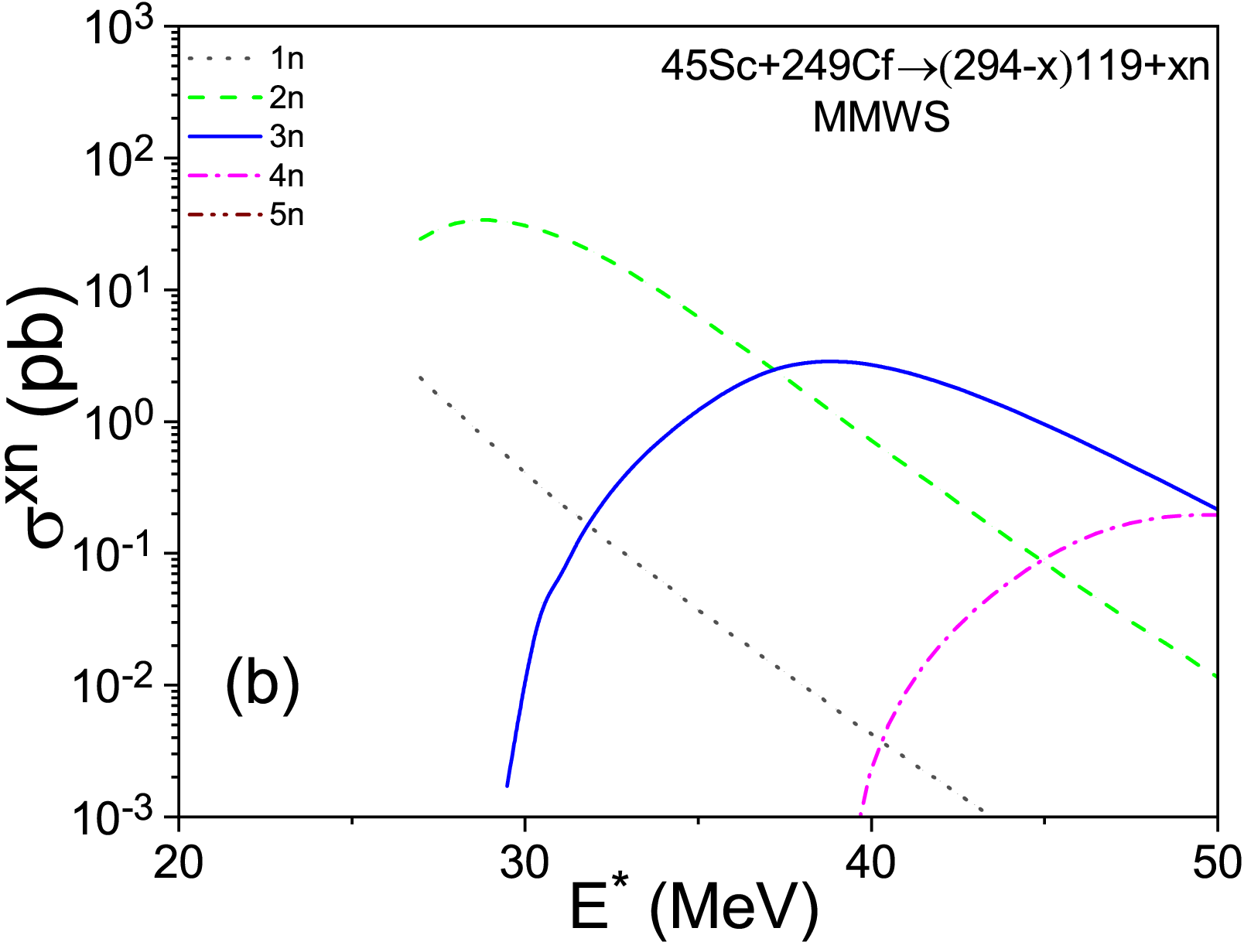}
		\caption{\label{fig10} The theoretical cross-sections of the SHN synthesis $\sigma^{xn}(E)$ for the reaction $^{45}$Sc$+^{249}$Cf$\rightarrow ^{294-x}$119+$xn$, which are calculated for the FRDM (a) \cite{frdm,frdm_fb} and MMWS (b) \cite{jks} fission barrier models.}
	\end{figure}
	
	\begin{figure*}
		\includegraphics[width=5.9cm]{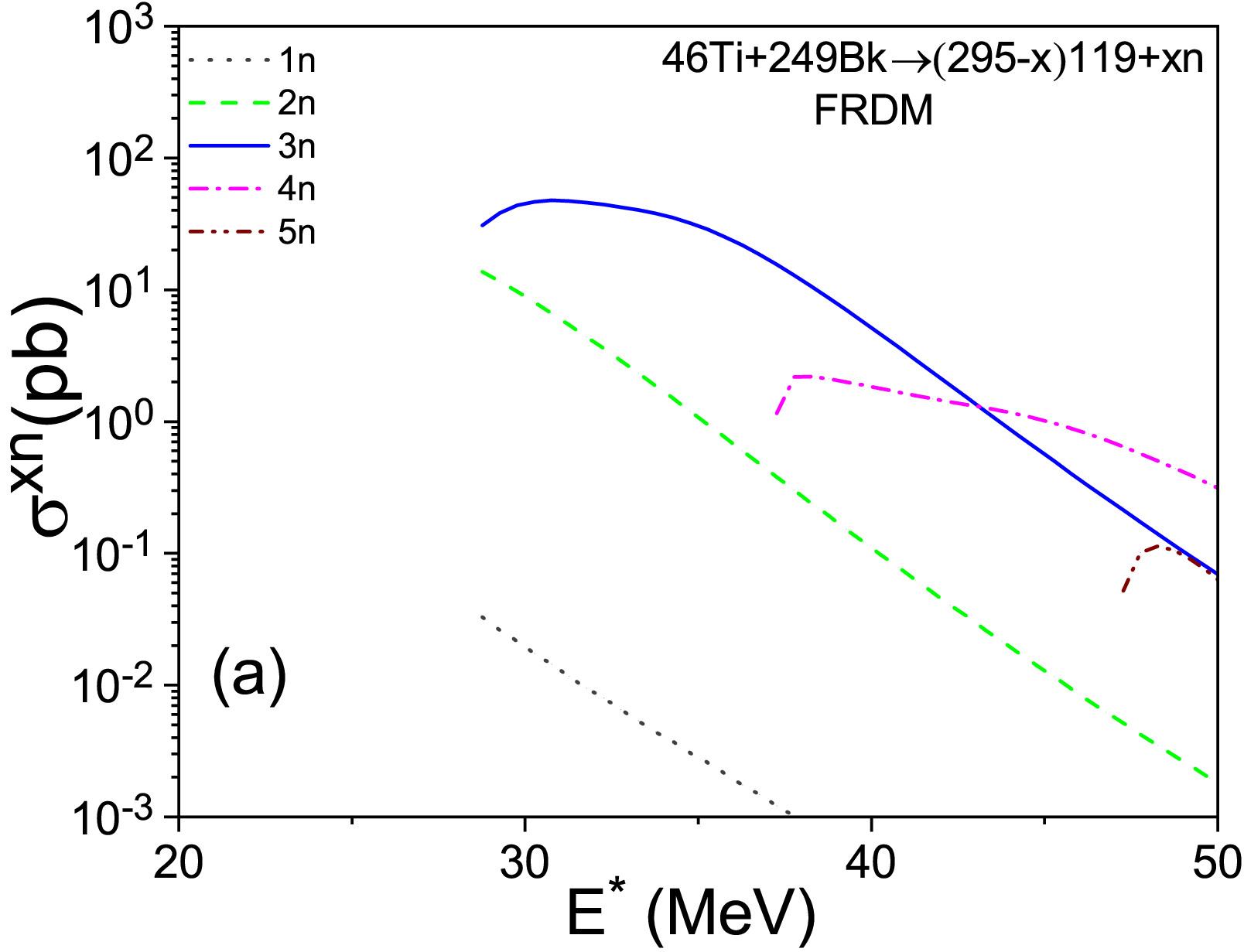}
		\includegraphics[width=5.9cm]{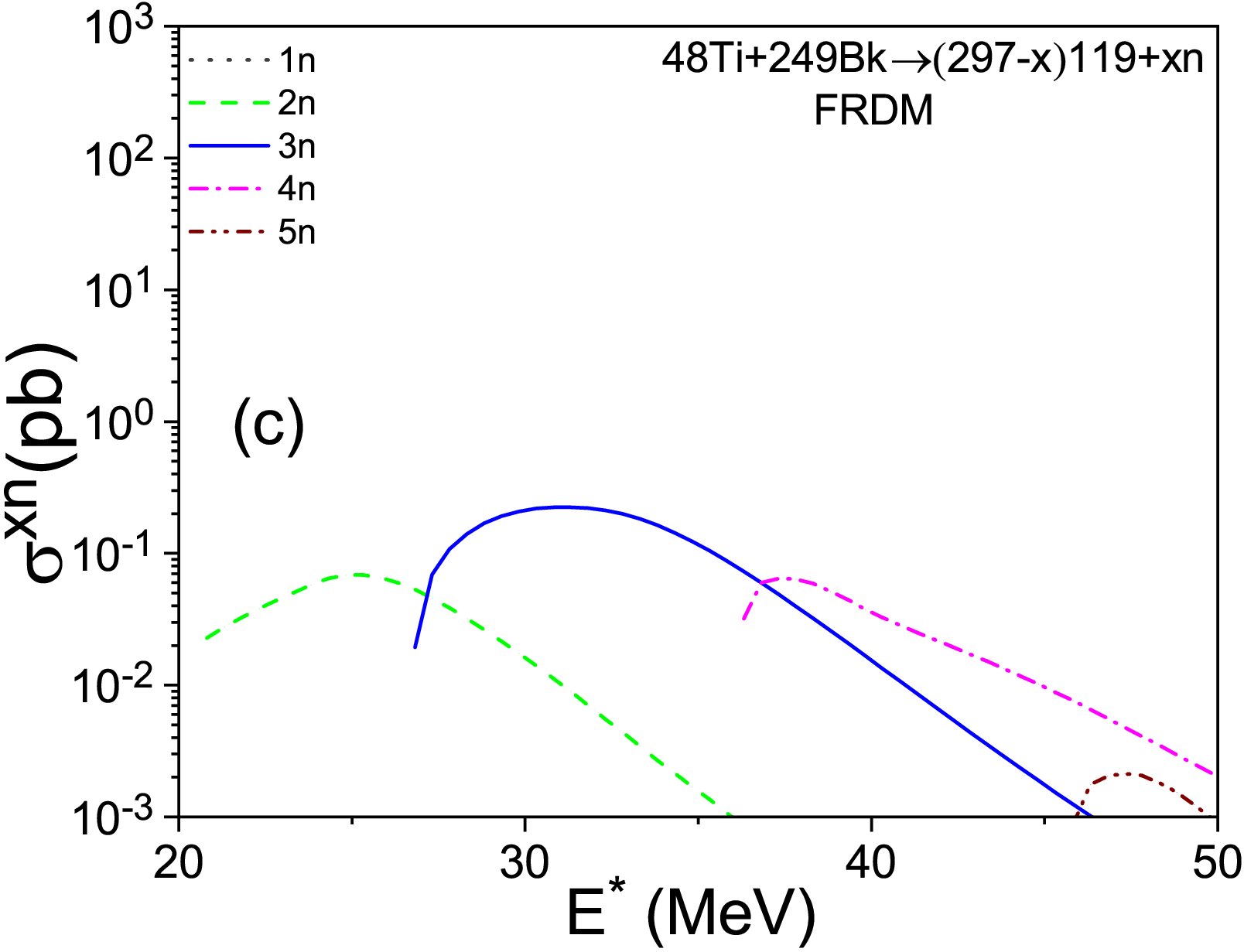}
		\includegraphics[width=5.9cm]{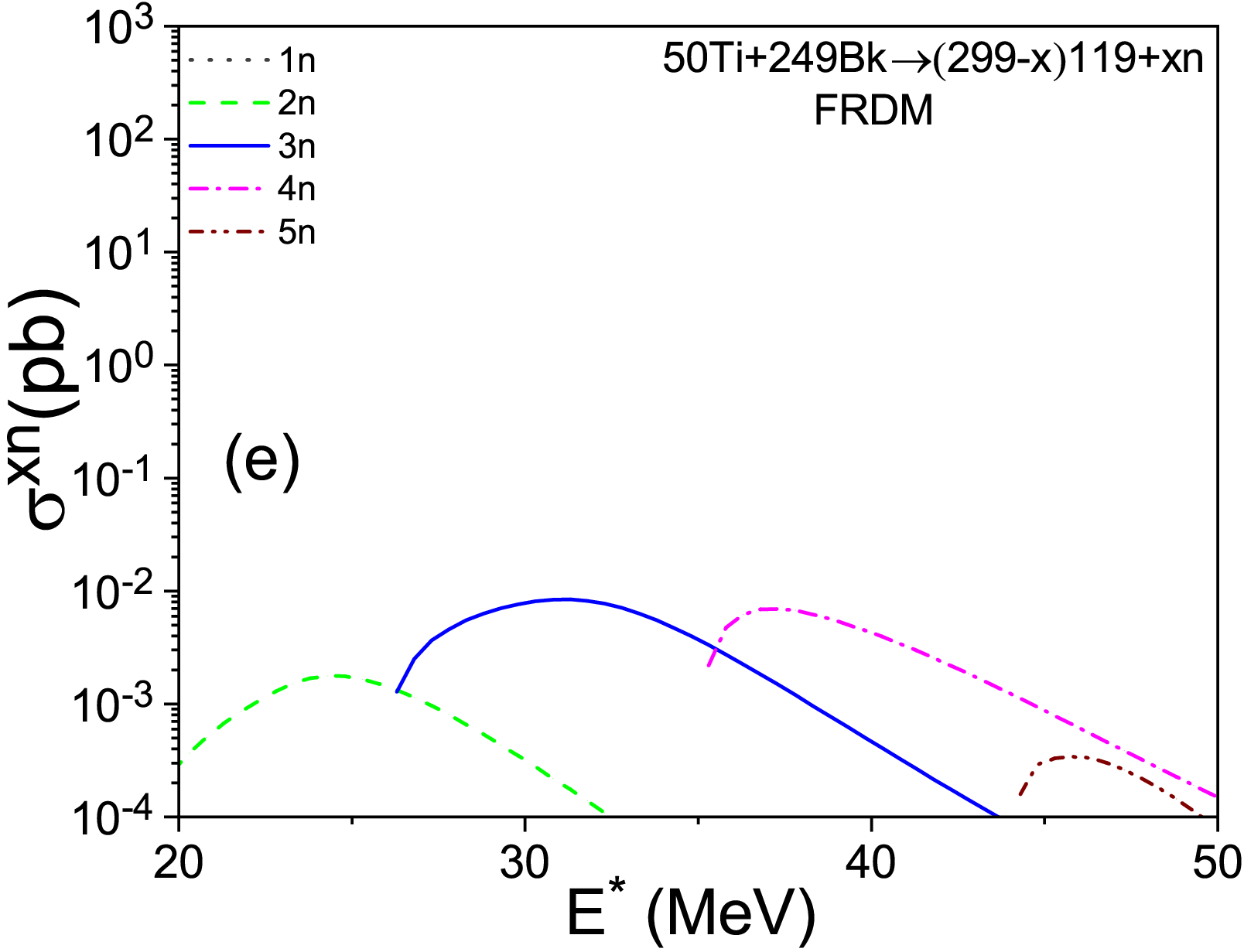}
		\includegraphics[width=5.9cm]{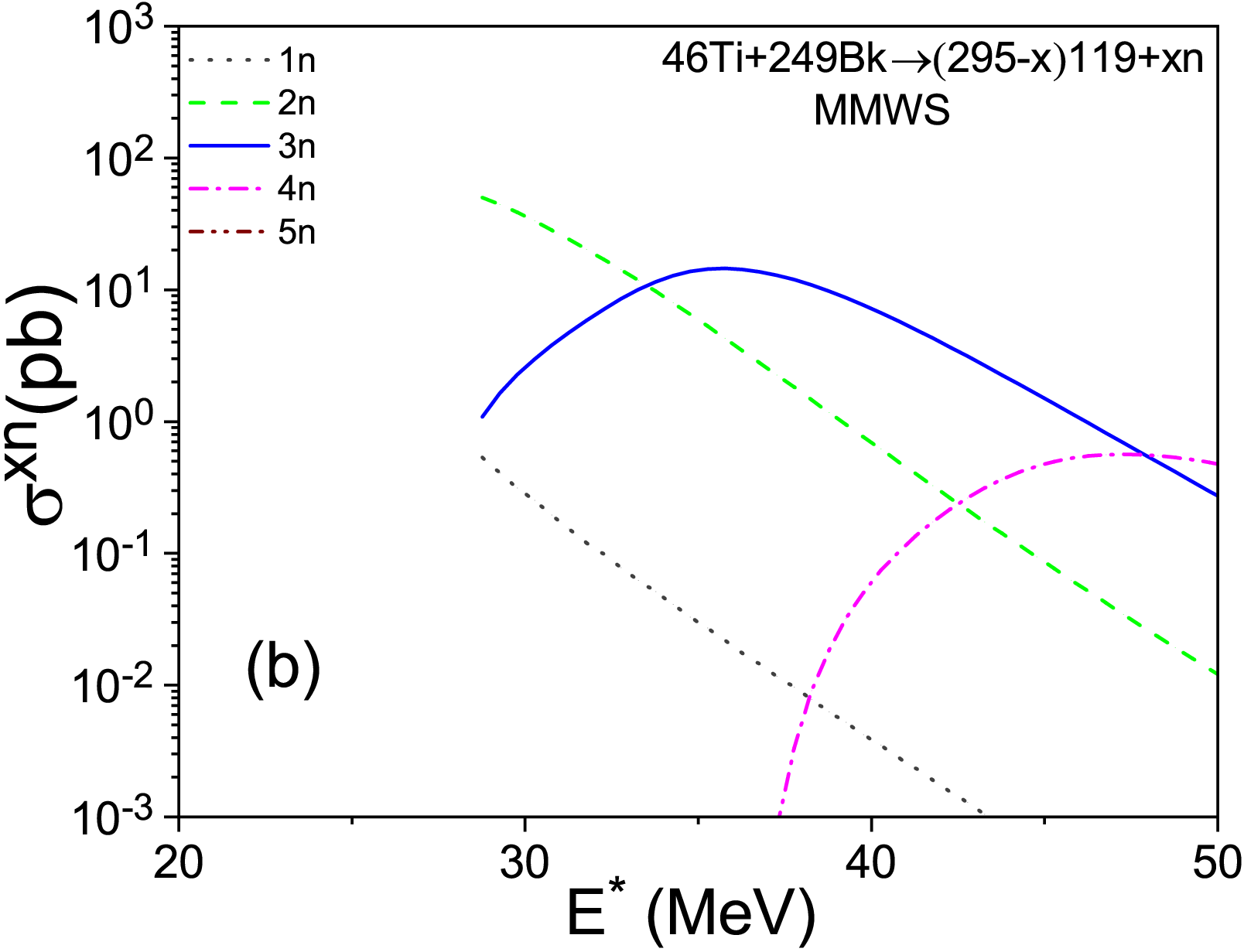}
		\includegraphics[width=5.9cm]{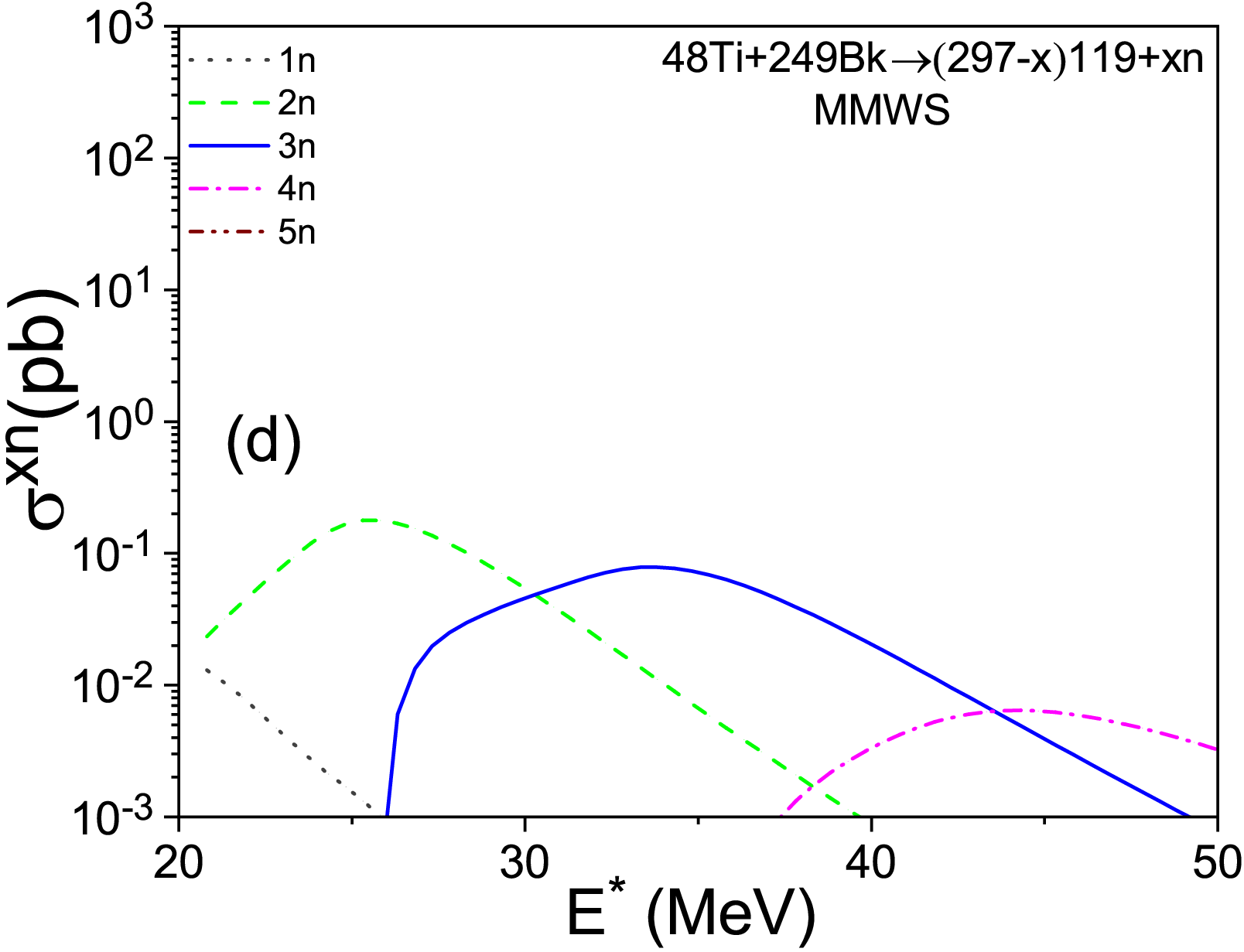}
		\includegraphics[width=5.9cm]{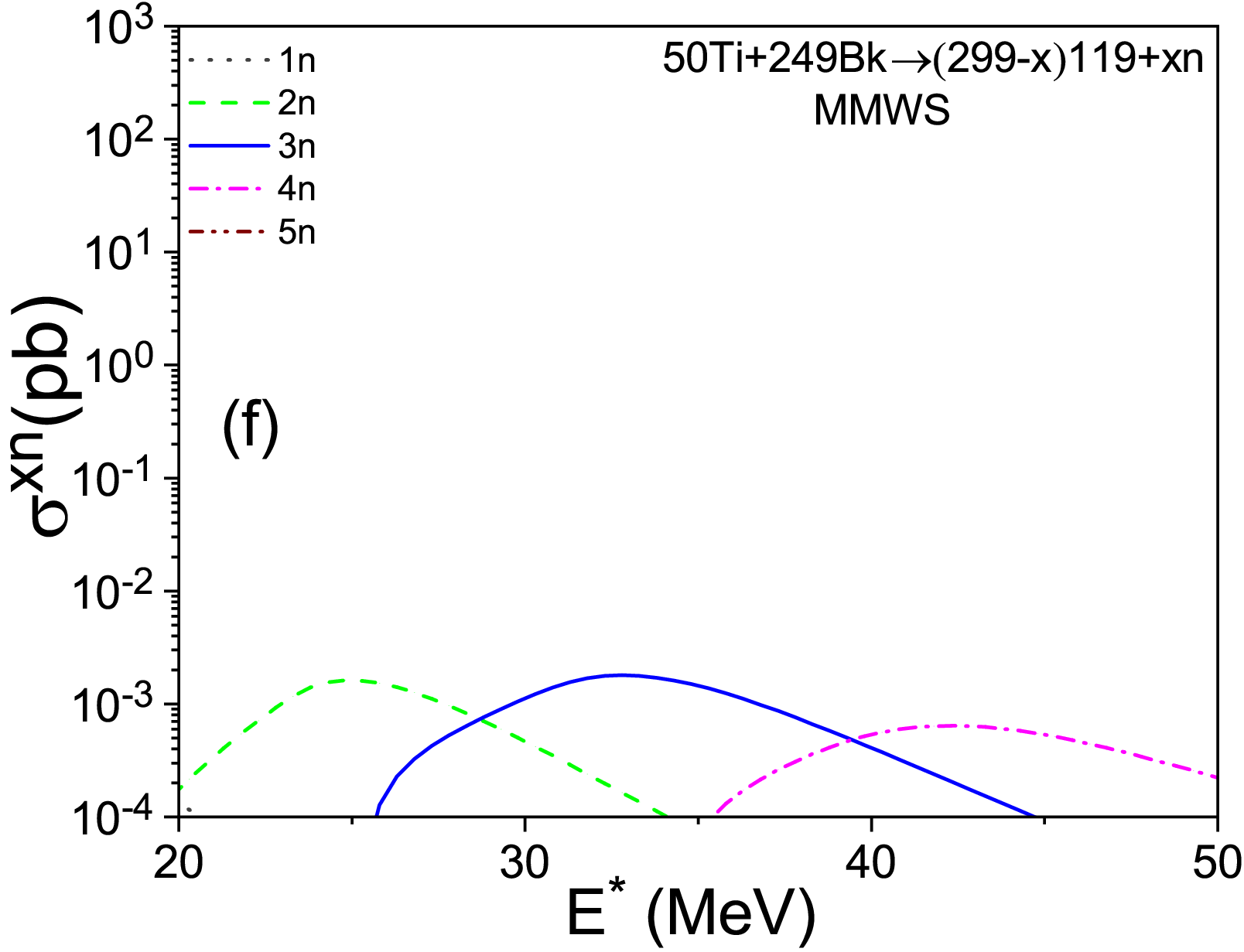}
		\caption{\label{fig11} The theoretical cross-sections of the SHN synthesis $\sigma^{xn}(E)$ for the reactions $^{46}$Ti$+^{249}$Bk$\rightarrow ^{295-x}$119+$xn$ (a,b), $^{48}$Ti$+^{249}$Bk$\rightarrow ^{297-x}$119+$xn$ (c,d), and $^{46,48,50}$Ti$+^{249}$Bk$\rightarrow ^{299-x}$119+$xn$ (e,f), which are calculated for the FRDM (a,c,e) \cite{frdm,frdm_fb} and MMWS (b,d,f) \cite{jks} fission barrier models.}
	\end{figure*}
	
	\begin{figure*}
		\includegraphics[width=5.9cm]{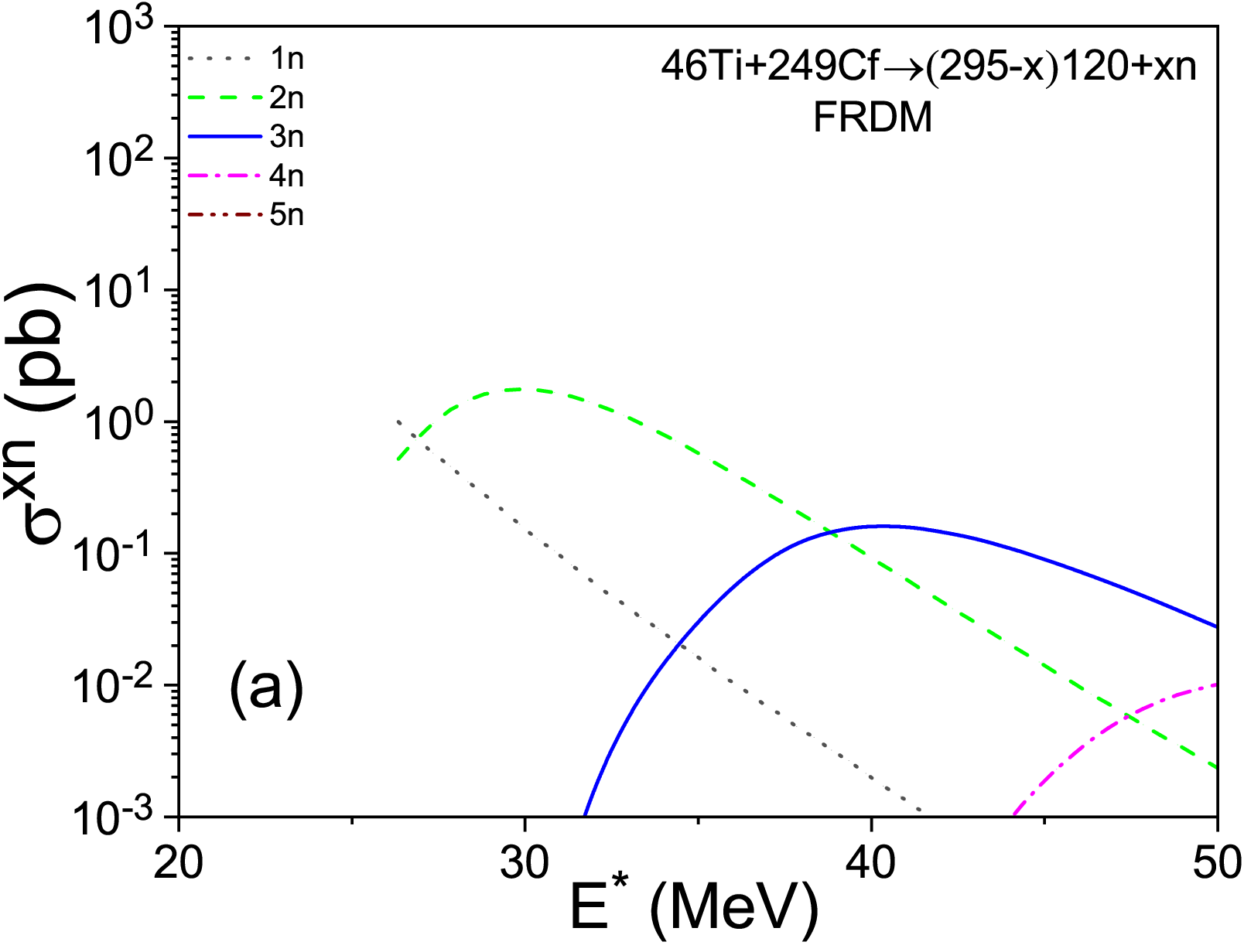}
		\includegraphics[width=5.9cm]{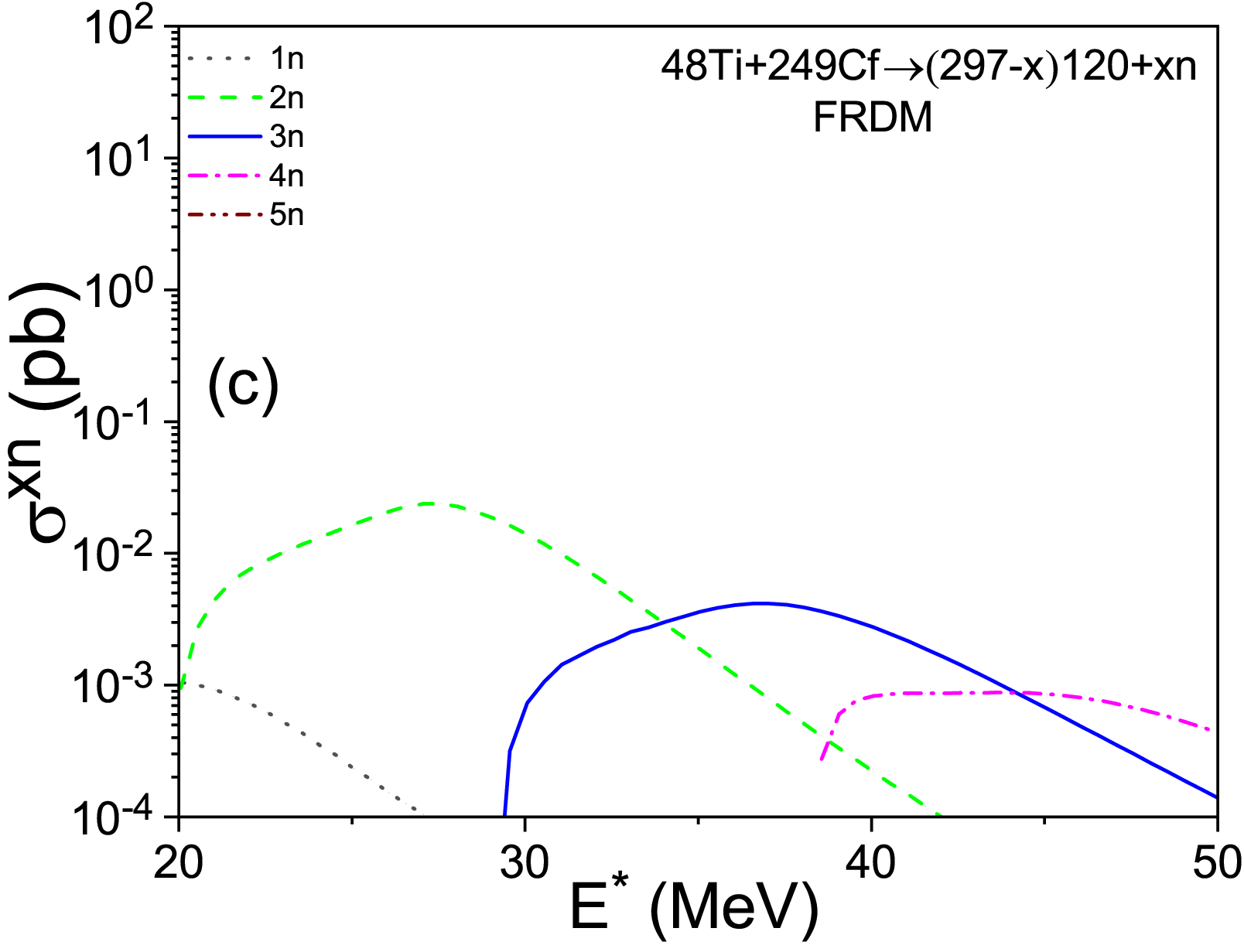}
		\includegraphics[width=5.9cm]{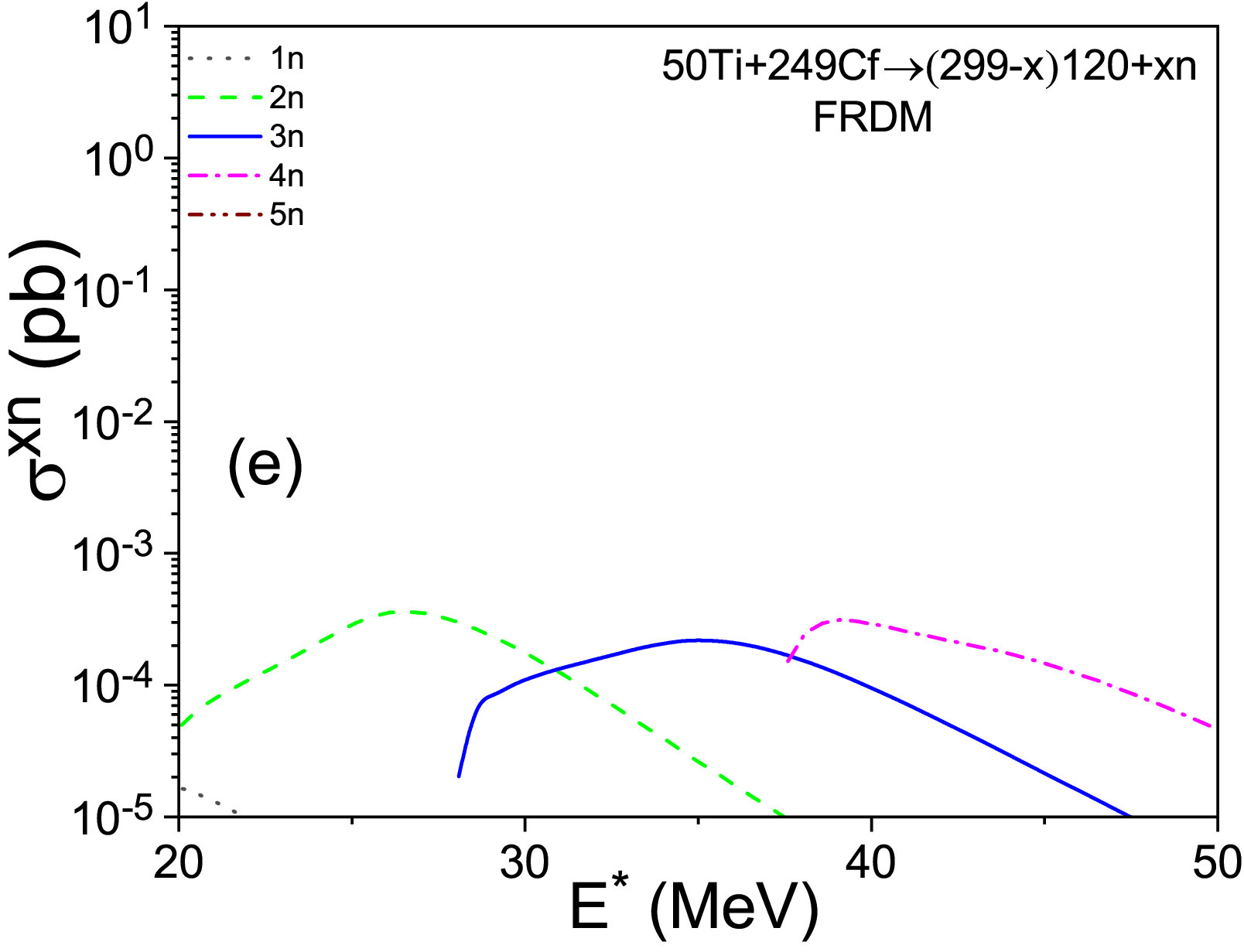}
		\includegraphics[width=5.9cm]{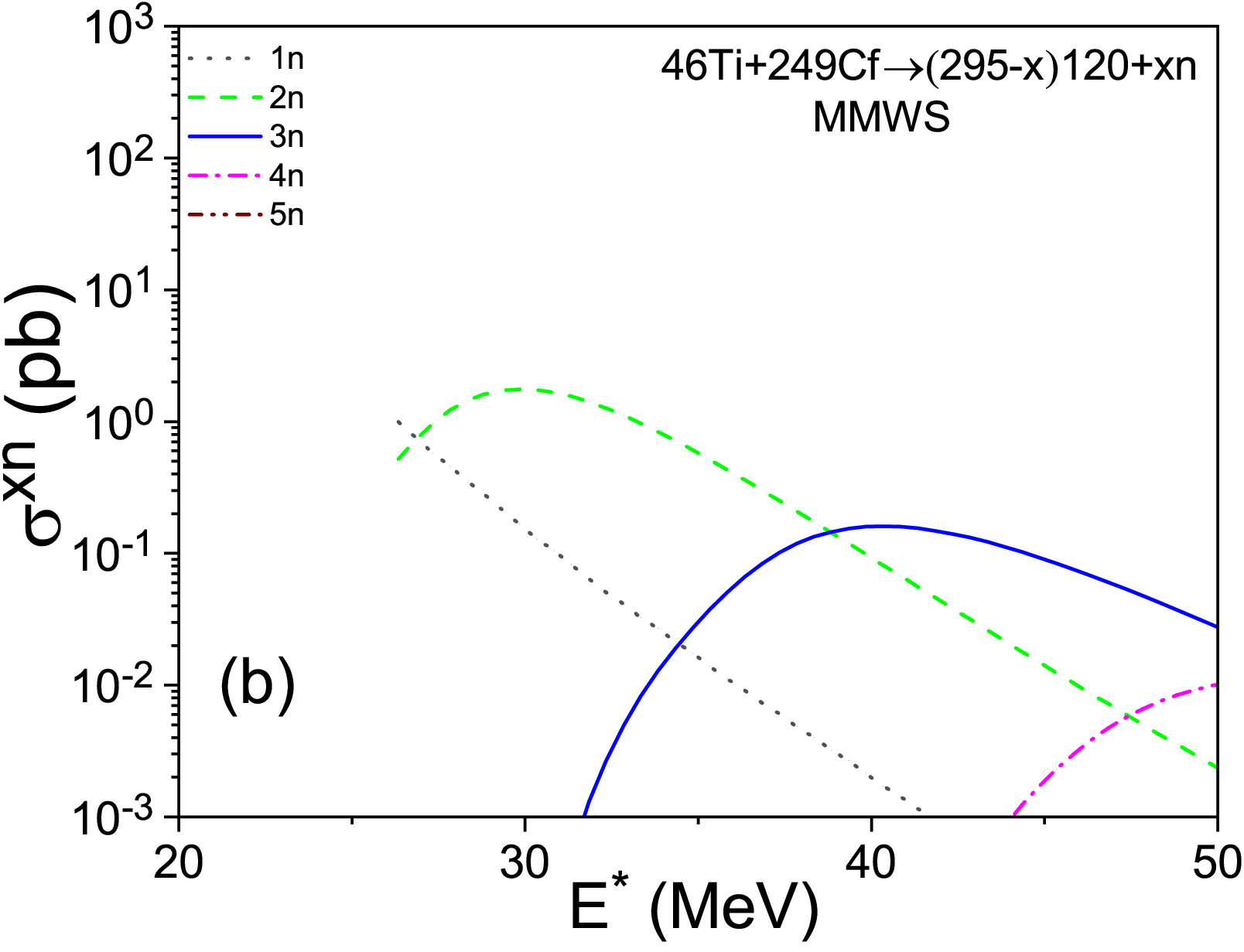}
		\includegraphics[width=5.9cm]{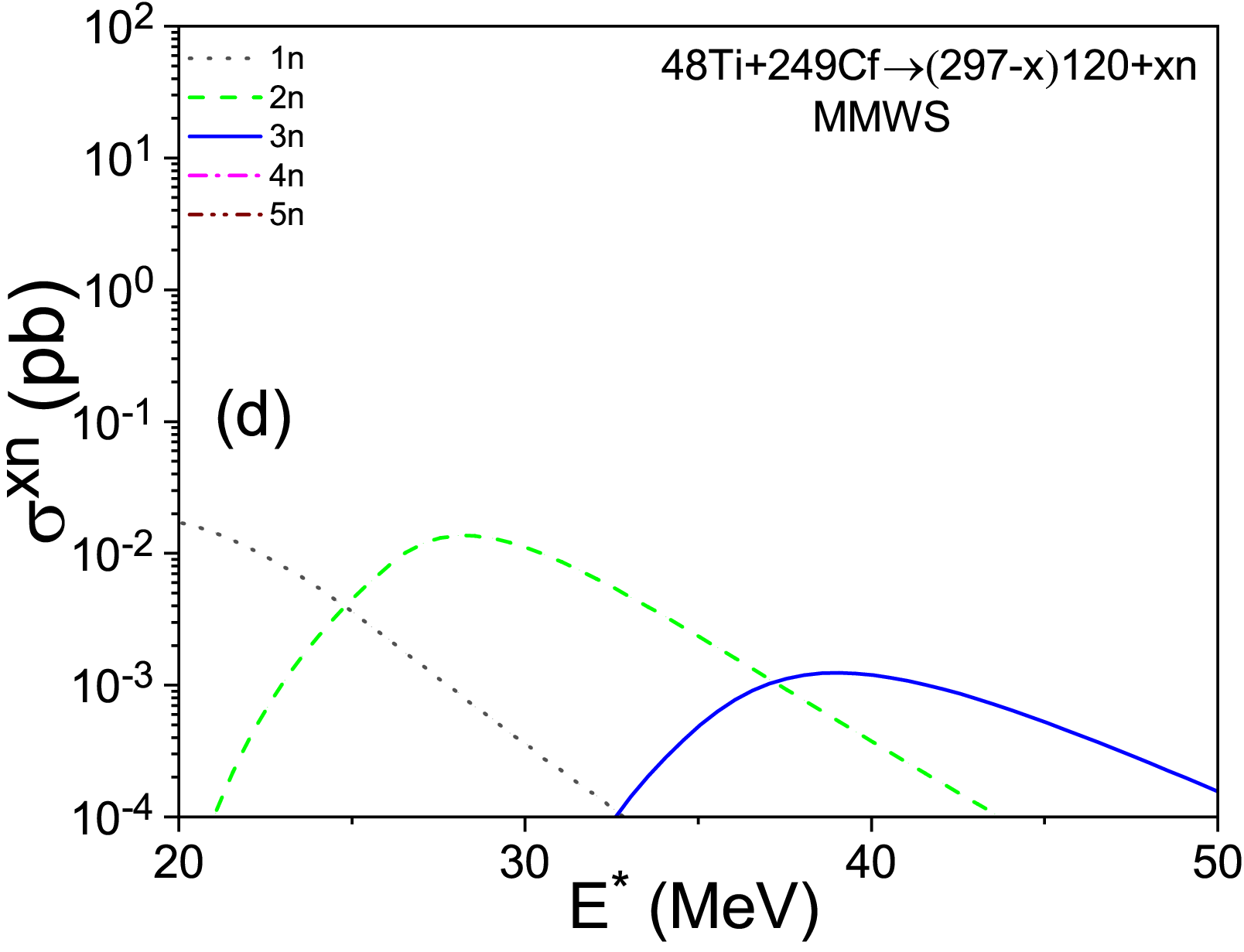}
		\includegraphics[width=5.9cm]{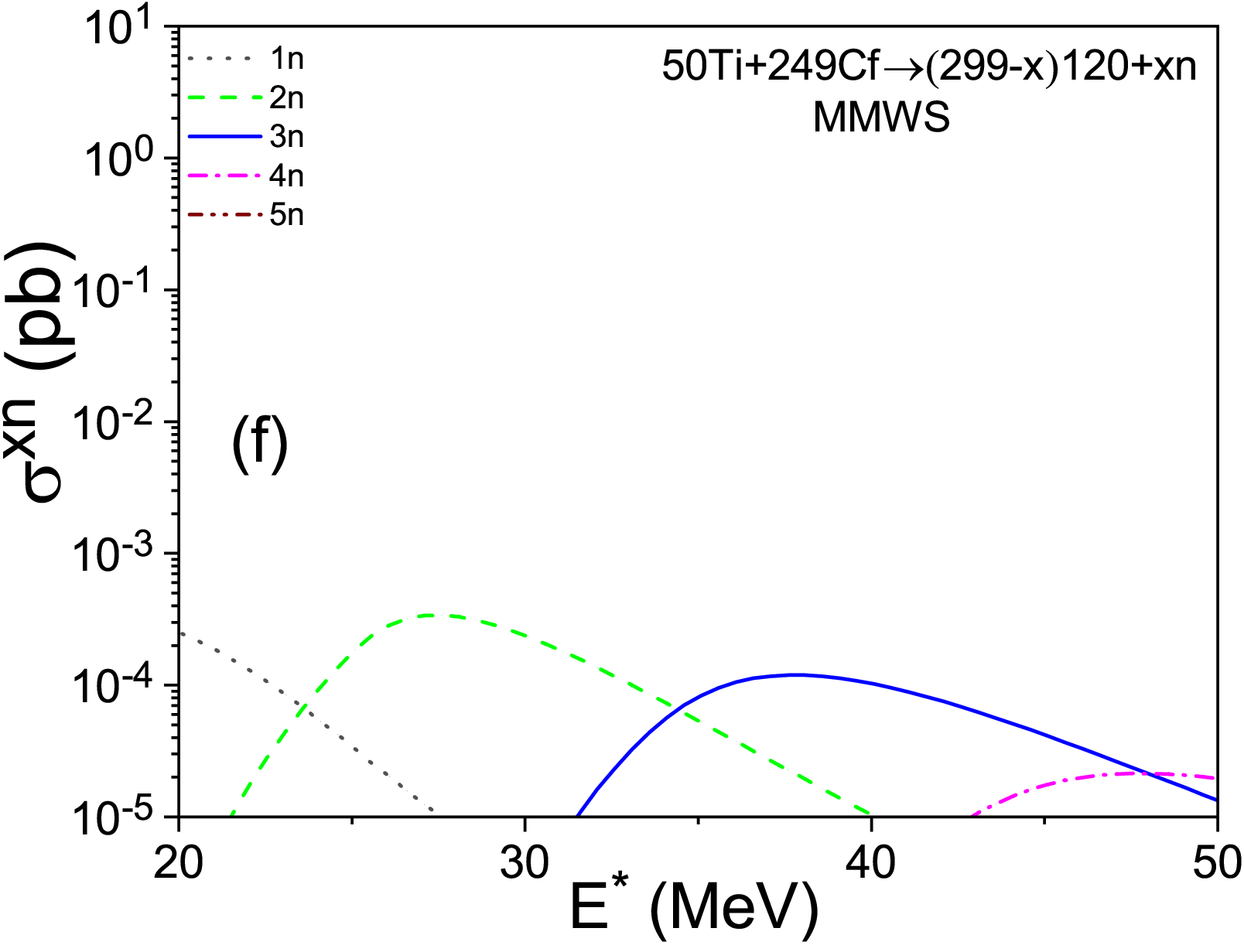}
		\caption{\label{fig12} The theoretical cross-sections of the SHN synthesis $\sigma^{xn}(E)$ for the reactions $^{46}$Ti$+^{249}$Cf$\rightarrow ^{295-x}$120+$xn$ (a,b), $^{48}$Ti$+^{249}$Cf$\rightarrow ^{297-x}$120+$xn$ (c,d), and $^{50}$Ti$+^{249}$Cf$\rightarrow ^{299-x}$120+$xn$ (e,f), which are calculated for the FRDM (a,c,e) \cite{frdm,frdm_fb} and MMWS (b,d,f) \cite{jks} fission barrier models.}
	\end{figure*}
	
	The results of the calculations are presented in Figs. 10 -- 12 show that the reactions $^{45}$Sc$+^{249}$Cf$\rightarrow ^{294-x}$119+$xn$, $^{46}$Ti$+^{249}$Bk$\rightarrow ^{295-x}$119+$xn$, and $^{46}$Ti$+^{249}$Cf$\rightarrow ^{295-x}$120+$xn$ can be used for the synthesis of the elements with 119 and 120 protons. The maximal values of cross sections of these reactions are around the picobarn levels for various versions of the fission barrier models.
	
	The reactions with $^{48,50}$Ti projectiles have lower values of cross sections than the ones for the $^{46}$Ti projectile, see Figs 11 -- 12. This is the effect of the Q-value. For example, the Q-value of the fusion reactions $^{46,48,50}$Ti$+^{249}$Cf$\rightarrow ^{295,297,299}$120 evaluated using the FRDM \cite{frdm,frdm_fb} (MMWS \cite{jks}) SHN binding energy models are, respectively, -183.7 (-187.3), -189.9 (-193.3), and -194.8 (-198.6) MeV. Due to Eqs. (25) and (26), the SHN cross-section exponentially decreases with the decrease of Q-value if the values of the fission barrier heights, separation energies, and quasi-elastic barrier are approximately similar for the reactions with different isotopes. 
	
	The height of the quasi-elastic barrier $B^{\rm qe}$ of the total potential (14), which separates the sticking and well-separated nuclei, depends on the nucleus-nucleus Coulomb and nuclear interactions and the deformation energies of both nuclei (15). The soft interacting nuclei can be stronger deformed at small distances between their surfaces than the stiff ones. Therefore, the barrier distance between the mass centers of the soft nuclei is larger than for the stiff ones. This leads to the reduction of the Coulomb energy and, as a result, the height of the quasi-elastic barrier $B^{\rm qe}$ \cite{ds21,d23d}. The reduction of the quasi-elastic barrier height leads to a decrease in the production cross-section, see Eqs. (26)-(27).
	
	The dependence of the SHN production cross sections on the stiffness of interacting nuclei and other parameters ($B^{\rm f}$, $B^{\rm cnf}$, $S_{\rm n}$) may be traced in the analysis of the reactions $^{44,48}$Ca+$^{208}$Pb$\rightarrow^{252,256-x}$No+$xn$ $^{48}$Ca+$^{206}$Pb$\rightarrow^{254-x}$No+$xn$. The nuclei $^{48}$Ca and $^{208}$Pb are stiffer than $^{44}$Ca and $^{206}$Pb because for $^{48}$Ca and $^{208}$Pb the energies of $2^+$ and $3^-$ surface oscillations are the largest and the oscillation amplitudes are smallest \cite{be2,be3}. Therefore, the reaction $^{48}$Ca+$^{208}$Pb$\rightarrow^{256-x}$No+$xn$ should have the highest values of the $B^{\rm qe}$ and, due to this, larger of both the probability of the compound nucleus formation and the super-heavy production cross-section. This agrees with the experimental values of the cross-section for the reactions $^{44,48}$Ca+$^{208}$Pb$\rightarrow^{252,256-x}$No+$xn$ $^{48}$Ca+$^{206}$Pb$\rightarrow^{254-x}$No+$xn$ \cite{capb}. 
	
	The dependence of nuclear surface stiffness is considered phenomenologically in the present model, see Eq. (15). The deviation of the stiffness from the shell correction value is related to the shell-correction contribution. The shell-correction energy decreases at high excitation energies. Therefore, the surface stiffness should be close to the liquid drop value at high excitation energies. However, it may happen that the surface stiffness is not close to the liquid drop value. Therefore, the cross-section calculations are done for the two times lower stiffness of the projectile nuclei too. This leads to the reduction of the cross-sections presented in Figs. 10 -- 12 on one order approximately. 
	
	So, the SHN production cross-section prediction is linked to many poorly defined quantities. Due to this, accurate prediction of the SHN production cross-section is impossible now.
	
	\section{Conclusion}
	
	The new model for the SHN production cross-section in the hot fusion reactions is presented. The available experimental data for the SHN production cross-section are well described in the model. It is shown, that the projectiles $^{45}$Sc and $^{46}$Ti can be used for the synthesis of the elements with 119 and 120 protons. 
	
	New qualitative expressions have been obtained for the probability of the formation of the compound nucleus and the cross-section of formation in reactions leading to SHN. 
	
	It is shown that the super-heavy nuclei production cross section is proportional to the transmission coefficient of the capture barrier, realization probability of the $xn$-evaporation channel, and exponentially depends on the quasi-elastic barrier, fusion reaction Q-value, compound nucleus formation barrier, neutron separation energies, and fission barrier heights. These quantities are important for the reaction choice for the synthesis of SHN.
	
	The various models for the fission barriers and SHN binding energy lead to very different values of the SHN production cross-section. This takes place due to dependencies of the compound-nucleus formation probability on these quantities and of the survival probability on the fission barriers.

	\section*{Acknowledgments}
	
	The author thanks the support of Professors Fabiana Gramegna, Enrico Fioretto, Giovanna Montagloli, and Alberto Stefanini.
	
	The author thanks for the support to Istituto Nazionale di Fisica Nucleare, Laboratori Nazionali di Legnaro of Istituto Nazionale di Fisica Nucleare, the National Academy of Sciences of Ukraine and Taras Shevchenko the National University of Kiev.


\begin{thebibliography}{999}
		
		\bibitem{epja5} Yu. Ts. Oganessian, et al., Eur. Phys. J. A 5, 63 (1999). 
		
		\bibitem{prc69} Yu. Ts. Oganessian, et al., Phys. Rev. C 69, 054607 (2004). 
		
		\bibitem{epja19} Yu. Ts. Oganessian, et al., Eur. Phys. J. A 19, 3 (2004). 
		
		\bibitem{prc70} Yu. Ts. Oganessian, et al., Phys. Rev. C 70, 064609 (2004). 
		
		\bibitem{prc74} Yu. Ts. Oganessian, et al., Phys. Rev. C 74, 044602 (2006). 
		
		\bibitem{prc76} Yu. Ts. Oganessian, et al., Phys. Rev. C 76, 011601(R) (2007). 
		
		\bibitem{epja32} S. Hofmann, et al., Eur. Phys. J. A 32, 251 (2007). 
		
		\bibitem{prl103} L. Stavsetra, et al., Phys. Rev. Lett. 103, 132502 (2009). 
		
		\bibitem{prl104} Ch. E. Dullmann, Phys. Rev. Lett. 104, 252701 (2010). 
		
		\bibitem{prl105} P. A. Ellison, Phys. Rev. Lett. 105, 182701 (2010). 
		
		\bibitem{prc83} J. M. Gates, et al., Phys. Rev. C 83, 054618 (2011). 
		
		\bibitem{epja48} S. Hofmann, et al., Eur. Phys. J. A 48, 62 (2012). 
		
		\bibitem{prc87_Am} Yu. Ts. Oganessian, et al., Phys. Rev. C 87, 014302 (2013). 
		
		\bibitem{prc87_Ra} Yu. Ts. Oganessian, et al., Phys. Rev. C 87, 034605 (2013).
		
		\bibitem{prc87_Bk} Yu. Ts. Oganessian, et al., Phys. Rev. C 87, 054621 (2013). 
		
		\bibitem{prc92} V. K. Utyonkov, et al., Phys. Rev. C 92, 034609 (2015). 
		
		\bibitem{npa953} U. Forsberg, et al., Nucl. Phys. A 953, 117 (2016). 
		
		\bibitem{jpsj86_Cm} D. Kaji, et al., J. Phys. Soc. Japan 86, 034201 (2017). 
		
		\bibitem{jpsj86_U} D. Kaji, et al., J. Phys. Soc. Japan 86, 085001 (2017). 
		
		\bibitem{prc97} V. K. Utyonkov, et al., Phys. Rev. C 97, 014320 (2018). 
		
		\bibitem{prc98} N. T. Brewer, et al., Phys. Rev. C 98, 024317 (2018). 
		
		\bibitem{prc99} J. Khuyagbaatar, et al., Phys. Rev. C 99, 054306 (2019). 
		
		\bibitem{prc106_Am} Yu. Ts. Oganessian, et al., Phys. Rev. C 106, L031301 (2022). 
		
		\bibitem{prc106_Pu_U} Yu. Ts. Oganessian, et al., Phys. Rev. C 106, 024612 (2022). 
		
		\bibitem{prc107} A. Samark-Roth, et al., Phys. Rev. C 107, 024301 (2023). 
		
		\bibitem{prc108} Yu. Ts. Oganessian, et al., Phys. Rev. C 108, 024611 (2023). 
		
		\bibitem{119} R. W. Lougheed, et al., Phys. Rev. C 32, 1760 (1985). 
		
		\bibitem{120} Yu. Ts. Oganessian, et al., Phys. Rev. C 79, 024603 (2009). 
		
		\bibitem{120a} E. M. Kozulin, et al., Phys. Lett. B 686, 227 (2010). 
		
		\bibitem{120b} S. Hofmann, et al., Eur. Phys. J. A 52, 180 (2016). 
		
		\bibitem{120c} F. P. Hessberger, D. Ackermann, Eur. Phys. J. A 53, 123 (2017). 
		
		\bibitem{120d} A. Di Nitto, et al., Phys. Lett. B 784, 199 (2018). 
		
		\bibitem{120e} K. V. Novikov, et al., Bull. Russ. Acad. Sci. Physics, 84, 495 (2020). 
		
		\bibitem{120f} J. Khuyagbaatar, et al., Phys. Rev. C 79, C 102, 064602 (2020). 
		
		\bibitem{119a} M. Tanaka, et al., J. Phys. Soc. Japan 91, 084201 (2022). 
		
		\bibitem{kaas} A. N. Kuzmina, G. G. Adamian, N. V. Antonenko, W. Scheid, Phys. Rev. C 85, 014319 (2012). 
		
		\bibitem{zwr} J. Zhang, C. Wang, Z. Ren, Nucl. Phys. A909,36 (2013). 
		
		\bibitem{zg} V. I. Zagrebaev W. Greiner, Nucl. Phys A 944, 257 (2015). 
		
		\bibitem{ss} K. P. Santhosh V. Safoora, Phys. Rev. C 95, 064611 (2017). 
		
		\bibitem{haa} J. Hong, G. G. Adamian, N. V. Antonenko, Phys. Rev. C 94, 044606 (2016). 
		
		\bibitem{aal} G. G. Adamian, N. V. Antonenko, H. Lenske, Nucl. Phys. A970, 22 (2018). 
		
		\bibitem{li} J. Li, et al., Phys. Rev. C 98, 014626 (2018). 
		
		\bibitem{sh} K. Sekizawa, K. Hagino, Phys. Rev. C 99, 051602(R) (2019). 
		
		\bibitem{swck} K. Siwek-Wilczyńska, T. Cap, M. Kowal, Phys. Rev. C 99, 054603 (2019). 
		
		\bibitem{b} X. J. Bao, Phys. Rev. C 100, 011601(R) (2019). 
		
		\bibitem{msmdss} H. C. Manjunatha, et al., Phys. Rev. C 102, 064605 (2020). 
		
		\bibitem{lv} X.-J. Lv, Z.-Y. Yue, W.-J. Zhao, B. Wang, Phys. Rev. C 103, 064616 (2021). 
		
		\bibitem{rkb} S. Rana, R. Kumar, M. Bhuyan, Phys. Rev. C 104, 024619 (2021). 
		
		\bibitem{lwz} J.-X. Li, W.-X. Wang, H.-F. Zhang, Phys. Rev. C 106, 044601 (2022). 
		
		\bibitem{kdck} N. Kumari, A. Deep, S. Chopra, R. Kharab, Phys. Rev. C 107, 014610 (2023)
		
		\bibitem{dz} X.-Q. Deng, S.-G. Zhou, Phys. Rev. C 107, 014616 (2023). 
		
		\bibitem{sg} X.-X. Sun, L. Guo, Phys. Rev. C 107, 064609 (2023). 
		
		\bibitem{sjks} H. Sharma, S. Jain, R. Kumar, M. K. Sharma, Phys. Rev. C 108, 044613 (2023). 
		
		\bibitem{zs} R. Zargini, S. A. Seyyedi, Phys. Rev. C 108, 034606 (2023). 
		
		\bibitem{kk} N. Yu. Kurkova, A. V. Karpov, Phys. At. Nucl. 86, No. 4, 31 (2023). DOI: 10. 1134/S1063778823040257 
		
		\bibitem{vh} R. Vandenbosch and J. R. Huizenga, Nuclear Fission (New York, Academic Press, 1973). 
		
		\bibitem{frdm} P. Moller, et al., At. Data Nucl. Data Tabl. 109-110, 1 (2016). 
		
		\bibitem{frdm_fb} P. Moller, et al., Phys. Rev. C 91, 024310 (2015). 
		
		\bibitem{jks} P. Jachimowicz, M. Kowal, J. Skalski, At. Data Nucl. Data Tabl. 138, 101393 (2021). 
		
		\bibitem{be} F. G. Kondev, et al., Chinese Phys. C 45, 030001 (2021). 
		
		\bibitem{ms} W. D. Myers, W. J. Swiatecki, Nuclear Physics A 601, 141 (1996). 
		
		\bibitem{ws4} N. Wang, M. Liu, X. Wu, J. Meng, Phys. Lett. B 734, 215 (2014). 
		
		\bibitem{fus_exp} E. M. Kozulin, et al., Phys. Rev. C 90, 054608 (2014).
		
		\bibitem{qelastbar} S. Mitsuoka, et al., Phys. Rev. Lett. 99, 182701 (2007).
		
		\bibitem{nrv} A. V. Karpov, et al., Phys. At. Nucl. 79, 749 (2016). DOI: 10.1134/S1063778816040141 (http://nrv.jinr.ru/nrv/)
		
		\bibitem{dhrs} M. Dasgupta et al., Annu. Rev. Nucl. Part. Sci. 48, 401 (1998).
		
		\bibitem{montstef} G. Montagnoli and A. M. Stefanini, Eur. Phys. J. A 53, 169 (2017).
		
		\bibitem{d22subfus} V. Yu. Denisov, Eur. Phys. J. A 58, 91 (2022).
		
		
		\bibitem{ds21} V. Yu. Denisov, I. Yu. Sedykh, Chinese Phys. C 45, 044106 (2021). 
		
		\bibitem{hm} S. Hofmann and G. Münzenberg. Rev. Mod. Phys. 72, 733 (2000).
		
		\bibitem{hofmann} S. Hofmann, Lect. Notes Phys. 764, 203 (2009). DOI: 10.1007/978-3-540-85839-3 6
		
		\bibitem{d23a} V. Yu. Denisov, Int. J. Mod. Phys. E 32, 2350005 (2023). 
		
		\bibitem{ahmed} Z. Ahmed, Phys. Lett. A 157, 1 (1991). 
		
		\bibitem{morse} P. M. Morse, Phys. Rev. 34, 57 (1929). 
		
		\bibitem{d23b} V. Yu. Denisov, Phys. Rev. C 107, 054618 (2023). 
		
		\bibitem{kemble} C. Kemble, Phys. Rev. 48, 549 (1935). 
		
		\bibitem{hw} D. L. Hill, J. A. Wheeler, Phys. Rev. 89, 1102 (1953). 
		
		\bibitem{d23d} V. Yu. Denisov, Phys. Rev. C (accepted for publication), arxiv 2309.14995. 
		
		\bibitem{dn} V. Yu. Denisov, W. Norenberg, Eur. Phys. J A 15, 375 (2002). 
		
		\bibitem{fl} P. Frobrich, R. Lipperheide, {\it Theory of nuclear reactions} (Clarendon Press, Oxford, 1996). 
		
		\bibitem{gk} D. H. E. Gross, H. Kalinowski, Phys. Rep. 45, 175 (1978). 
		
		\bibitem{frobrich} P. Frobrich, Phys. Rep. 116, 337 (1980). 
		
		\bibitem{bw} N. Bohr, J. A. Wheeler, Phys. Rev. 56, 426 (1939).
		
		\bibitem{strut4} M. Brack, et al., Rev. Mod. Phys. 44, 320 (1972). 
		
		\bibitem{bsfgm} W. Dilg, et al., Nucl. Phys. A 217, 269 (1973). 
		
		\bibitem{ripl3} R. Capote et al., Nucl. Data Sheets 110, 3107 (2009). 
		
		\bibitem{ist} A. V. Ignatyuk, G. N. Smirenkin, A. S. Tishin, Yad. Fiz. 21, 485 (1975) [Sov. J. Nucl. Phys. 21, 255 (1975)]. 
		
		\bibitem{mn} A. Mengoni, Y. Nakajima, J. Nucl. Sci. Tech. 31, 151 (1994). 
		
		\bibitem{vmk} D. A. Varshalovich, A. N. Moskalev, V. K. Khersonsky, {\it Quantum Theory of Angular Momentum: Irreducible Tensors, Spherical Harmonics, Vector Coupling Coefficients, 3nj Symbols} (World Scientific, Singapore, 1988).
		
		\bibitem{bm} A. Bohr, B. Mottelson, {\it Nuclear structure}, Vol. 2 (W. A. Benjamin Inc., New York, Amsterdam, 1974).
		
		\bibitem{wong68} C. Y. Wong, Nucl. Data A 4, 271 (1968).
		
		\bibitem{d2022f} V. Yu. Denisov, Eur. Phys. J. A 58, 188 (2022).
		
		\bibitem{spg} A. Sandulescu, D. N. Poenaru, W. Greiner, Fiz. Elem. Chastits At. Yadra 11, 1334 (1980) [Sov. J. Part. Nucl. 11, 528 (1980)].
		
		\bibitem{rgd} G. Royer, R. K. Gupta, V. Yu. Denisov, Nucl. Phys. 632, 275 (1998).
		
		\bibitem{mirea} M. Mirea, R. Budaca, A. Sandulescu, Ann. Phys. 380, 154 (2017).
		
		\bibitem{wzr} M. Warda, A. Zdeb, L. M. Robledo, Phys. Rev. C 98, 041602(R) (2018).
		
		\bibitem{matheson} Z. Matheson, et al., Phys. Rev. C 99, 041304(R) (2019).
		
		\bibitem{jack} J. D. Jackson, Can. J. Phys. 34, 767 (1956). 
		
		\bibitem{strut1} V. M. Strutinsky, Sov. J. Nucl. Phys. 3, 449 (1966). 
		
		\bibitem{strut2} V. M. Strutinsky, Nucl. Phys. A 95, 420 (1967). 
		
		\bibitem{strut3} V. M. Strutinsky, Nucl. Phys. A 122, 1 (1968). 
		
		\bibitem{ach} G. D. Adeev, P. A. Cherdantsev, Yad. Fiz. 18, 741 (1973) [Sov. J. Nucl. Phys. 18, 381 (1974)]. 
		
		\bibitem{bq} M. Brack, Ph. Quentin, Phys. Scripta 10 A, 163 (1974). 
		
		\bibitem{dah} M. Diebel, K. Albrecht, R. W. Hasse, Nucl. Phys. A 355, 66 (1981). 
		
		\bibitem{lpc} Z. Lojewski, V. V. Pashkevich, S. Cwiok, Nucl. Phys. A 436, 499 (1985). 
		
		\bibitem{snp} J. A. Sheikh, W. Nazarewicz, J. C. Pei, Phys. Rev. C 80, 011302 (2009). 
		
		\bibitem{pnsk} J. C. Pei et al., Nucl. Phys. A 834, 381c (2010). 
		
		\bibitem{dh} V. Yu. Denisov, S. Hofmann, Phys. Rev. C 61, 034606 (2000).
		
		\bibitem{ds18} V. Yu. Denisov, I. Yu. Sedykh, Phys. Rev. C 98, 024601 (2018). 
		
		\bibitem{ds18gg} V. Yu. Denisov, I. Yu. Sedykh, Eur. Phys. J. A 54, 231 (2018). 
		
		\bibitem{dds22} O. I. Davydovska, V. Yu. Denisov, I. Yu. Sedykh, Phys. Rev. C 105, 014620 (2022). 
		
		\bibitem{be2} B. Pritychenko, et al., At. Data Nucl. Data Tabl. 107, 1 (2016).
		
		\bibitem{be3} T. Kibedi, R. H. Spear, At. Data Nucl. Data Tabl. 80, 35 (2002).
		
		\bibitem{capb} A.V. Belozerov et al., Eur. Phys. J. A 16, 447 (2003).
		
	\end{thebibliography}
\end{document}